\documentclass[structabstract]{aa}   
\usepackage{longtable,lscape}
\usepackage{natbib}
\usepackage{graphicx}
\def\m2s2{\,m$^{2}$\,s$^{-2}$} 

\begin{document}
\title{The VLT/NaCo Large program to probe the occurrence of exoplanets and brown dwarfs in wide orbits\thanks{Based on observations collected at La Silla and Paranal Observatory, ESO (Chile)
using FEROS, HARPS, and NaCo
Based on observations collected with the CORALIE echelle spectrograph mounted on the 1.2 m 
Swiss telescope at La Silla Observatory.
Based on the All Sky Automated Survey (ASAS) photometric data.
}
\subtitle{I Sample definition and characterization}
}

\author{S. Desidera  \inst{1} 
\and E. Covino \inst{2} 
\and S. Messina \inst{3}
\and J. Carson \inst{4,5}
\and J. Hagelberg \inst{6}
\and J.E. Schlieder  \inst{5}
\and K. Biazzo \inst{3} 
\and J.M. Alcal\'a \inst{2}
\and G. Chauvin \inst{7}
\and A. Vigan \inst{8}
\and J.L. Beuzit \inst{7}
\and M. Bonavita \inst{1}
\and M. Bonnefoy \inst{5}
\and P. Delorme \inst{7}
\and V. D'Orazi \inst{9,1}
\and M. Esposito \inst{10,2}
\and M. Feldt \inst{5}
\and L. Girardi \inst{1}
\and R. Gratton \inst{1} 
\and T. Henning \inst{5}
\and A.M. Lagrange \inst{7}
\and A.C. Lanzafame \inst{3,11}
\and R. Launhardt \inst{5}
\and M. Marmier \inst{6}
\and C. Melo \inst{12}
\and M. Meyer \inst{13}
\and D. Mouillet \inst{7}
\and C. Moutou \inst{8}
\and D. Segransan \inst{6}
\and S. Udry \inst{6}
\and C.M. Zaidi \inst{7}         
}

  \institute{
INAF-Osservatorio Astronomico di Padova, Vicolo dell'Osservatorio 5,  35122 Padova, Italy 
\and INAF-Osservatorio Astronomico di Napoli, Salita Moiariello 16, 80131, Napoli, Italy
\and INAF-Osservatorio Astrofisico di Catania, Via S. Sofia 78, 95123 Catania, Italy
\and Department of Physics \& Astronomy, College of Charleston, 58 Coming St. Charleston, SC 29424, USA
\and Max-Planck-Institut f\"ur Astronomie, K\"onigstuhl 17, 69117, Heidelberg, Germany
\and Geneva Observatory, University of Geneva, Chemin des Mailettes 51, 1290 Versoix, Switzerland
\and Institut de Plan\'etologie et d'Astrophysique de Grenoble, UJF, CNRS, 414 rue de la piscine, 38400 Saint Martin d'H\`eres, France
\and Aix Marseille Universit\'e, CNRS, LAM (Laboratoire d'Astrophysique de Marseille) UMR 7326, 13388, Marseille, France
\and Department of Physics and Astronomy, Faculty of Science, Macquarie University, Sydney, NSW, 2109, Australia
\and Instituto de Astrofisica de Canarias, C/ Via Lactea, s/n, E38205 - La Laguna (Tenerife), Spain
\and Universit\'a di Catania, Dipartimento di Fisica e Astronomia, Via S. Sofia 78, 95123 Catania, Italy
\and European Southern Observatory, Casilla 19001, Santiago 19, Chile
\and Institute for Astronomy, ETH Zurich, Wolfgang-Pauli Strasse 27, 8093 Zurich, Switzerland
}

\date{}
 
\abstract
{Young, nearby stars are ideal targets to search for planets using the direct imaging technique.
 The determination of stellar parameters is crucial for the interpretation of imaging survey results, 
particularly since the luminosity of substellar objects has a strong dependence on system age.}
{We have conducted a large program with NaCo at the VLT in order to search for planets and brown dwarfs in wide orbits
around 86 stars.
A large fraction of the targets observed with NaCo were poorly investigated in the literature. We performed a study to
characterize the fundamental properties (age, distance, mass) of the stars in our sample.  To improve target age determinations, we compiled and analyzed  a complete set of age diagnostics.}
{We measured spectroscopic parameters and age diagnostics using dedicated observations acquired with FEROS and CORALIE 
spectrographs at La Silla Observatory. We also made extensive use of
archival spectroscopic data and results available in the literature. 
Additionally, we exploited photometric time-series, available in ASAS and Super-WASP archives, to derive rotation period for a large fraction of our program stars. }
{We provided updated characterization of all the targets observed in the VLT NaCo Large program, a survey designed to probe the occurrence
of exoplanets and brown dwarfs in wide orbits. The median distance and age of our program stars are 64 pc and 100 Myr,
respectively. Nearly all the stars have masses between 0.70 and 1.50~$M_{\odot}$, with a median value of 1.01~$M_{\odot}$.
The typical metallicity is close to solar, with a dispersion that is smaller than that of samples usually observed
in radial velocity surveys. Several stars are confirmed or proposed here to be members of nearby young
moving groups. Eight spectroscopic binaries are identified. 
 }
{} 

\titlerunning{The VLT/NaCo Large program to probe the occurrence of exoplanets in wide orbits}

\maketitle

\section{Introduction}
\label{s:intro}

High-angular resolution deep imaging in the near-infrared maximizes the capability 
to detect faint circumstellar bodies amidst the large brightness of the central star \citep{2010A&ARv..18..317A}.
The first detections of comoving companions with planetary mass were obtained using deep VLT/NaCo 
infrared imaging, with contrast up to 6-8 mag
at $>0.5$ arcsec separation, around very young objects such as 2MASSWJ\,1207334$-$393254 
\citep{2004A&A...425L..29C} and AB\,Pic \citep{2005A&A...438L..29C}.

After these earliest discoveries, improvements in data acquisition and data analysis techniques 
\citep[e.g. the introduction 
of angular differential imaging,][]{2006ApJ...641..556M} allowed the detection of
the first objects likely formed in protoplanetary disks, such as 
the 4-planet system orbiting HR 8799 \citep{2008Sci...322.1348M,2010Natur.468.1080M} and the giant planet 
orbiting at 8 AU from the well known star-disk system $\beta$ Pic \citep{2010Sci...329...57L}. 
In the last year, several additional discoveries of directly-imaged planets have been reported
\citep{2013ApJ...772L..15R,2013ApJ...774...11K,2014ApJ...780L..30C}.

Several surveys were also carried out, with published detection limits for hundreds of objects (e.g. 
\citealt{2005ApJ...625.1004M,2007ApJ...670.1367L,2010A&A...509A..52C}). 
These observations allowed a first estimate of the frequency of giant planets in wide orbits 
\citep{2010ApJ...717..878N}. There are exciting prospects for the detection of planets 
of lower masses (below $1 M_{J}$), and of smaller semimajor axes, 
thanks to the availability of new instrumentation, such as SPHERE
at VLT \citep{2010ASPC..430..231B}; GPI at Gemini-South \citep{2008SPIE.7015E..31M};
SCExAO at Subaru \citep{2009SPIE.7440E..20M} and LBTI \citep{2008SPIE.7013E..67H,2011SPIE.8149E...1E}.

As the luminosity of substellar objects is strongly dependent on age \citep{1997ApJ...491..856B,2003A&A...402..701B},
the determination of the mass of a detected planet, or the derivation of detection limits in terms
of planetary mass, requires a careful evaluation of the stellar age.
Other parameters, such as stellar mass, metallicity, distance to the system, multiplicity and membership 
to associations are also required for a proper interpretation of the results, in terms of both positive detections 
and statistics of the whole survey.  Such comprehensive knowledge allows for meaningful 
comparisons with the outcomes of other planet-search projects 
using different techniques \citep[e.g.][]{2005ApJ...622.1102F,2011arXiv1109.2497M} as well as outcomes 
from planet formation models \citep{2009A&A...501.1139M,2012A&A...547A.111M,1997Sci...276.1836B,2013MNRAS.433.3256V}.

The determination of stellar age is a challenging task, especially for isolated field objects 
\citep{2010ARA&A..48..581S}. A variety of indicators can be used such as isochrone fitting, lithium content, 
chromospheric and coronal emission, stellar rotation, and kinematics.
The various indicators have complementary sensitivities for stars of different spectral types and different ages. 
Furthermore, some diagnostics like those related to rotation and magnetic activity might give spuriously young ages 
as e.g., in the case of tidally-locked binaries.
Therefore, the best results are obtained from a complete examination of all possible indicators and by a careful 
and suitable combination of their results.

The NaCo Large Programme (NaCo-LP, PI. J.L.~Beuzit) is a coordinated European effort, aimed at conducting 
an homogeneous and systematic study of the occurrence of extrasolar giant planets (EGPs) 
and brown dwarfs (BDs) in wide orbit (5-500 AU) around young, nearby stars. 
First epoch observations were performed in 2009-2010 and the follow-ups for confirmation of physical association
of the candidates to their central stars were completed in March 2013.
First results include the characterization of the debris disk around HD~61005 \citep{2010A&A...524L...1B}
and the study of the astrophysical false alarm represented by the white dwarf around HIP~6177,
which mimics a substellar object at near-infrared (NIR) wavelengths \citep{2013A&A...554A..21Z}.

The goal of this paper is to provide detailed stellar characterization of the targets of the NaCo-LP, 
based on high-resolution spectroscopic data acquired for this purpose, as well as spectroscopic 
and photometric data available from public archives, complemented by information from the literature.
The companion paper (Chauvin et al.~2013, submitted) presents the deep imaging observations and their results.
Results from the NaCo-LP sample will be merged with those of other surveys of comparable depth to produce a 
comprehensive statistical analysis of the occurrence of giant planets in wide
orbits around FGK stars (Vigan et al.~2014, in preparation) as well as a complementary study of the brown dwarf population
(Reggiani et al.~2014, in preparation).

The layout of this paper is the following. 
In Sect.~\ref{s:sample} we introduce the target sample while in Sect.~\ref{s:spec_data} 
we present the spectroscopic data-set and describe the data reduction procedures;
in Sect.~\ref{s:spec} we present
the spectroscopic analysis performed to determine radial and projected rotational 
velocities, to measure lithium line 6707.8~\AA~strength, and to derive effective temperature; 
in Sect.~\ref{s:rotation} we present the analysis of photometric time series to derive
the rotation period;
in Sect.~\ref{s:activity} we examine magnetic activity diagnostics Ca II H\&K and X-ray emission; 
in Sect.~\ref{s:companions} we report the presence of possible stellar companions;
in Sect.~\ref{s:kin} we analyze the kinematic parameters supporting
the membership to known nearby associations; 
in Sect.~\ref{s:fundamental} we evaluate distances and metallicities of the targets and infer
 stellar masses and ages; in Sect.~\ref{s:conclusion} we summarize our results.
In Appendix A we report the compilation of data available in the literature for our targets;
in Appendix B we describe individual targets; in Appendix C we provide our calibration of
choromospheric emission for FEROS spectra; 
while in Appendix D the details of our search for rotation period on photometric time series
are presented.

\section{Sample selection}
\label{s:sample}

The target selection was based on an initial compilation of about one thousand young nearby stars, 
based on various literature sources and dedicated observations, assembled in preparation for
the SPHERE GTO survey \citep{2010lyot.confE..50M}.  
The youth diagnostics include high lithium content, 
Ca\,II H\&K line chromospheric emission, H$\alpha$ emission, X-ray activity, 
rotation rate and kinematics.
From that list, objects within a distance horizon of 100\,pc, younger that 200\,Myr\footnote{Three objects 
slightly older than 200 Myr were observed in the July 2010 run, because of the lack of suitable 
targets in the original target list.}, 
and a magnitude limit of approximately $R<9.5$ were selected\footnote{More precisely, the sample selection is based 
on the expected sensitivity limits for the SPHERE adaptive optics sensor, which also depends on the color of the star.}.
Furthermore, known spectroscopic and close visual (separation $<6$ arcsec) binaries were rejected, 
to  have a sample as free as possible from dynamical perturbations to planets in wide orbits
and to avoid potential problems with the quality of adaptive optics (AO) wavefront sensing, posed by the presence of a close stellar companion.
For spectroscopic binaries, there are no technical difficulties for AO observations, 
as the components are not spatially resolved, but we decided nevertheless to focus our attention
on single stars and on individual components of wide binaries. This allows us to get more statistically 
significant inferences and to have a sample more suited for comparison with the frequency of giant
planets in wide orbits as derived from radial velocity (RV) surveys. 
The study of the frequency of circumbinary planets is deferred for a separate study
(Bonavita et al.~2014, in preparation). 

The resulting list was cross-correlated with that of recent direct imaging surveys 
achieving similar sensitivities to our project. 
The 90 objects that were already observed with adequate depth were excluded 
and the final sample of the NaCo-LP is formed by about 110 targets never observed 
with deep imaging instruments.
Eighty-six of these originally selected 110 stars were observed at the telescope and will form the sample studied in this paper
(Table 9). 
The remaining 25 were not observed due to weather or technical losses.  
The previously published wide companions of the targets were also considered in our study even when 
they were not observed as part of NaCo-LP, because of the constraints they can provide 
on the parameters of the system (e.g. age, distance).
The pending statistical analysis on the frequency of planetary-mass objects
in wide orbits will be based on the whole sample of about 200 objects, which combines targets observed by NaCo-LP as 
well as observational results from the literature. 
(Vigan et al.~2014, in preparation).

Final values for the stellar parameters derived in this paper might lie outside of the original 
selection criteria due to the availability of
new observational determinations, improvements in
adopted calibrations or a better understanding of the physical nature of the systems being investigated.

\section{Spectroscopic Data }
\label{s:spec_data}

For the spectroscopic characterization of our targets, we used data acquired
with several instruments. Some of the spectra were
specifically acquired by us to characterize the targets. Other spectra were accessed through the NaCo-LP collaboration 
as well as from public archives.

\subsection{FEROS Spectra}

High-resolution spectra were collected using FEROS as part of 
preparatory observations for the SPHERE GTO survey.
Observations were performed during several runs from November 2008 to March 2010 
as part of Max Planck Institut f\"ur Astronomie (MPIA) observing time. The OBJ+SKY mode of FEROS was preferred
to avoid contamination by Th-Ar lines, as we are more interested in having clean spectra 
for stellar characterization purposes rather than attaining the highest RV precision.
Integration times were selected to obtain a S/N ratio of about 100 for stars
brighter than V$\sim$9.
In addition, we exploited FEROS spectra available from the ESO public data archive\footnote{\tt archive.eso.org}. 
These latter data were taken with various FEROS instrument set-ups and have a variety of S/N ratios.
Altogether, we considered in this paper 77 spectra for 64 stars.

The FEROS spectra extend between 3600 \AA\ and 9200 \AA\ with a resolving power 
R\,=\,48000 \citep{1999Msngr..95....8K}. 
All the data were reduced using a modified version of the FEROS-DRS pipeline 
(running under the ESO-MIDAS context FEROS), which yields the wavelength-calibrated, 
merged, normalized spectrum. The reduction steps were the following:
bias subtraction and bad-column masking; definition of the echelle orders on
flat-field frames; subtraction of the background diffuse light; order extraction;
order-by-order flat-fielding; determination of wavelength-dispersion solution 
by means of ThAr calibration-lamp exposures; 
order-by-order normalization; rebinning to a linear wavelength-scale 
with barycentric correction; and merging of the echelle orders. 

We also considered FEROS spectra of our targets obtained by \citet{2006A&A...460..695T}, reduced with the instrument pipeline 
and kindly provided to us by the SACY team. We refrained from systematically repeating the measurement of the spectroscopic
parameters, but we considered the SACY spectra for in-depth evaluation of individual cases 
(e.g. suspected spectroscopic binaries), as reported in App.~B.

\subsection{HARPS Spectra}

For our investigation, we also considered the spectra obtained with HARPS at the
ESO 3.6m telescope in La Silla \citep{2003Msngr.114...20M} as part of planet search 
surveys by the Geneva team \citep{2011arXiv1109.2497M} and the \cite{2009A&A...495..335L} RV survey of early type stars. 
We also retrieved additional HARPS spectra available in the ESO archive. 
Most of these come from the RV survey for planets around young stars by \cite{2007astro.ph..1293G},
with a few individual spectra from other projects.
Overall, we have considered HARPS spectra for 13 objects, with the number of spectra per object ranging from 1 to 77.

\subsection{CORALIE Spectra}

Observations of 11 targets were performed with the CORALIE spectrograph at the Euler 1.2m telescope
in La Silla from January 2012 to August 2013 with the main goal of checking the binarity of the objects in our
sample through RV determination and line profile evaluation.
The spectra were reduced with the instrument on--line pipeline.
Spectra of six additional stars observed as part of the CORALIE exoplanet search survey
\citep{2000A&A...356..590U} were also considered in our analysis.

\subsection{Other data}

Two spectra of HIP 46634 (kindly provided by the SOPHIE Consortium) obtained with the 
SOPHIE spectrograph at the Haute Provence 1.93m telescope
were considered to derive the star's activity level.
We also analyzed the CES spectra of the components of HD 16699 A and B
obtained by \citet{1999A&AS..138...87C} (kindly provided by L.~Pastori) to understand the origin
of discrepancies between results from the literature and results from our own analysis, 
which turned out to be due to an exchange of binary system components (see App.~B for details).

\section{Determination of spectroscopic parameters }
\label{s:spec}

We exploited the FEROS, HARPS and CORALIE spectra to measure several
spectroscopic parameters useful for stellar characterization and age determination.
Our determinations (i.e., heliocentric RV, $v \sin i$, effective temperature, 
lithium equivalent width, S-index and $\log R_{HK}$) are listed in Tables \ref{tab:spec_feros},
\ref{t:coralie}, \ref{t:coralie_new}, and \ref{t:harps3} for the different datasets
and the procedure to measure them are described below.
The final adopted values including literature measurements into account are listed in Table
\ref{t:kinparam} for RV and Table 11 
for the other spectroscopic parameters.

\subsection{Projected rotational velocity and radial velocity  }

\subsubsection{FEROS}

The heliocentric RV and the projected rotational velocity ($v \sin i$) 
were determined by cross-correlating the object spectra with  low-$v \sin i$ template 
spectra (HD\,80170, K5~III-IV, RV $\sim$ 0\,km/s) 
obtained with the same instrument and reduced in the same way. 
The normalized target spectra were preliminarily rebinned to a logarithmic wavelength scale 
and split into six wavelength-ranges 
that were free of emission lines and of telluric absorptions \citep[cf.~][]{2007MNRAS.376.1805E}.
A Gaussian profile was fitted to the peak of the cross-correlation function (CCF), 
computed in each of the six distinct spectral ranges. The resulting RV values were 
averaged with appropriate weights \citep{2009A&A...501.1013S}. 
The projected rotational velocity $v \sin i$ was derived from the Full Width Half Maximum
(FWHM) of 
the cross-correlation peak in each of the six ranges above.
The relation FWHM-$v \sin i$ was obtained by convolving the template spectra
with rotational profiles of $v \sin i$ from 1 to 100 km/s.
Our measurements of heliocentric RV and $v \sin i$
are listed in Table \ref{tab:spec_feros}.

\subsubsection{HARPS and CORALIE}

RV and FWHM of CCF are directly delivered by the instrument pipelines, which
are based on the Weighted CCF technique \citep{2002A&A...388..632P}.
For moderately fast rotators, the correlation was performed by increasing the width of the
region used in the CCF fit according to the line width.
For some very fast rotators ($v \sin i \sim 50$ km/s) observed with CORALIE, we simply measured 
the positions of a few very strong lines in the spectra to derive the RVs.
For fast-rotating A and F type stars observed with HARPS, we also obtained differential 
RV using the SAFIR code \citep{2005A&A...443..337G}, 
which is optimized for this kind of analysis.
The transformation between FWHM and $v \sin i$ was performed as described in \citet{2002A&A...392..215S}.
Spectroscopic parameters derived from CORALIE and HARPS spectra are listed in Tables \ref{t:coralie}, 
\ref{t:coralie_new}, and \ref{t:harps3}.

\subsection{Lithium Equivalent Width}
\label{s:LiEW}

The equivalent width (EW) measurements of the Li 6707.8~\AA\ resonance line were obtained 
by Gaussian fit and the results are reported in Tables~\ref{tab:spec_feros} and \ref{t:harps3}
for FEROS and HARPS spectra respectively.
The listed EW does not include corrections for blends with close lines, which occur for stars with
significant rotation.
Typical measurement errors are of the order of $\le$ 5~m\AA\ while for fast rotators they are up to $\sim$~50~m\AA.

\subsection{T$_{\rm eff}$ determinations}
\label{s:teff}

\begin{figure}[h]
\includegraphics[width=8cm]{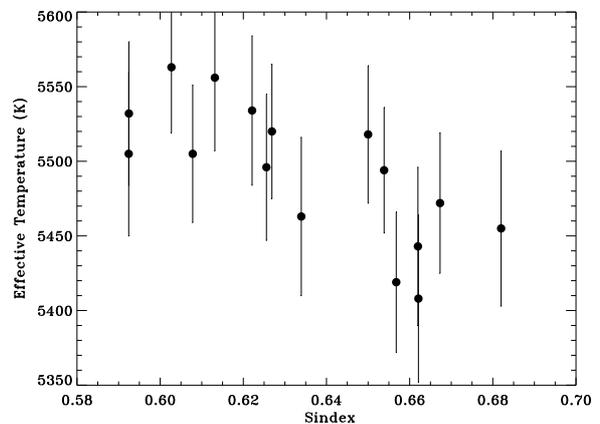}
\caption{Variations in effective temperature as measured by applying ARES on HARPS spectra of HIP 32235. The temperature is displayed as a function of S index chromospheric emission.}
\label{f:sindex_teff}
\end{figure}

The determinations of the spectroscopic effective temperature in Tables~\ref{tab:spec_feros} and \ref{t:harps3} for FEROS and
HARPS spectra respectively were obtained 
through the method based on EW ratios of spectral absorption lines.  These were acquired by means of 
the ARES\footnote{\tt http://www.astro.up.pt/\~ sousasag/ares/} automatic code 
\citep{2007A&A...469..783S}, using the calibration for FGK dwarf stars by 
\cite{2010A&A...512A..13S}.
The quoted errors are the internal ones and do not include calibration uncertainties.
When multiple spectra are available, we list the mean values and the error of the mean, which
are below 10~K, well below systematics.
For rotational velocity above $\approx 20$\,km/s, the number of measured line-ratios drops 
drastically with correspondingly larger errors, due to increased line blending. The temperature determinations become less reliable for $v\sin i$ larger than 30\,km/s 
\citep{2011A&A...529A..54D}.
For five stars with both FEROS and HARPS spectra available, the spectroscopic temperatures resulting from
the two instruments are fully consistent (mean difference $\Delta T_{\rm eff (FEROS-HARPS)}= -4$~K with an r.m.s. dispersion of 34~K).
Four stars have more than 10 spectra with HARPS, enabling an estimate of the intrinsic variability of effective
temperature, which results in 20-50~K r.m.s. In some cases, as shown in Fig.~\ref{f:sindex_teff} for HIP 32235,
temperature variations show significant correlations with chromospheric activity. 
This is due to large spots covering the stellar photosphere when the star is active.
This issue will be further examined in a forthcoming study (Messina et al.~2014; in preparation).
Larger temperature variations ($\ge$ 100~K) are expected to be present for the stars in our sample with very high 
levels of magnetic activity and photometric variability 
 \citep[see e.g. ][]{2005A&A...432..647F}.

\begin{figure}[h]
\includegraphics[width=8cm]{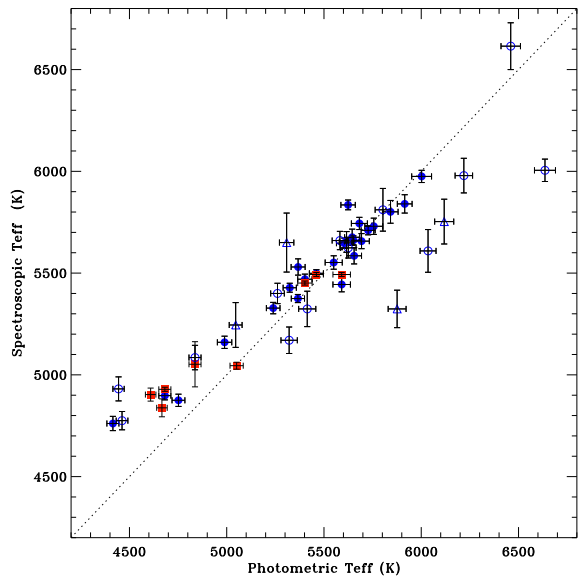}
\caption{Spectroscopic vs photometric effective temperatures. Blue filled circles: stars with FEROS spectra and $v \sin i < 15$ km/s; 
blue empty circles: stars with FEROS spectra and $15 < v \sin i < 30$ km/s; blue empty triangles: stars with FEROS spectra and $v \sin i > 30$ km/s;
red filled squares: stars with HARPS spectra.}
\label{f:teff}
\end{figure}

Photometric temperatures were obtained from B-V, V-I and V-K colors using the
calibrations by \citet{2010A&A...512A..54C} (adopting metallicity from the literature when available 
[see Sect.~\ref{s:metal}] and assuming solar metallicity otherwise).
The comparison between photometric and spectroscopic temperatures (Fig.~\ref{f:teff}) shows that
stars with faster rotation show larger dispersion, as expected for an increasing role of line blending.
Large scatter is also observed for F-type stars. This is likely due to the moderately fast rotation of most of these
stars in our sample and also due to an effective temperature which is close to the limit of the \cite{2010A&A...512A..13S}
calibration or even beyond it in some cases. The available spectral types further support the 
photometric temperatures in these discrepant cases.

When considering only stars with $v \sin i < 15$ km/s and FEROS spectra (24 stars),  
the mean difference (photometric minus spectroscopic temperatures) is $-52$~K, with an rms 
dispersion of 111~K.
A systematic trend, with spectroscopic temperature being warmer, appears to be present below 5300 K, while
results for solar-type stars agree very well (mean difference $-2$~K with r.m.s. dispersion of 79~K for 16 stars
with $v \sin i < 15$ km/s and $T_{\rm eff}>5400$~K).

Small amounts of interstellar or circumstellar absorption might be present in some individual cases, therefore biasing
the photometric temperature, but this is not a general explanation for the systematic offset observed for the coolest stars.
A similar discrepancy in effective temperature for cool stars was observed by \citet{2008A&A...487..373S},
when comparing their spectroscopic temperatures with those based on the Infrared Flux Method as 
implemented by \citet{2006MNRAS.373...13C}.
This was interpreted by \citet{2008A&A...487..373S} as being due to the increasing errors in the spectroscopic 
analysis at cooler temperatures, as the analysis is differential with respect to the Sun.

Considering the additional uncertainties that are specific to active young stars, such as photometric and spectroscopic
variability, the effects of magnetic activity on the stellar atmosphere structure, and a slightly lower gravity for the 
youngest stars, we consider the agreement between photometric and spectroscopic temperature satisfactory.
A full understanding of the contribution of these source of errors on spectroscopic and photometric temperatures 
is beyond the scope of this paper.

\begin{figure}[h]
\includegraphics[width=8cm]{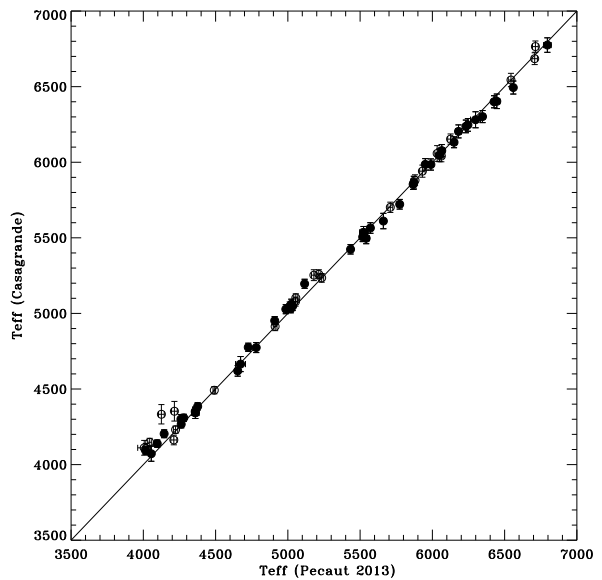}
\caption{Comparison between photometric effective temperatures obtained using \citet{2010A&A...512A..54C} calibration for
B-V, V-I and V-K color to the effective temperatures of the 5-30 Myr stars considered in \citet{2013ApJS..208....9P}
(their Table 9, BT-Settl models). Close visual binaries and spectroscopic binaries are plotted as open circles; 
single stars and components of wide binaries as filled circles.}
\label{f:pecaut}
\end{figure}

The temperature calibration by  \citet{2010A&A...512A..54C} is mostly based on old stars. Its application to 
our sample  should then be checked. 
\citet{2013ApJS..208....9P} have recently developed an effective temperature scale which is optimized for 
5-30 Myr old stars, then applicable to a significant fraction of our targets.
We considered the stars in their Table 9 with effective temperature between 4000 to 7000 K and compared to the 
effective temperatures derived from \citet{2010A&A...512A..54C} calibration using B-V, V-I and V-K colors. 
The results are shown in Fig.~\ref{f:pecaut}. 
It results that there is no significant offset between the two temperature scales (mean difference of +3~K 
and median -3 K for single stars), with a scatter of 34~K, with a possible trend 
vs temperature.
We also compared the effective temperatures for Sco-Cen members from \citet{2012ApJ...746..154P} finding comparable results.

Therefore, the differences of effective temperatures from colors (at least for the combination of colors we used) 
are quite small. This is consistent with the difference of 40 K with respect to \citet{2011A&A...530A.138C} for six stars in 
common quoted in \citet{2013ApJS..208....9P}. The difference is instead much larger for temperatures derived 
from spectral types. 

There is another concern in the application of \citet{2013ApJS..208....9P} relations to our sample stars: 
if the origin of the differences of temperature scale of main sequence and pre-main sequence stars is linked to 
the lower stellar gravity at the youngest age, its use should be limited to stars younger than about 30 Myr 
(and possibly even younger for F-G stars). 
If instead is due to the effects of phenomena related to the magnetic activity, as plages and star spots 
(cfr. their sect. 5), then the calibration should be applied also to stars up to 100-120 Myr old, which have 
similar activity levels of $\beta$ Pic and Tuc-Hor MGs members, and to old tidally-locked binaries. 

Henceforth, we will adopt the photometric temperatures using the \citet{2010A&A...512A..54C} scale, as these are homogeneously 
available for all targets and are likely more accurate than the spectroscopic temperature for cool stars and objects 
with high $v \sin i$, as discussed above.

\section{Rotation periods}
\label{s:rotation}

\subsection{Data}
In the present study, we use photometric time-series data  to investigate the variability of our targets, 
and specifically, to infer their rotation periods, brightest (least spotted) magnitudes, and levels of magnetic
activity at photospheric level.\\
\indent
The photometric data were retrieved mainly from three public archives: the All Sky Automated Survey 
\citep[ASAS\footnote{\tt www.astrouw.edu.pl/asas/},][]{1997AcA....47..467P,2002AcA....52..397P}, 
the Wide Angle Search for Planets
\citep[SuperWASP\footnote{\tt www.wasp.le.ac.uk/public/},][]{2006PASP..118.1407P,2010A&A...520L..10B}, and the  Hipparcos
catalogue\footnote{\tt vizier.u-strasbg.fr/viz-bin/VizieR?-source=I\%2F239} \citep{1997A&A...323L..49P}.

\subsubsection{The ASAS photometry}

The ASAS project monitors all stars brighter than V=14 mag at declinations $\delta <$ +28$^{\circ}$, 
with a typical sampling of 2 days. Since the ASAS archive contains data from 1997 to date,
the time-series it provides are particularly suitable for investigating variability 
over different time scales, specifically for probing the activity-induced rotational modulation.
ASAS provides aperture photometry through five apertures. For each star in our sample, we adopted
the aperture that gave the highest photometric precision (i.e., the minimum average magnitude uncertainty), 
which is the range 0.02-0.03 mag for target stars under analysis.

\subsubsection{The SuperWASP photometry}
The public SuperWASP light curves are available for a time interval
from 2004 to 2008 for both the northern and southern observatories. 
Although temporal and spatial coverage are irregular, SuperWASP stars at the most favourable sky declinations have observation sequences up to nine hours long with a sampling of about ten minutes, making the 
data more sensitive than that of ASAS for revealing periods $<$ 1\,d. 
Our analysis is based on the processed flux measurements obtained through application of the SYSREM
algorithm \citep{2005MNRAS.356.1466T}. 
The typical photometric accuracy for the targets under study is of the order of 0.01-0.02 mag.
Owing to differences with respect to the standard Johnson V-band ASAS photometry, 
we decided to analyse the SuperWASP data-set independently, without merging it with ASAS data.
SuperWASP observations from 2004 to 2005 were collected without a filter, with the spectral transmission 
being defined by the optics, detector, and atmosphere.
In the subsequent years, observations were collected through a wide-band filter (400-700 nm).

\subsubsection{The Hipparcos photometry}
For a limited number of targets, Hipparcos photometry was also available. 
However, owing to very irregular sampling and low photometric precision, we were able to derive 
rotation periods for only a handful of these stars.

\subsubsection{Data from the literature}
For all of our targets, we searched the literature for published period determinations. We focused on published measurements that used the same time-series as our own measurements, i.e. the ASAS Catalogue of Variable Stars (ACVS), 
and the search for variable stars from SuperWASP \citep{2007A&A...467..785N}. 
We also retrieved 17 rotation periods from Messina et al. (2010, 2011), and 22 rotation periods from other sources 
(see Appendix B). For most of the published periods, we were able to confirm the values with our own analyses.
We relied solely on literature values only for HIP 46637 and HIP 71908.

\subsection{Photometry rotation period search}

We used the Lomb-Scargle periodogram method \citep{1992nrfa.book.....P,1982ApJ...263..835S,1986ApJ...302..757H} 
and the Clean algorithm \citep{1987AJ.....93..968R}
to search for the rotation periods. As demonstrated in \cite{2010A&A...520A..15M,2011A&A...532A..10M}, 
the most accurate approach to infer the rotation period of late-type stars from the light rotational modulation 
is to perform the period search on the whole time-series and also on associated time segments when a sufficiently long 
time-series is available (generally longer than one observing season). 
This approach not only speeds up the period detection process, but also prevents other phenomena, e.g. the 
growth and decay of active regions, whose typical timescales are longer than the period of the rotational modulation, 
from distorting period determinations.

For the present investigation, we had at our disposal ASAS time-series  up to 10\,yr long and SuperWASP time-series  
up to 4\,yr long. 
We therefore segmented both datasets into sub-time-series, each corresponding typically to one yearly 
observation season. This approach was unsuitable for Hipparcos data, owing to low and uneven sampling; 
in that case the analysis was only performed on the whole time-series.

A detailed description of the period search method is given in \cite{2010A&A...520A..15M}. Here we briefly recall 
that the false alarm probability (FAP) associated with the period determination is computed by using Monte Carlo bootstrap methods on 1000 simulated light curves; the FAP is the probability that a peak 
of a given height in the periodogram is caused simply by statistical variations, i.e., due to Gaussian noise.
 For each complete time-series and
for each time segment, a corresponding FAP was computed. Only periods with FAP $\le$ 1\% were retained for 
subsequent analysis. The uncertainty associated with each period was computed from the FWHM of the main power peak in the window function, following the methods of \cite{2004A&A...417..557L}.

\subsection{Period results}

The results of our period search are summarized in Table 11, where we list for each target the
rotation period, its uncertainty, and a flag on the confidence level of the determination.
On a sample of 99 targets, we were able to determine the rotation period for 70 objects. Specifically, 27 periods are
determined in this study for the first time, while the remaining 43 are retrieved from the literature \citep[17 out of 43 
from ][]{2010A&A...520A..15M,2011A&A...532A..10M}. For 25 out of these 43 targets, we were able to retrieve either ASAS or SuperWASP 
photometry, carry out our independent analysis, and confirm the literature values.
As described in \cite{2011A&A...532A..10M}, periods derived in at least five segments, or in less than five but 
with an independent confirmation from the literature, and that are consistent with the $v \sin i$, will be referred to 
as {\it confirmed periods} (C). 
Periods derived from less than five time segments, but are consistent with $v \sin i$, and that produce mean sinusoidal fit residuals 
smaller than the light-curve amplitude in at least five segments, will be referred to as {\it likely 
periods} (L). Periods derived from less than five segments that produce mean sinusoidal fit residuals smaller than 
the light-curve amplitude in less than five segments, or with only 1-2 determinations in the literature, 
or that are inconsistent with v$\sin i$, will be referred to as {\it uncertain periods} (U).
We have 40 \it confirmed \rm periods, 14 \it likely \rm periods, 14 \it uncertain\rm, periods and 2 taken from 
the literature alone.\\
\indent
We could not determine the rotation periods for 29 out of 99 stars in our sample.
Sixteen of the 29 stars with unknown periods have spectral types earlier than F8. These stars
are expected to have very shallow external convection zones and, consequently, very low activity levels. 
Therefore, any eventual rotational modulation is unlikely to be detected by analysis of ground-based data\footnote{An exception 
is represented by HD 32981 (F8V). F-type stars occasionally exhibit spots sufficiently large enough 
to induce light modulation with amplitude of a few hundredths of a magnitude \citep[see e.g.][]{2011A&A...533A..30G}, 
thus enabling detection from ground-based observatories}. 
In the case of HIP 71908 (F1Vp), a 
low-amplitude ($\sim$ 5 mmag) modulation was detected by space-based observations \citep{2009MNRAS.396.1189B}. 
The remaining 13 stars, although of late spectral type,
exhibited non-periodic variability. In several cases, they are members of binary systems which are
not spatially resolved in ASAS and Super-WASP images; ASAS and Super-WASP have pixel scales of 15.5 and 13.7 arcsec respectively
\citep{2001ASPC..246...53P}.
Individual cases are discussed in Appendix \ref{a:notes}.
A summary of the results of our period search for the individual seasons and additional details 
are provided in Appendix \ref{a:rot}.

\section{Activity diagnostics}
\label{s:activity}

Magnetic activity and related phenomena are known to depend on rotation period
and also on stellar age for stars of spectral types from mid-F to mid-M. 
In this section we considered indicators of chromospheric and coronal activity.

\subsection{Ca\,II H\&K emission}
\label{s:HK}

Chromospheric emission in CaII H\&K lines was determined by measuring the $S$-index on 
FEROS spectra as described in \citet{2006A&A...454..553D}.
Details on the calibration of our instrumental $S$-index into the standard Mt.~Wilson scale
are provided in Appendix \ref{a:sindex}.
The procedures adopted for HARPS and CORALIE are described in \citet{2011arXiv1107.5325L} and \cite{marmier} respectively.
The activity measurements obtained with the three instruments are presented in Tables
\ref{tab:spec_feros}, \ref{t:coralie}, \ref{t:harps3}.
Most of the measurements obtained with FEROS are single-epoch while those from the CORALIE
planet search and some of those from HARPS archive are multi-epoch over months or years, then partially averaging
the variability of the chromospheric emission on timescales of the rotation period and the activity cycle of the star. 
The activity measurement of HIP 46634 was obtained by the SOPHIE Consortium following the
procedures described in \citet{2010A&A...523A..88B}.
When available, we also considered additional data from the literature (Appendix A).

\subsection{X-ray emission}
\label{s:X-ray}

The target list was cross-correlated with the ROSAT all-sky survey (RASS) catalog 
\citep{1999A&A...349..389V,2000IAUC.7432R...1V}, adopting a matching radius of 30 arcsec.
X-ray fluxes were derived using the calibration by \cite{1999A&AS..135..319H}.
For wide multiple systems that were unresolved by ROSAT (see Table \ref{t:widebin}), a suitable correction 
to the flux was included on the basis on the bolometric flux ratio of the components.
In a couple of cases (HD 16699 A and B; HIP 72399/72400), the availability of XMM observations at higher 
spatial resolution enabled us to identify the main source of the X-ray emission in a binary system.
For the systems HIP 35564 and HIP 25434/6, whose primaries are early-F stars, we assumed that the whole
X-ray emission comes from the secondary. However, this assumption is uncertain, as both primaries are
spectroscopic binaries and then the additional late-type star might contribute significantly to the integrated
X-ray flux. 

\begin{figure}[h]
\includegraphics[width=8cm]{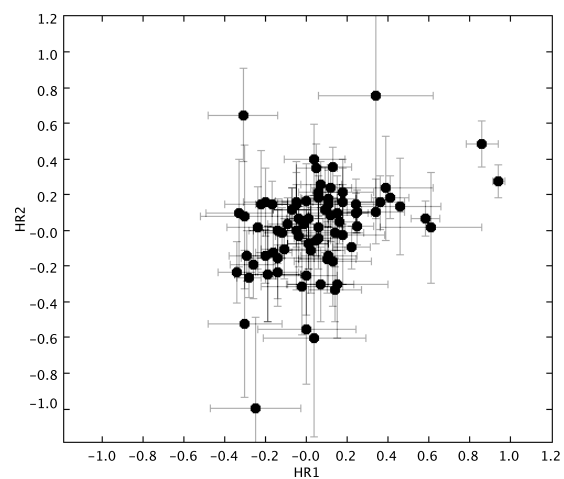}
\caption{Hardness-ratios diagram for the NaCo-LP targets.
The two sources with HR1 closest to 1 are TYC\,7743-1091-1 and TYC\,6781-0415-1.}
\label{f:HR1_HR2}
\end{figure}

The hardness-ratios 
HR1=(B-A)/(B+A) and HR2=(D-C)/(D+C), where 
A, B, C and D are the integrated counts in the energy range 0.11-0.41~keV, 0.52-2.01~keV,
 0.52-0.90~keV, and 0.91-2.01~keV, respectively, are plotted in Fig.~\ref{f:HR1_HR2}. 
The two stars TYC\,7743-1091-1 and TYC\,6781-0415-1 are found to be hard X-ray sources with HR1 close to 1.
Three stars, HIP 71933, TYC\,6818-1336-1 and TYC\,5736-0649-1, are moderately hard sources (HR1 $\sim$ 0.4 - 0.6).
The hardness-ratios can be used to derive clues on stellar age, as the members of young moving groups 10-30 Myr old
typically have soft X-ray emission while hard emission is observed for classical T Tauri stars 
\citep{1995A&A...297..391N,2003ApJ...585..878K}. 
Hard X-ray emission might in principle be due to interstellar
or circumstellar absorption \citep{2003ApJ...585..878K}, but the aforementioned hard X-ray sources likely exhibit emission due to flare activity
\citep[see][and references therein]{2000A&A...356..949S}.

For stars which are not detected by ROSAT, upper limits on X-ray luminosity were derived considering
the integration time of the nearest X-ray source in the ROSAT Faint Source Catalog.

To estimate the ratio of X-ray to bolometric luminosity $R_{\rm X}= \log (L_{X} / L_{bol})$, we derived  
the luminosities of the targets using the apparent V magnitude, 
the bolometric correction from \cite{1996ApJ...469..355F} and the distances 
derived in Sect.~\ref{s:dist}.
We converted the bolometric magnitude, defined as $M_{\rm bol}=V + BC_V + 5\log{\pi} + 5$,
to $L/L_{\odot}$ using $M_{{\rm bol} \odot} = 4.73$, as recommended by \cite{2010AJ....140.1158T}.
The interstellar extinction was assumed to be negligible since all targets are relatively nearby,
except for a few individual cases where there existed evidence of modest reddening 
(see Sect.~\ref{s:dist} and App.~B).
X-ray luminosities and $\log (L_{X} / L_{bol})$ are listed in Table 11. 

\subsection{Correlations of activity indicators and rotation}
\label{s:corr}

\begin{figure}[h]
\includegraphics[width=8cm]{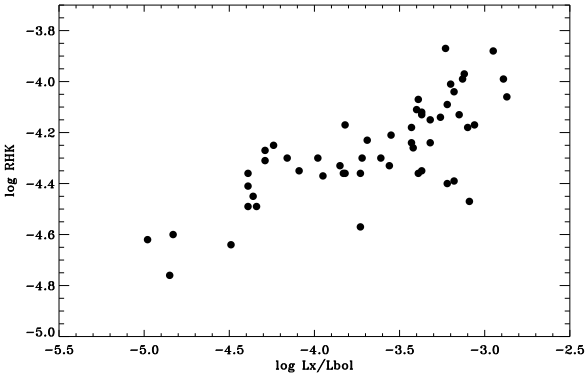}
\caption{$\log R_{HK}$ vs $\log L_{X}/L_{bol}$ for the stars in our sample} 
\label{f:xhk}
\end{figure}

Coronal and chromospheric emissions are in general expected to show correlation with the
rotation period, as their origin is tied to the rotation itself \citep[see, e.g., ][]{2003A&A...397..147P}.
On the other hand, at very fast rotational velocities and activity levels, 
correlations are no longer observed, as the star enters the so-called ''saturated regime''.
To evaluate the status of our targets and also to check for peculiar discrepant cases
that might distort the observed quantities (e.g.~aliases of the true rotation period),
we evaluate here the correlation between activity and rotation.
Fig.~\ref{f:xhk} shows the correlation between $\log R_{HK}$ vs $\log L_{X}/L_{bol}$ for the stars in our sample.
A similar slope is observed for the whole range of emission levels.

\begin{figure}[h]
\includegraphics[width=8cm]{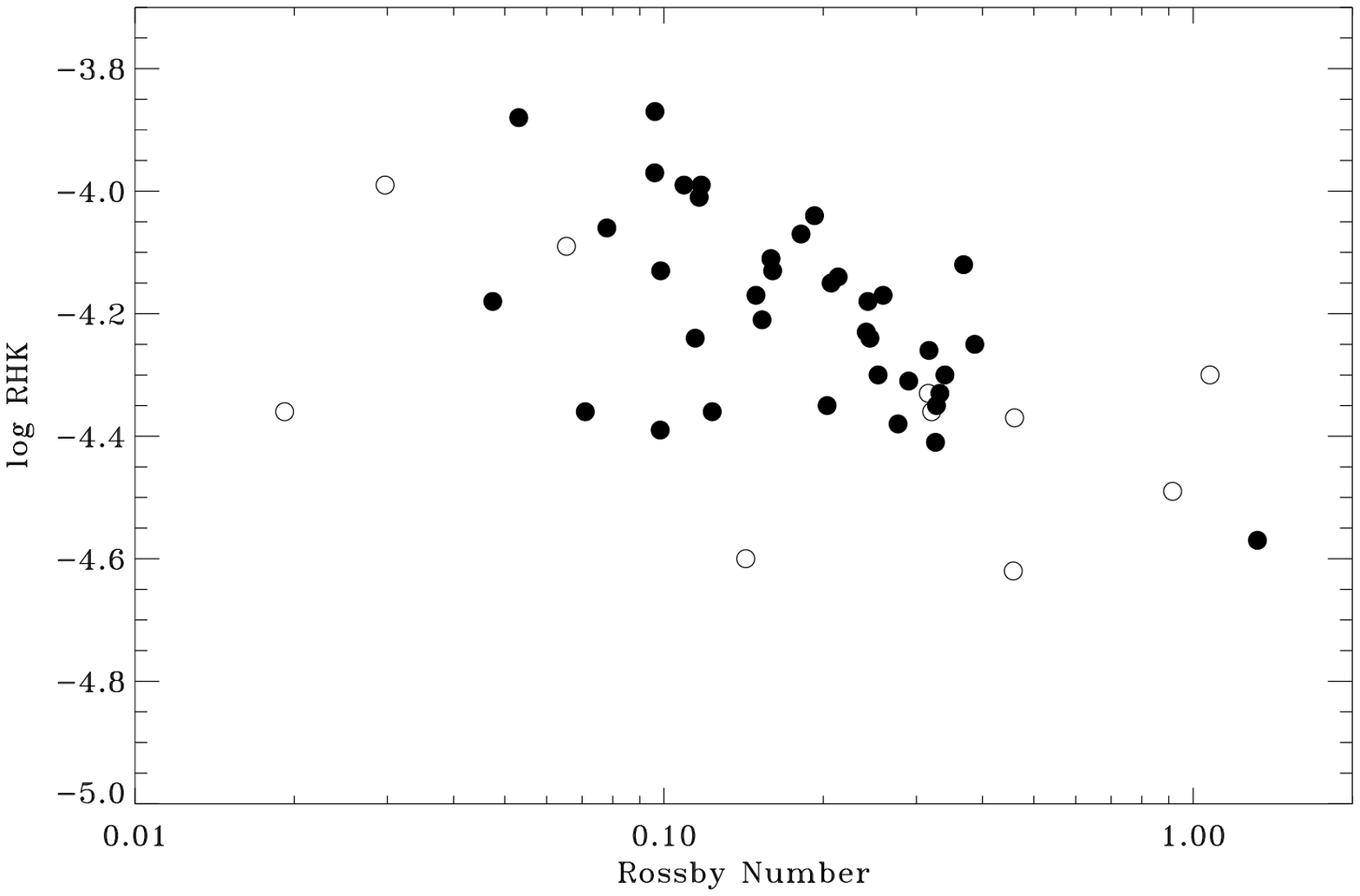}
\includegraphics[width=8cm]{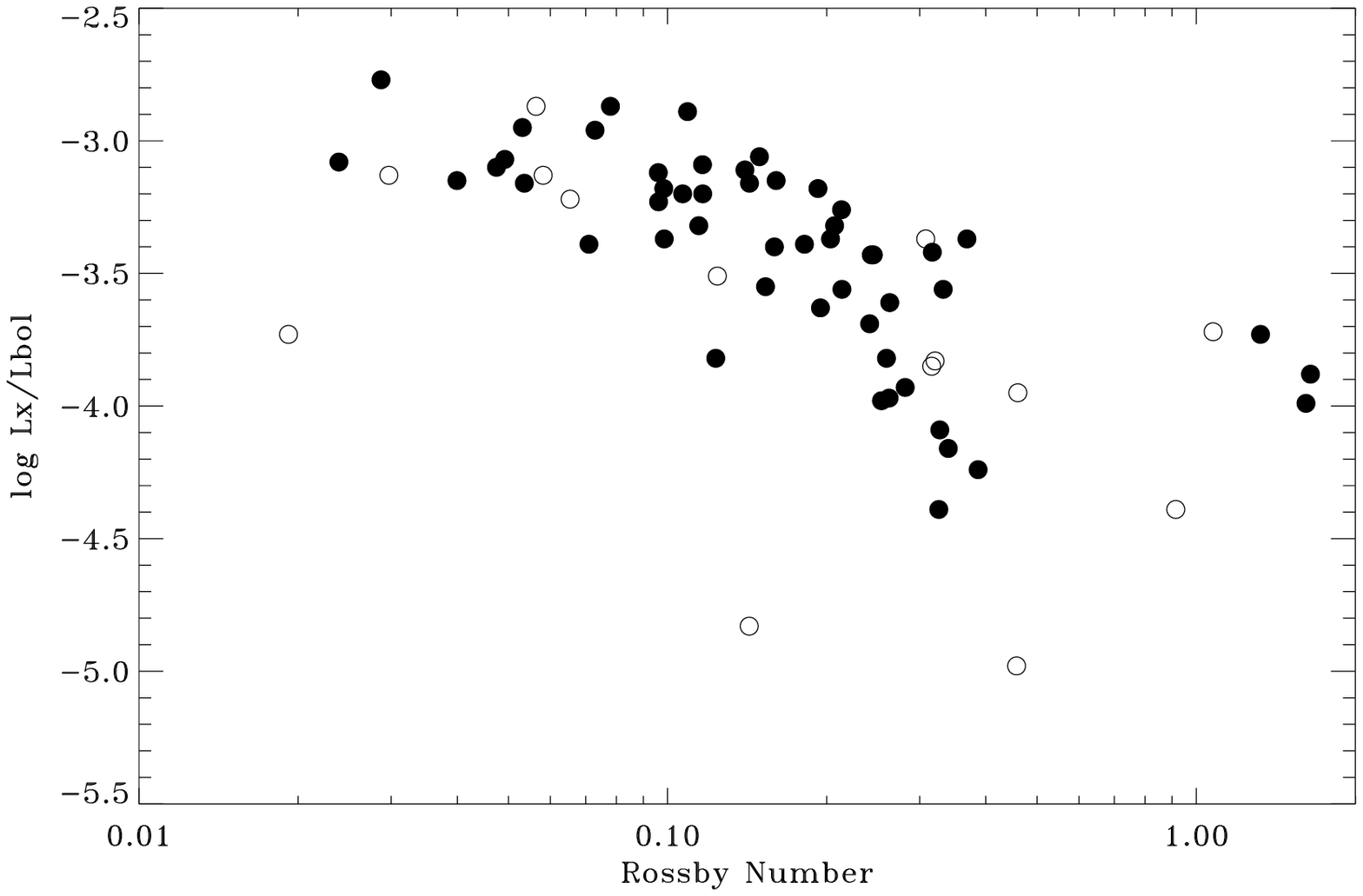}
\caption{$\log R_{HK}$ and $\log L_{X}/L_{bol}$ (upper and lower panels respectively) vs Rossby number
for the stars in our sample. Filled circles represent stars with {\it confirmed} or {\it lilely} rotation period.
Empty circles represent stars with {\it uncertain} rotation period.} 
\label{f:rossby}
\end{figure}

Fig.~\ref{f:rossby} shows the Rossby number \citep[used here to take the dependence of the rotation period
on color into account, ][]{1984ApJ...279..763N} vs $\log R_{HK}$ and $\log L_{X}/L_{bol}$.
The change in slope due to the saturation effects is clearly seen in the $\log L_{X}/L_{bol}$ plot.

\begin{figure}[h]
\includegraphics[width=8cm]{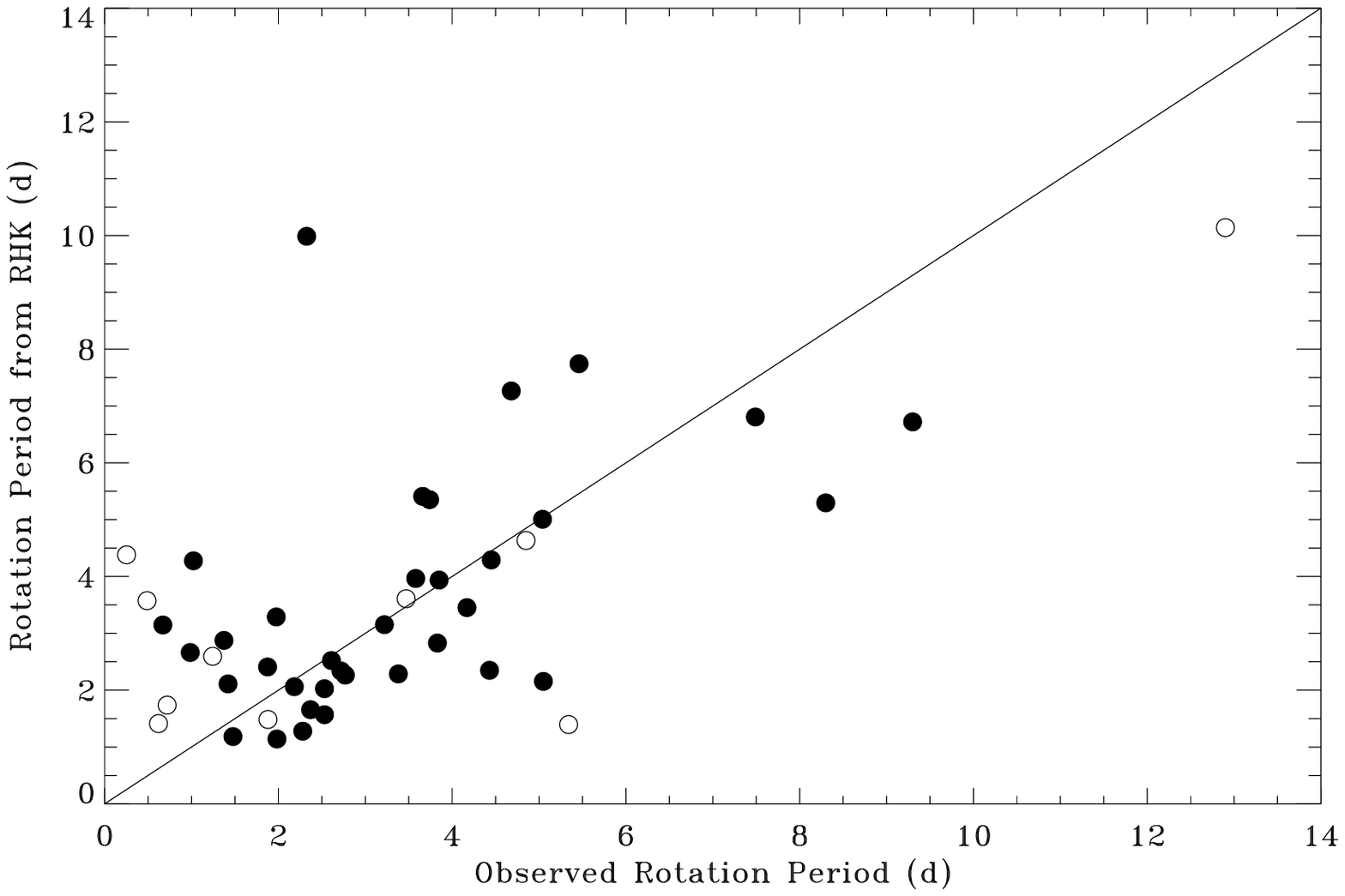}
\includegraphics[width=8cm]{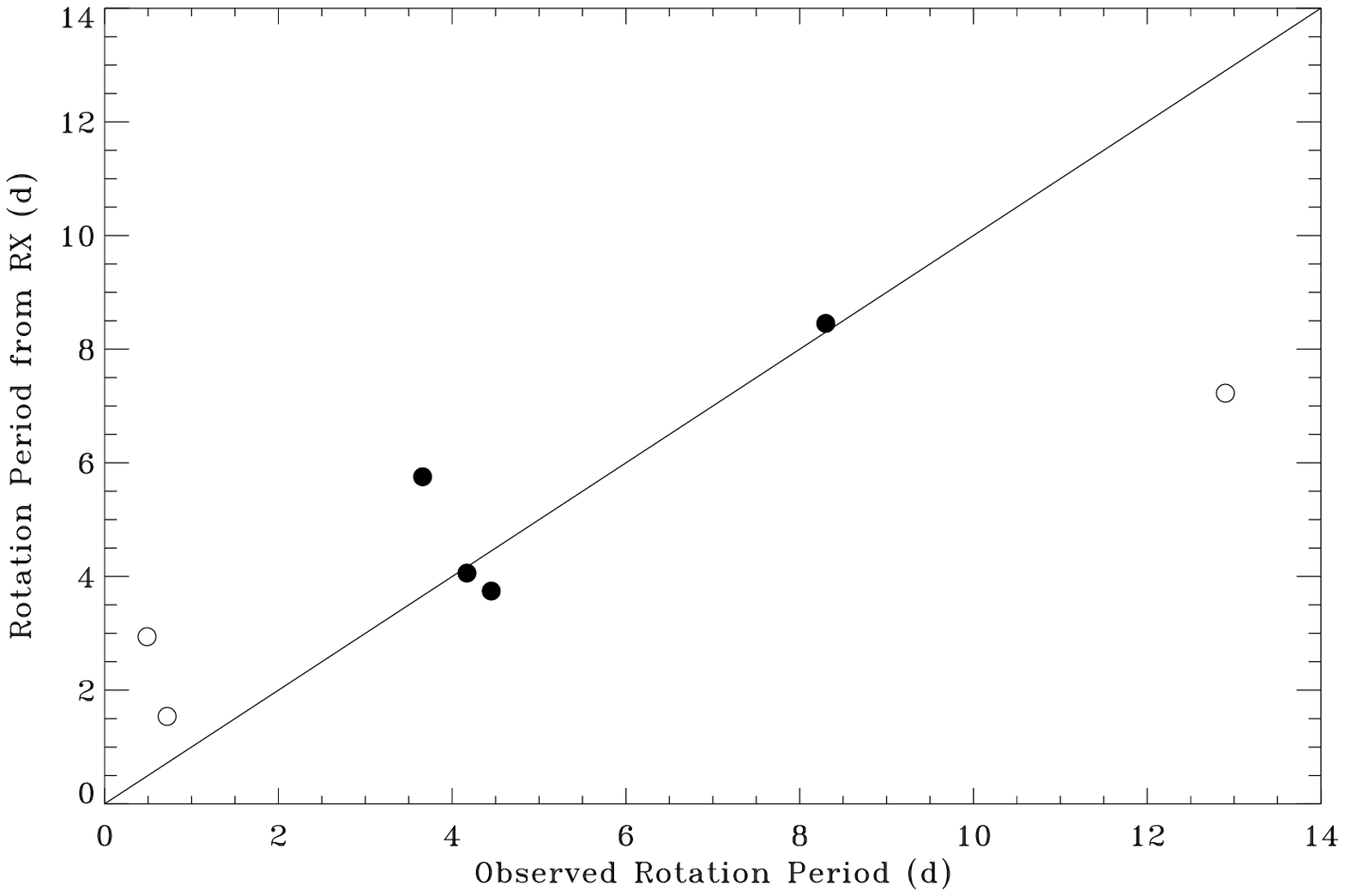}
\caption{Observed rotation period vs the expected rotation period from the value of $\log R_{HK}$ (upper panel)
and $R_{X}$ (lower panel),
derived using the Mamajek \& Hillenbrand (2008) calibrations. Filled circles represent stars with {\it confirmed} or {\it lilely} rotation period.
Empty circles represent stars with {\it uncertain} rotation period.
The outlier in the upper panel is represented
by HIP 79958. As the measured rotation period appears to be robust, the discrepancy may be due to poor
sampling of H\&K variability or due to calibration issues resulting from the low temperature of the star. See App.~B for
further details.} 
\label{f:prot_calc}
\end{figure}

We also compared the measured rotation period with the expected one based on the Ca II H\&K
and X-ray coronal emission and calibrations by \citet{2008ApJ...687.1264M}. In the first case (Fig.~\ref{f:prot_calc}; upper panel),
the correlation between predicted and observed rotation period is low for stars with fast rotation. This effect was also noted
by \citet{2008ApJ...687.1264M}, who propose that this may be explained by 1) a larger intrinsic variability compared to stars
with slower rotation rates, 2) a sparseness of activity measurements, and 3) a spread in convective turnover times due 
to a mixing of main sequence and pre-main sequence stars.
One should also consider that a proper calibration of the S index into the standard Mt.~Wilson scale is challenging at high
activity levels, due to the paucity of suitable calibrators and the presence of intrinsic variability of the chromospheric emission.
Therefore, mixing results from independent sources (as done here) might result in an increased scatter.
Furthermore, for projected rotational velocities larger than about 40~km/s, the emission of the Ca II H\&K cores 
is spread out beyond
the $\sim 1$ \AA~window typically used for the measurement of the S-Index \citep{2006A&A...454..553D}.
X-ray derived periods (Fig.~\ref{f:prot_calc}; lower panel) can be derived only for stars outside the saturated regime
\citep{2008ApJ...687.1264M}. The plot shows the expected correlation for unsaturated stars. 

\subsection{Ultraviolet flux}
\label{s:uv}

Ultraviolet NUV and FUV magnitudes from the GALEX satellite \citep{2005ApJ...619L...1M}
were checked to identify or confirm WD companions
\citep[see][for the case of HIP 6177]{2013A&A...554A..21Z} and to
indirectly evaluate the level of magnetic activity when direct
measurements were not available.
GALEX magnitudes were available for 34 objects in our sample.
With the exception of HIP 6177, we did not identify additional objects
with peculiar FUV and NUV fluxes.

\section{Multiplicity, close companions and discs}
\label{s:companions}

We summarize here the evidence for additional stellar companions either at wide or close separations.
A complete census of stellar multiplicity is important for several reasons.
For instance, the presence of a close companion can bias the derivation of stellar parameters, as is the case for photometric distance. A wide companion may also be used
to provide additional diagnostics to constrain the properties of the target.
More generally, knowledge of target environment is important for the proper interpretation of the astrophysical results
of our NaCo imaging program, and for comparisons with other surveys. 
Beside stellar multiplicity, the presence of disks is also relevant in this context.
There are four stars with resolved debris disks in our sample: HIP 11360, HD 61005, HIP 76829, and HIP 99273.
They are discussed in App.~B.

\subsection{Visual binaries}
\label{s:visbin}

Several new close binaries were discovered as part of the NaCo-LP.
The details of their discovery are reported in the companion paper presenting the direct imaging results 
(Chauvin et al.~2013).
We summarize here their properties, as the binarity has to be taken into account to correct photometric
distances and temperatures; binarity is also relevant for a proper interpretation of the age indicators.
Three close binaries were previously known: HIP 108422 \citep[][physical association confirmed by our NaCo observations]{2003A&A...404..157C}, 
TYC 7835-2569-1 \citep{1996A&A...307..121B}, and TYC 6786-0811-1 \citep{2000A&A...356..541K}.
Three of the newly discovered companions are proposed members of the AB Dor moving group (MG) and one belongs 
to the $\epsilon$ Cha group.  These discoveries
contribute to the census of stellar multiplicity at young ages in loose associations \citep{2013ARA&A..51..269D}.

\setcounter{table}{4}
\begin{table}[h]
\caption{Close visual binaries discovered or confirmed by NaCo-LP AO observations. We list the name of the star,
the projected separation measured with NaCo (from the companion paper Chauvin et al. 2013), 
in mas and in AU, the magnitude difference in H band.} 
\label{t:newbin}
\begin{center}       
\begin{tabular}{cccc} 
\hline
Target               &  Sep        & Sep &   $\Delta$ H  \\ 
                	 & (mas)       & (AU) &	(mag) 	 \\ 
\hline
TYC 0603-0461-1   &    74$\pm$14 &   4.3 &  0.1$\pm$0.2 \\ 
HIP 6177          &  1566$\pm$6  &  52.6 &  7.1$\pm$0.2 \\ 
HIP 8038          &   437$\pm$7  &  12.8 &  2.5$\pm$0.2 \\ 
TYC 8484-1507-1   &    60$\pm$14 &   3.6 &  0.2$\pm$0.2 \\ 
TYC 9181-0466-1   &  1891$\pm$7  & 146.9 &  1.5$\pm$0.2 \\ 
TYC 8927-3620-1   &    87$\pm$14 &   7.1 &  0.5$\pm$0.2 \\ 
TYC 9231-1566-1   &  1975$\pm$7  & 193.6 &  3.0$\pm$0.2 \\ 
TYC 8989-0583-1   &  2584$\pm$8  & 169.0 &  2.6$\pm$0.2 \\ 
TYC 9010-1272-1   &   262$\pm$8  &  24.0 &  1.0$\pm$0.2 \\ 
TYC 7835-2569-1   &   901$\pm$7  &  63.3 &  0.1$\pm$0.2 \\ 
TYC 6786-0811-1   &   120$\pm$10 &   9.4 &  0.3$\pm$0.2 \\ 
HIP 80290         &  3340$\pm$2  & 278.2 &  1.9$\pm$0.2 \\ 
HIP 94235         &   506$\pm$7  &  31.0 &  3.8$\pm$0.3 \\ 
HD 199058         &   471$\pm$7  &  31.1 &  2.5$\pm$0.2 \\ 
HIP 107684        &   326$\pm$7  &  29.4 &  2.8$\pm$0.3 \\ 
HIP 108422        &   170$\pm$7  &   9.9 &  3.0$\pm$0.2 \\ 

\hline
\end{tabular}
\end{center}
\end{table}

Several targets have wide companions (defined here as those with projected separation larger than 6 arcsec)
whose presence and association were either already established in the literature or are proposed here.   
They are listed in Table~\ref{t:widebin}.  These wide companions are outside NaCo's field of view so they did not impact target selection.
To estimate the reliability of these wide pairs as true bound systems,
we followed the methods of \cite{2007AJ....133..889L} to calculate the separation, projected separation 
(assuming identical distance as the brighter companion), and proper motion difference for all candidate wide binaries.
We used relations in \cite{2012ApJS..201...19D} and \cite{2007AJ....133..889L} to place statistical constraints 
on the likelihood that binary pairs are real and physically bound.
From the fractional proper motion difference of the components,  $fr_{\mu} = \Delta_{\mu}/avg_{\mu}$,
where $\Delta_{\mu}$ is the proper motion difference and $avg_{\mu}$ is the average proper motion of the
components, \cite{2012ApJS..201...19D}
show that a $fr_{\mu} \leq 0.2$ is indication that the pair is likely physically bound.
We also used the statistic $\Delta_{X}$, from \cite{2007AJ....133..889L}, which uses a power law to describe the
relationship between separation, $\Delta_{\mu}$, and $avg_{\mu}$. 
If $\Delta_{X} < 1.0$, the pair is likely bound.  This statistic
is calibrated only for stars with 150 mas/yr $< \mu <$ 450 mas/yr.  Thus, it could be less useful for stars
with smaller proper motion where the relationship may not follow a power law.  

\setcounter{table}{5}
\begin{table*}[h]
\caption{Wide binaries in the NaCo-LP sample. We list the name of the star observed with NaCo, the name of the wide
companion, the projected separations in arcsec and in AU, the average proper motion of the components, the 
proper motion difference, the magnitude difference $\Delta V$, the fractional proper motion difference, 
the $\Delta_{X}$ index, the absolute value of the RV difference and the number of components. For high-multiplicity systems,
additional remarks are listed.} 
\label{t:widebin}
\begin{center}       
\begin{tabular}{llccccccccl} 
\hline
NaCo LP	 &	 Companion   &	   Sep.	  &  Sep. & $avg_{\mu}$ &  $\Delta_{\mu}$  &	$\Delta V$   &	$fr_{\mu}$ &  $\Delta_{X}$  & $| \Delta RV |$ & $N_{comp}$ \\
Target	 &		     &	 (arcsec) & (AU)  & (mas/yr)  &  (mas/yr)  &  (mag)         &	         &	        &	(km/s)        &	      \\
\hline
HIP 8038	& HIP 8039	  &   15.0 &    435 & 51.7  &   7.0 & 1.88 &	0.14  &	2.7  &  0.75 & 3 \\
TYC 8484-1507-1	& HIP 12326	  &    8.7 &    535 & 87.1  &   7.9 & 0.51 &	0.09  &	0.7  &  0.00 & 4 \\
TYC 8484-1507-1 & HIP 12361       &  213.2 &  12557 & 88.4  &   6.7 & 1.58 &    0.072 & 3.0  &  1.10 & -- \\
HIP 25434	& HIP 25436	  &   12.0 &    976 & 27.1  &   3.7 & 0.80 &	0.14  &	6.2  &  0.60 & 3 \\
HIP 35564	& HD 57853 	  &    9.2 &    275 & 151.6 &  47.9 & 0.58 &	0.37  &	0.9  &  3.20 & 5 \\
TYC 8128-1946-1	& HIP 36312	  &   81.7 &   6853 & 51.9  &   3.6 & 1.05 &	0.07  &	4.4  &  2.20 & 2 \\
HIP 37923	& HIP 37918	  &   23.1 &    836 & 166.2 &  15.4 & 0.08 &	0.09  &	0.5  &  0.10 & 2 \\
HIP 46634	& HIP 46637	  &   14.0 &    508 & 202.7 &   5.3 & 0.18 &	0.03  &	0.02 &  1.33 & 3 \\
HIP 47646	& HIP 47645	  &   16.5 &   1161 & 123.2 &   3.2 & 2.78 &	0.03  &	0.3  &   --  & 2 \\
HIP 58240	& HIP 58241	  &   18.6 &    649 & 174.9 &  34.5 & 0.03 &	0.2   &	0.5  &  0.90 & 2 \\
HIP 71908	& SAO 252852	  &   15.8 &    263 & 302.6 &   0.7 & 5.22 &	0.0024&	0.03 &  0.50 & 2 \\
HIP 72399	& HIP 72400	  &   11.0 &    609 & 142.7 &  12.2 & 1.06 &	0.09  &	0.4  &  4.50 & 3 \\
TYC 6786-0811-1	& HIP 77235	  &   50.5 &   3588 & 64.5  &  10.0 & 2.38 &	0.15  &	3.3  &  3.30 & 2-5? \\
HD 199058	& TYC 1090-0543-1 &  245.1 &  16200 & 68.0  &   3.3 & 2.92 &	0.05  &	4.0  &  0.70 & 3 \\
TYC 9339-2158-1	& HIP 116063	  &   36.3 &   1109 & 220.4 &   4.2 & 1.89 &	0.02  &	0.2  &  0.50 & 2 \\
\hline
\end{tabular}
\end{center}
Note on high-multiplicity systems: HIP 8038: close visual binary (VB);  TYC 8484-1507-1: close VB and SB2 (likely same companion); HIP 12361: wide companion to
TYC 8484-1507-1 and HIP 12326 triple system; HIP 25436: spectroscopic and astrometric binary; HIP 35564: SB; HD 57853: SB3; 
HIP 46637: SB; HIP 72399: SB; TYC 6786-0811-1: close VB; HIP 77235: triple system but physical link to TYC 6786-0811-1 
needs confirmation; HD 199058: close VB. See App.~B for further details and references.
\end{table*}

The analysis shows that most of the wide pairs in Table~\ref{t:widebin} are likely bound.
Measured parameters such as
age, distance, and moving group membership can therefore be shared between the components.   
All stars for which $\Delta_{X} > 1.0$ are stars that have $avg_{\mu} < 150$ mas/yr, implying that this statistic 
is less useful.  For these stars, the $fr_{\mu}$ is probably a better indication of a physically 
bound nature, although it is not as strong a constraint as $\Delta_{X}$.
Finally, we note that several of the proposed binaries are actually hierarchical multiples, 
as revealed by imaging of tight visual companions and  observed RV variations (see App.~B).

\subsection{Spectroscopic and astrometric binaries}
\label{s:sb}

We summarize in Table \ref{t:specbin} the presence of spectroscopic binarity among
our targets, considering both RV variability and the presence of multiple components
in the spectra (SB2).
Two critical issues prevent firm conclusions in some specific cases.
1) For single-epoch spectra, line-profile analysis of young, active stars
often reveals deformation due to star spots crossing the stellar disk. It is difficult
to distinguish
 between such effects and the presence of a companion.
2) The evaluation of the significance of the proposed RV variations depends
on the adopted error bars, which are often not available in the literature data.
Furthermore, there are cases where significant RV variations, of amplitudes up to a few km/s,
can be caused by stellar or circumstellar phenomena, such as variable obscuration from
a circumstellar disk \citep[see e.g.][for the case of T Cha]{2009A&A...501.1013S}
or stellar oscillations in the case of early type stars.
A couple of suspected astrometric binaries is also included in Table \ref{t:specbin}.

\setcounter{table}{6}
\begin{table}[h]
\caption{Spectroscopic and astrometric binaries in the NaCo-LP sample. See notes in Appendix B
for further details.} 
\label{t:specbin}
\begin{center}       
\begin{tabular}{cl} 
\hline
Target           & Remarks \\
\hline
\multicolumn{2}{c}{confirmed SB} \\
\hline
TYC 5839-0596-1  & SB2  \\
HIP 3924         & SB2  \\
HIP 72399        & SB1 \\    
TYC 7835-2569-1  & SB2 + close VIS \\ 
HIP 35564        & SB1 \\  
HIP 36414        & SB1 \\ 
TYC 7188-575-1   & SB1  \\
TYC 5206-0915-1  & SB1   \\  
\hline
\multicolumn{2}{c}{suspected SB} \\
\hline
TYC 7796-2110-1  & SB2?  \\ 
TYC 9010-1272-1  & SB2? + close VIS \\ 
TYC 8128-1946-1  & RV Var ?  \\   
HIP 71933        & RV var ?; astr. acceleration \\  
TYC 6815-0874-1  & SB2 ?  \\  
HIP 13008        & astrom. acceleration \\
TYC 9245-0617-1  & RV var \\
\hline
\end{tabular}
\end{center}
\end{table}

\section{Kinematics and membership to nearby associations}
\label{s:kin}

\subsection{Kinematic parameters and space velocities}

Proper motions were taken from Tycho2 \citep{2000A&A...355L..27H} and \citet{2007A&A...474..653V} when available. For HIP 46634/7 and HIP 58240/1, we instead adopted the
original Hipparcos values \citep{1997A&A...323L..49P} (see Sect.~\ref{s:dist}).
For HIP 25434/6, we adopted the long term proper motion from  Tycho2 as there are indications that
the Hipparcos astrometry is biased by the presence of an additional component (see App.~B).
Absolute RVs were derived from our own spectra and extensive 
literature compilation (see Table \ref{t:kinparam}).
A minimum error bar of 0.5 km/s was adopted to take into account the uncertainties
in the absolute RV and the uncertainties in the zero-point systematics between different instruments. 
The determination of distances is discussed in Sect.~\ref{s:dist}.

The space positions and space velocities and their uncertainties 
were derived using relations from \citet{1987AJ.....93..864J}.  
In  these  calculations,  we  have  used  the  J2000
transformation   matrix, taken   from  the
introduction to the Hipparcos catalog, to convert  to  Galactic   coordinates.

\subsection{Membership to nearby associations}
\label{s:groups}

Table \ref{t:mgprop} presents the list of moving
groups, associations, and solar-neighborhood open clusters 
considered in our kinematic analysis. Adopted parameters are described for each group.
If the uncertainties in the U, V, and W space velocities of the associations were quoted  in the 
literature,  we included
those  in the  analysis.   If no  uncertainties  were quoted,  we used  a
conservative uncertainty estimate of $\pm$1 km/s for moving groups and
$\pm$2 km/s  for open clusters. 
We considered the Young, Nearby Population to be represented by
stars within the  U,V, and W boundaries defined by \citet{2004ARA&A..42..685Z},
as also shown in Fig.~\ref{f:ageindicators}.  
\citet{2004ARA&A..42..685Z} noted that the majority of solar-neighborhood stars 
within $\sim$60 pc that are younger than 50 Myr
have kinematic parameters within these boundaries.
However, it should be considered that old stars in the same kinematic space outnumber
true young stars. Therefore, independent indicators of youth are required for a
firm inference on stellar age. 
HIP 71908 is an example of a moderately old star with kinematics similar to
those of young stars  (see App.~B).
The adopted ages are presented in Sect.~\ref{s:absage}.

\setcounter{table}{7}
\begin{table*}[h]
\caption{Properties of young moving groups (MGs). We list the MG name and the U, V, W space velocity with
corresponding literature source. We give the most probable age, and minimum and maximum avalues and the associated literature source.  
The final column gives the number of members in the NaCo-LP sample (wide companions not observed with NaCo are not included).} 
\label{t:mgprop}
\begin{center}       
\begin{tabular}{lccclccclr} 
\hline
MG                                 &  U   & V    &  W   &  Ref & Age & Age & Age & Ref & N \\
                                   & km/s & km/s & km/s &      & best & min & max &  &  \\ 
\hline
TW Hydrae Association              &  -9.87$\pm$4.15 & -18.06$\pm$1.44  &  -4.52$\pm$2.80 & Mal13       &   8 &   6 &  12 & ZS04  &  0 \\
$\beta$  Pictoris  Moving  Group   & -10.94$\pm$2.06 & -16.25$\pm$1.30  &  -9.27$\pm$1.54 & Mal13       &  16 &  10 &  22 & Des13 &  3 \\
Tucana/Horologium Association      &  -9.88$\pm$1.51 & -20.70$\pm$1.87  &  -0.90$\pm$1.31 & Mal13       &  30 &  20 &  40 & Tor08 &  5+3? \\
Columba Association                & -12.24$\pm$1.03 & -21.32$\pm$1.18  &  -5.58$\pm$1.89 & Mal13       &  30 &  20 &  40 & Tor08 &  5+1? \\
Carina  Association                & -10.50$\pm$0.99 & -22.36$\pm$0.55  &  -5.84$\pm$0.14 & Mal13       &  30 &  20 &  40 & Tor08 &  1 \\
Argus Association                  & -21.78$\pm$1.32 & -12.08$\pm$1.97  &  -4.52$\pm$0.50 & Mal13       &  45 &  30 &  55 & Des13 &  2 \\
AB Doradus  Moving  Group          &  -7.12$\pm$1.39 & -27.31$\pm$1.31  & -13.81$\pm$2.16 & Mal13       & 100 &  50 & 150 & Des13 &  8+1? \\ 
Octans Association                 &  -14.5$\pm$0.9  &   -3.6$\pm$1.6   &  -11.2$\pm$1.4  & Tor08       &  20 &  10 &  30 & Tor08 &  0 \\
$\epsilon$ Chamaeleon              &  -11.0$\pm$1.2  &  -19.9$\pm$1.2   &  -10.4$\pm$1.6  & Tor08       &   6 &   4 &   8 & Tor08 &  2 \\
$\eta$ Chamaeleon                  &  -12.0$\pm$1.0  &  -19.0$\pm$1.0   &  -10.0$\pm$1.0  & ZS04        &   8 &   6 &  10 & ZS04  &  0 \\
Cha-Near Association               &  -11.0$\pm$1.0  &  -16.0$\pm$1.0   &   -8.0$\pm$1.0  & ZS04        &  10 &   6 &  15 & ZS04  &  0 \\ 
Carina-Near  Moving  Group         &  -25.9$\pm$1.0  &  -18.1$\pm$1.0   &   -2.3$\pm$1.0  & Zuk06       & 200 & 150 & 300 & Zuk06 &  4+1? \\
Ursa Majoris Moving  Group         &   14.0$\pm$1.0  &    1.0$\pm$1.0   &   -9.0$\pm$1.0  & ZS04        & 500 & 400 & 600 & Kin03 &  0 \\ 
Hercules-Lyra  Association         &  -13.2$\pm$2.5  &  -20.6$\pm$1.6   &  -11.9$\pm$1.0  & LoS06       & 200 & 150 & 300 & LoS06 &  1 \\ 
Corona  Australis Association      &   -3.8$\pm$1.2  &  -14.3$\pm$1.7   &   -8.3$\pm$2.0  & Qua01       & 10  &   6 &  20 & Qua01 & 3? \\
IC2391  Open Cluster               &  -20.6$\pm$2.0  &  -15.7$\pm$2.0   &   -9.1$\pm$2.0  & Mon01       & 45  &  30 &  55 & Des13 & 0 \\
Hyades Open Cluster                &  -40.0$\pm$2.0  &  -17.0$\pm$2.0   &   -3.0$\pm$2.0  & ZS04        & 625 & 575 & 675 & Per98 & 0 \\ 
Pleiades  Open  Cluster            &  -12.0$\pm$2.0  &  -21.0$\pm$2.0   &  -11.0$\pm$2.0  & ZS04        & 125 & 110 & 140 & Sta98 & 0 \\
Upper Scorpius Association         &   -6.4$\pm$0.5  &  -15.9$\pm$0.7   &   -7.4$\pm$0.2  & Che11       &  11 &   5 &  14 & Pec12 & 2? \\
Upper Centaurus  Lupus Ass.        &   -5.1$\pm$0.6  &  -19.7$\pm$0.4   &   -4.6$\pm$0.3  & Che11       &  16 &  10 &  18 & Pec12 & 1? \\
Lower Centaurus Crux Ass.          &   -7.8$\pm$0.5  &  -20.7$\pm$0.6   &   -6.0$\pm$0.3  & Che11       &  17 &  10 &  19 & Pec12 & 3? \\
\hline
\end{tabular}
\end{center}
References: Mal13: \citet{2013ApJ...762...88M};
            Tor08: \citet{2008hsf2.book..757T}; 
            ZS04: \citet{2004ARA&A..42..685Z};
            Mon01: \citet{2001MNRAS.328...45M};
            Zuk06: \citet{2006ApJ...649L.115Z}; 
            Kin03: \citet{2003AJ....125.1980K};
            LoS06: \citet{2006ApJ...643.1160L};
            Qua01: \citet{2001ASPC..244...49Q};
            Per98: \citet{1998A&A...331...81P};
            Sta98: \citet{1998ApJ...499L.199S};
            Che11: \citet{2011ApJ...738..122C};
            Pec12: \citet{2012ApJ...746..154P};
            Des13: this paper
\end{table*}

We then calculated the  significance in the difference ($Sig D$)  between U, V,
and W galactic velocities, calculated for each NaCo-LP target and each
association in  the above  list. As  an example, the  $Sig D$ for  the U
velocity is defined as:

\begin{equation}
SigD_{U} = \frac{ ( U_{star} - U_{group} ) }  {\sqrt{ (dU_{star})^2 + (dU_{group})^2 }}
\end{equation}

\noindent
The $SigD$ value represents how many sigma separate the space velocity of the target 
from that of the group being considered.
In our analyses of NaCo-LP targets, we considered U, V, and W space velocities to be consistent with
membership if it had a $SigD \le  2.0$ in each of the
U, V  and W  velocities. 
This is  only one indication  that a
star may belong to a given association, but it serves as a first step in
the process to define bona fide membership.  
We further checked moving group membership by calculating Bayesian membership probabilities using the 
on-line tools BANYAN I and II \citep{2013ApJ...762...88M, 2014ApJ...783..121G}. The on-line versions of these
tools use only the available kinematics for this star as input and do not consider the star's age (e.g. CMD position)
We used all available kinematic information for each star as input (coordinates and our adopted proper motions,
radial velocities, and distances).
In several cases, the two versions of BANYAN yield highly discrepant results. This is likely caused 
by changes to prior, UVWXYX distributions, and the definition of the Galactic field between the first and
second iterations of BANYAN \citep{2014ApJ...783..121G}. A detailed comparison of the differences 
between the two tools is beyond the scope of this paper. 
We indicate the targets where discrepancies arise along with further details regarding target ages and
group membership in Appendix B.

In most cases we confirm the membership assignments by \citet{2008hsf2.book..757T} and
\citet{2013ApJ...762...88M} and we identify some possible additional members. 
Discrepancies for individual cases
are mostly due to differences in the kinematic data (e.g. RV in the case of HIP 11360).
Different membership criteria adopted in the various studies also play a role.
A large number  of  the NaCo-LP  targets  are found  to be  kinematically
consistent  with the  Young, Nearby  Population.

\subsection{Absolute ages of nearby Moving Groups}
\label{s:absage}

Our age determination is mostly based on comparison
of indirect indicators as measured for individual NACO-LP targets
to those of members of groups and clusters, such as $\beta$ Pic, Tucana, Columba, Carina,
AB Dor, Pleiades, and Hyades. While the age ranking of these groups is mostly defined,
the absolute values of their ages is widely debated in the literature.
A full assesment of the issue of the absolute ages is beyond the scope of this paper but we
present the motivation of our adopted ages for the youngest groups for which 
members are found within the NACO-LP sample.
We anticipate that for some of the groups considered below ($\beta$ Pic, IC 2391, Pleiades)
there is a systematic discrepancy between ages obtained using the lithium depletion boundary
of low mass stars with respect to isochrone fitting, with the first method yielding older ages.
This trend is common to several other open clusters \citep[e.g.][]{2010MNRAS.409.1002D}.

\subsubsection{$\beta$ Pic, $\epsilon$ Cha, and Sco-Cen groups}

The most widely adopted age for the $\beta$ Pic is 12 Myr, based on
the isochrone fitting by \citet{2001ApJ...562L..87Z} ($12^{+8}_{-4}$ Myr) and of kinematic 
traceback analysis by \citet{2002ApJ...575L..75O} and \citet{2003ApJ...599..342S}.
\citet{2008hsf2.book..757T} adopt instead 10 Myr.
Very recently, a significantly older age ($21\pm4$ Myr) has been derived by \citet{2013MNRAS.tmpL.192B}
using the lithium depletion boundary of low mass stars. 
An age around 20 Myr is only marginally compatible with the isochrone fitting results by \citet{2001ApJ...562L..87Z}

The kinematic traceback age have nominally a small error bar (about 1 Myr)  but is based on the list of members 
known in 2002-2004.
The current enlarged list of members and the improvement in the kinematic parameters, due to the release of
the revised Hipparcos parallaxes by \citet{2007A&A...474..653V}, additional RV data, and more complete census of
multiplicity, indicate the need of updating the analysis. 
The study by Mamajek and collaborators \citep[in preparation, preliminary results presented in ][]{soderblom13} claims that
the group was not smaller in the past, ruling out a 12 Myr age at $>3$ sigma.

An older age for $\beta$ Pic would allow to reconcile the result by \citet{2012AJ....144....8S}
that stars in LCC and UCL are slightly younger than those of $\beta$ Pic MG (from the
comparison of Lithium EW as a function of color) to the absolute ages of UCL and LCC
of 16 and 17 Myr respectively, derived using isochrone fitting, including the recent work
by \citet{2012ApJ...746..154P}.

Performing isochrone fitting to $\beta$ Pic MG members with trigonometric parallax,
adopting the photometric temperatures as in Sect.~\ref{s:teff} and 
using \citet{2012MNRAS.427..127B} models as described in Sect.~\ref{s:isoc}, yields
an age of 16$\pm$2 Myr (internal error only) midway between the original 12 Myr estimate and the recent one based on 
lithium depletion boundary. 
To check the reliability of our result, we considered the F-type stars in Sco-Cen region recently studied by 
\citet{2012AJ....144....8S} and we got very similar results, our ages being older by about 1 Myr.

Considering the uncertianties related to the various techniques we adopt $16\pm6$ Myr as our age for the $\beta$ Pic MG.
This estimate has the additional advantage of being homogeneous with the age determination from isochone fitting
performed for our program stars in Sect.~\ref{s:isoc}.
For the Sco-Cen subgroups the ages by \citet{2012ApJ...746..154P} were adopted, with lower limits to the ages
proposed by \citet{2012AJ....144....8S} for UCL and LCC and by \citet{2002AJ....124..404P} and 
\citet{2008hsf2.book..235P} for US.

In our sample, there are also two members of the $\epsilon$ Cha group. 
\citet{2008hsf2.book..757T} adopt an age of 6 Myr. As we are adopting older ages for
young groups such as $\beta$ Pic MG and Upper Scorpius, an upward revision of the age of this group
is also possible. The analysis by \citet{2009A&A...508..833D} based on lithium shows that
$\epsilon$ Cha is younger than TW Hya, which is younger than  $\beta$ Pic MG.
Therefore, we adopt 10 Myr as the age for the $\epsilon$ Cha group, allowing for 
the younger age proposed by \citet{2008hsf2.book..757T} as lower limit.

\subsubsection{Tucana, Columba, Carina, and Argus}

The most widely adopted age for Tucana, Columba, Carina is 30 Myr \citep{2004ARA&A..42..685Z,2008hsf2.book..757T}.
Argus and IC 2391 open cluster are considered to be slightly older (35-40 Myr).
The absolute age of IC 2391 open cluster was estimated to be 40 Myr using
isochrone fitting \citep{2007A&A...461..509P} and 50 Myr using the lithium depletion boundary \citep{2004ApJ...614..386B}.
The very recent determination of the lithium depletion boundary for Tucana from \cite{2014arXiv1403.0050K}
yields an age of about 40 Myr.
Our isochrone fitting yields an age of 35 Myr for the members of these latter groups, compatible with the previously 
adopted age.

Considering that Tucana, Columba, and Carina appears to be younger than Argus from Li EW vs effective temperature
and Li depletion boundary \citep{2009A&A...508..833D,2014arXiv1403.0050K}
and the absolute age determination of IC 2391 appear sound within 10 Myr, there are only limited margins for any upward
revision of age of these groups with respect to the values adopted in \citet{2008hsf2.book..757T}.
Additional evidence for a young age of the Columba association comes 
from the interferometric measurement of the proposed member HR 8799 \citep{2011ApJ...732...61Z}, where the radius of the
star is consistent with an age $\sim$30 Myr \citep{2012ApJ...761...57B}\footnote{See \citet{2010ApJ...716..417H} for a different view on Columbs membership.}.
Dynamical stability studies of the planetary system orbiting HR 8799 provide additional indirect evidence for a young system age. 
An older stellar age implied larger planet masses that would lead to a dynamically unstable systems
\citep{2011ApJ...729..128C,2012ApJ...755...38S,2013A&A...549A..52E}.

Therefore, we adopt 30 Myr for Tucana, Columba, and Carina and 45 Myr for Argus (the mean between
isochrone and lithium depletion boundary ages).

\subsubsection{AB Dor moving group}

The age of AB Dor MG is highly debated in the literature, ranging from 50 Myr in the first
paper presenting the association \citep{2004ApJ...613L..65Z} to about 150 Myr. 
\citet{2007A&A...462..615J} found an age of 50-100 Myr for the AB Dor quadruple system while
\citet{2011A&A...533A.106G} favour an age of 40-50 Myr from the size of AB Dor A derived 
through interferometric measurements.
\citet{2005ApJ...628L..69L} suggest a link to the Pleaides open cluster with a common origin and a similar age.
\citet{2008hsf2.book..757T} and \citet{2009A&A...508..833D} adopt 70 Myr, equal to their adopted age for the Pleiades.
\citet{2013ApJ...766....6B} cast dubth of the existence of the group as a whole, finding that
only half of the proposed members might share a common origin and set an age lower limit at 110 Myr
from the position of K dwarfs on color-magnitude diagram.
\citet{2010A&A...520A..15M} noted the similar distribution of rotation periods
among AB Dor and Pleiades members, while that of the $\alpha$ Per open cluster is clearly different, indicating its younger age. 
On the other hand, the comparison of Li EW of AB Dor and Pleiades members show that the median values for AB Dor are above 
those of Pleiades for B-V between 0.7 to 0.9 (Fig.~\ref{f:ageindicators}), suggesting a slightly younger age for AB Dor.
X-ray luminosity is also on average slightly larger for AB Dor members.
Adopting 125 Myr as absolute age for the Pleiades \citep{1998ApJ...499L.199S}, we assume in the following 100 Myr as the age for 
AB Dor MG, with the caveat
that stars with slightly different ages but sharing kinematics may be mixed in the adopted target list.

\section{Fundamental parameters}
\label{s:fundamental}

\subsection{Photometric and trigonometric distances}
\label{s:dist}

Our adopted distances are based on trigonometric parallaxes from \cite{2007A&A...474..653V} 
when available. For multiple systems where separate parallaxes were available, their weighted average
was adopted. For HIP 25434/6, HIP 46634/7 and HIP 58240/1, the weighted averages of parallaxes from
the original Hipparcos catalog were adopted, as they lead to more consistent positions in the color-magnitude
diagrams. However, formal parallax errors for these pairs may be underestimated.
For TYC~8484-1507-1, TYC 8128-1946-1, and TYC~9339-2158-1, we used in our analysis the trigonometric parallaxes of 
their bound wide companions, though these companions were not a part of our target sample. 
For TWA 21, parallax is from \citet{2013ApJ...762..118W}.

\begin{figure}[h]
\includegraphics[width=8cm]{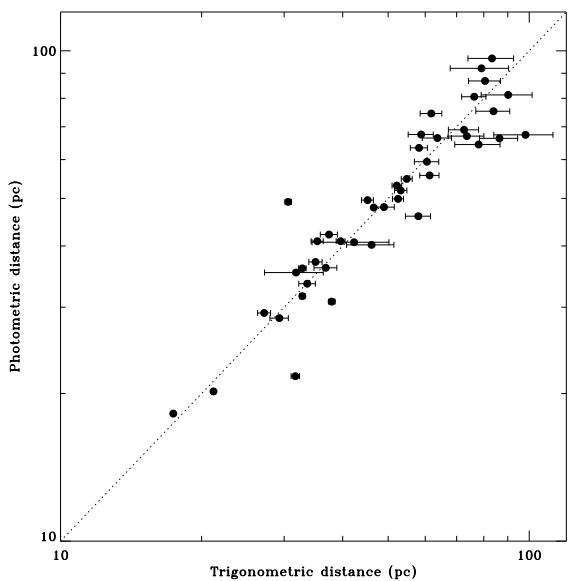}
\caption{Comparison of trigonometric and photometric distances for the stars in the sample with
available trigonometric parallax. The rms scatter in the difference is  17\%. }
\label{f:checkdist}
\end{figure}

For stars without trigonometric parallax, photometric distances were obtained using
BVIJHK photometry (Table 9). 
Reference sequences at various ages were built using members of moving groups
with available trigonometric parallaxes. Target distance was then derived based on the multi-band photometry, adopted age, and the reference calibration.
Typical scatter in these standard relations
implies errors of about 15-20\% in the photometric distances (Fig.~\ref{f:checkdist}).
The presence of known companions was taken into account including correction to the unresolved photometry
when applicable.
Interstellar reddening was checked by comparing photometric temperature, spectroscopic
temperatures when available, and spectral types from the literature.
In a few cases of stars with distance close or slightly above 100 pc, the presence of 
small amounts of reddening ($E(B-V) \le 0.1$ mag) was suggested.
Photometric distances have larger errors for very young stars (age 5-20 Myr).  In these instances 
the absolute magnitude and  derived distances depend sensitively on the adopted ages; the derived values would be systematically
underestimated for unidentifed cases of evolved  RS CVn variables.

\subsection{Metallicity}
\label{s:metal}

Metallicity has been found to have a strong correlation with planet frequency, at least for RV detections of giant planets around solar type stars
 \citep{2005ApJ...622.1102F}.
Whether the planet metallicity connection extends from the separations 
accessible with RV surveys to those considered in direct imaging surveys
is highly uncertain, and depends
on the main formation and migration mechanisms.

The determination of the metallicity of young stars, such as those of our sample,
is made difficult by the high level of magnetic activity, the fast
rotational velocities and the possibility of veiling in the spectra.  These phenomena 
cause larger errors in the abundance analysis with respect to
older stars of similar spectral types. Differences in the atmospheric structure
of young, active stars with respect to old, inactive main sequence stars might introduce
spurious effects \citep[see e.g.~][]{1996ApJ...471..847P}.

In spite of these challenges, 
a number of studies have addressed the issue of the metallicity of nearby 
young stars, with a rather general agreement on a typical 
solar or slightly sub-solar metallicity and the lack of metal-rich very young stars
\citep[see e.g.][]{2006A&A...446..971J,2008A&A...480..889S,2011A&A...525A..35B}.

The study of the chemical composition of several
stars in young associations resulted in no significant scatter between 
the individual members \citep{2009A&A...501..965V,2009A&A...501..973D,2013MNRAS.431.1005D}.
Therefore, mean values (derived from stars both inside and external to our sample) were adopted for all association members.

In order to have
the largest number of associations analyzed in an homogeneous way, we adopted
the metallicities by \cite{2009A&A...501..965V} for $\beta$ Pic, Tuc, Col, Car,
AB Dor and $\epsilon$ Cha MGs\footnote{\cite{2009A&A...501..965V} noted a trend 
of metallicity with effective temperatures, possibly due to the effects of stellar 
activity. Removing it by assuming that no correction is needed for a solar type
star would increase the metallicities by about 0.1 dex. Since the origin of this
trend is uncertain and the assumption of its null value for solar temperature
is questionable, we do not apply such shifts.}.
Metallicity of the Carina-Near MG is from \cite{2012MNRAS.427.2905B}. Metallicity of the Her-Lyr MG is from \cite{2004AN....325....3F}.

A few other stars have spectroscopic metallicity determinations in the literature.
Photometric metallicities based on Str\"omgren photometry were not considered,
except for the few cases of moderately old stars, 
as the strong magnetic activity is known to bias, toward lower values, the metallicity derived from this method \citep{1996A&AS..119..403M}.

\subsection{Isochrone fitting, stellar masses and radii}
\label{s:isoc}

Stellar masses and isochrone ages were derived through interpolation of stellar models
by \cite{2012MNRAS.427..127B}, which include in an homogeneous way pre and post-main sequence
evolutionary phases.
We used the tool {\it param} \citep{2006A&A...458..609D}\footnote{http://stev.oapd.inaf.it} adapted to  use the 
\citet{2012MNRAS.427..127B} stellar evolutionary models. {\it Param} performs a Bayesian determination of the most 
likely stellar intrinsic properties, weighting  all the isochrone sections compatible with the measured 
parameters (metallicity, effective temperature, parallax, apparent magnitude) and their errors. It
adopts a flat distribution of ages, within a given interval, as a prior. 
We found that \citet{2012MNRAS.427..127B} models are very similar to those by \citet{2011A&A...533A.109T}
and \citet{2008ApJS..178...89D}, at least for effective temperatures warmer than 4500 - 4000 K. 
Differences at $T_{\rm eff} < 4000$~K are not of concern for the analysis of the stars in our sample.

\begin{figure}[h]
\includegraphics[width=8cm]{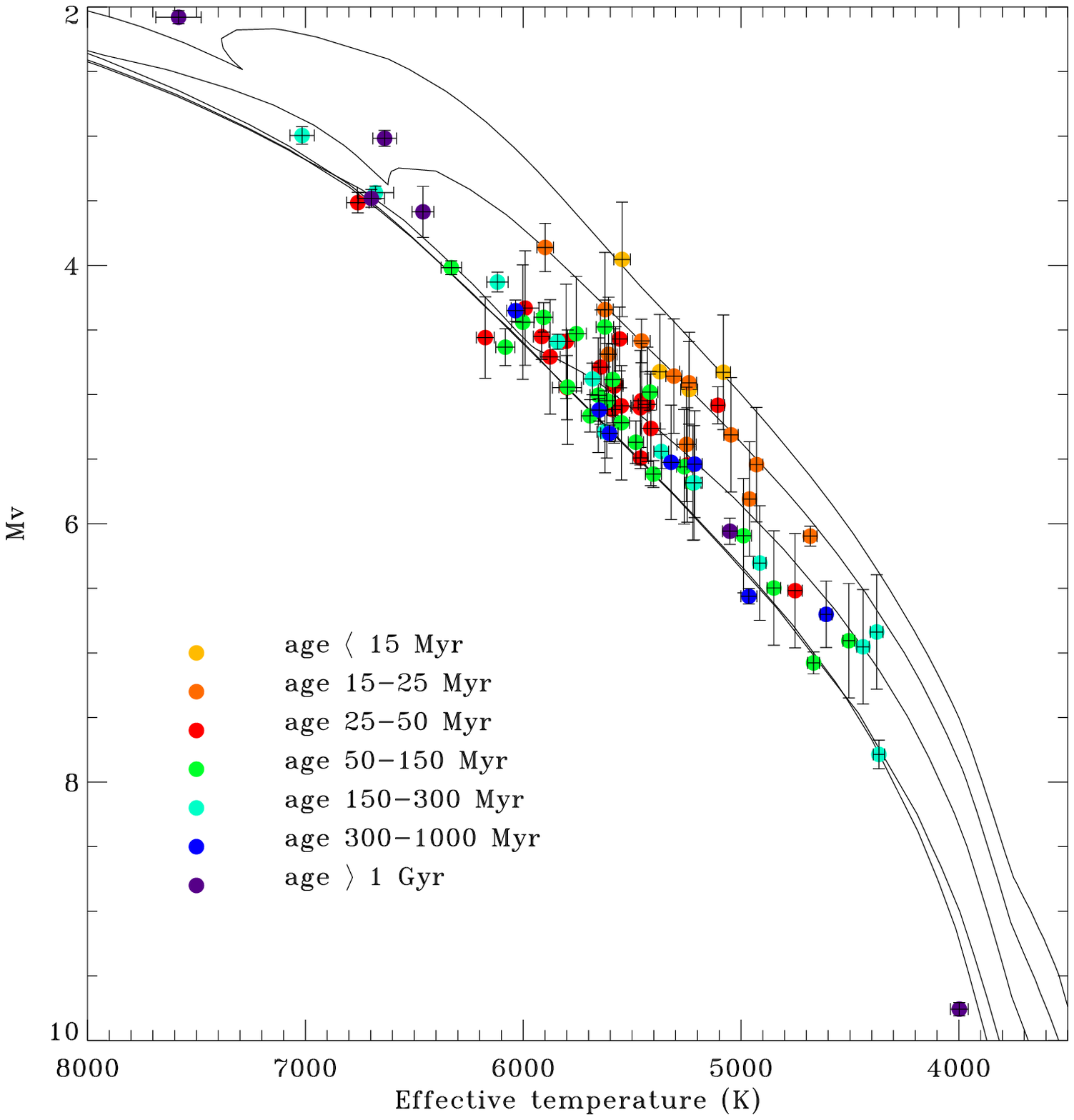}
\includegraphics[width=8cm]{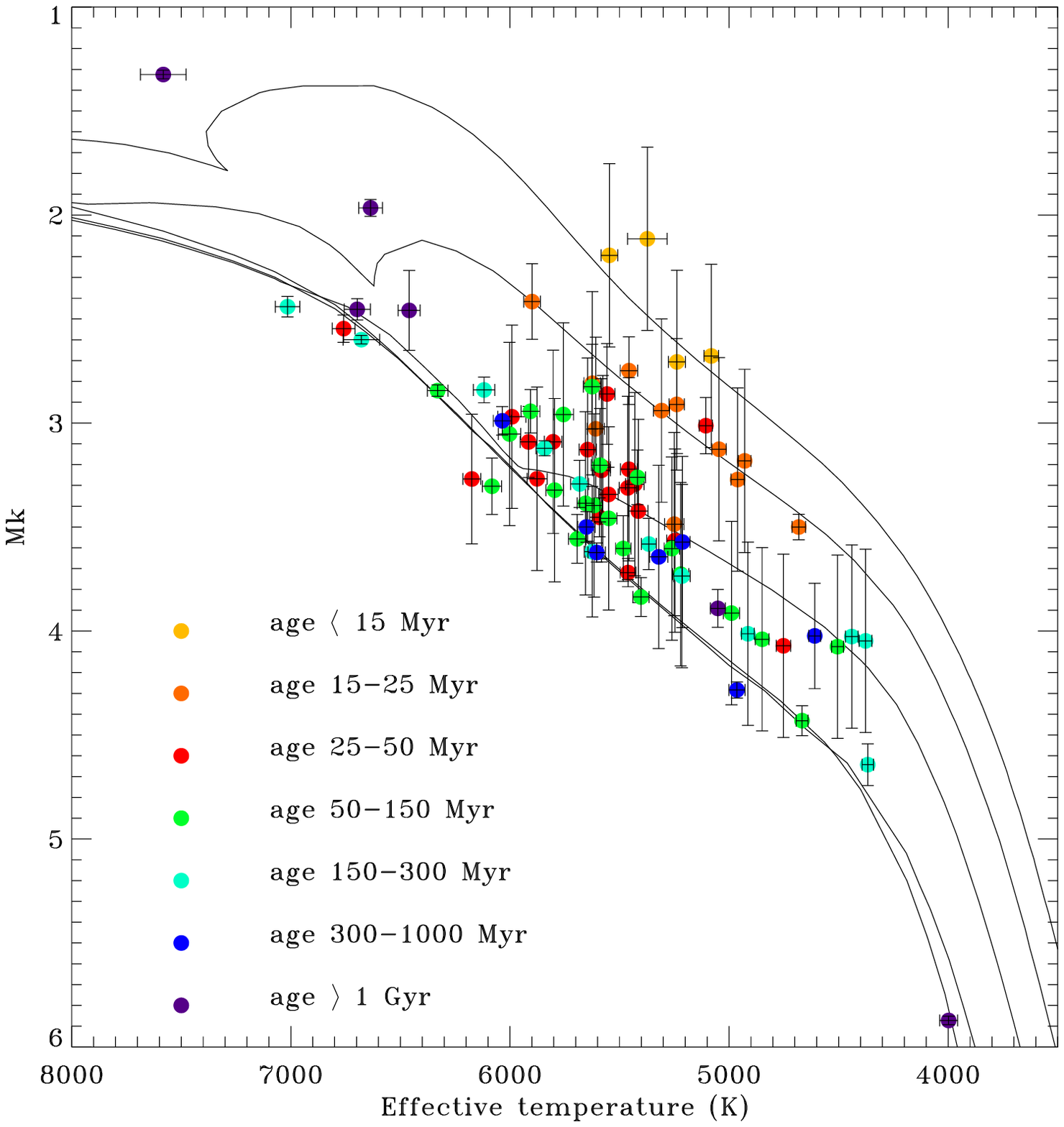}
\caption{HR diagram (upper panel: effective temperature vs $M_{V}$; lower panel: effective temperature vs $M_{K}$) 
for stars in the NaCo-LP sample. Stars with different ages are plotted with different colors. Overplotted are
pre-main sequence, solar-metallicty isochrones for 5, 12, 30, 70, and 200 Myr \citep[from][]{2012MNRAS.427..127B}.
The two B-type stars HIP 10602 and HIP 12394 and the Li-rich giant HD 99409 are outside plot limits.}
\label{f:cmd}
\end{figure}

In most cases, isochrone fitting yields very poor constraints on stellar ages (e.g. errors of a few Gyrs),
which is expected for stars close to the main sequence (Fig.~\ref{f:cmd}). Tighter constraints are derived from indirect indicators 
such as lithium and rotation-activity phenomena (see Sect.~\ref{s:age}).   
The stellar masses and ages resulting from isochrone fitting, within the whole range of input stellar 
parameters, are correlated, with older ages yielding lower stellar masses.
To infer the stellar masses corresponding to the plausible ages of
our targets, we run the {\it param} interface a second time for these targets, 
adopting the minimum and maximum ages from Table \ref{t:agetable}, and keeping fixed the other input parameters. 
Differences in stellar masses with respect 
to the unconstrained run are typically of about $0.02-0.04~M_{\odot}$, with larger masses occurring when adopting the
age limits from the indirect indicators. The {\it param} tool also provides the stellar radii, which were coupled
with the measured $v \sin i$ to estimate the reliability of detected rotation periods (see App.~B for discussion
on individual objects).
The isochrone ages of the few targets for which this method contributes significantly to determination of stellar ages
are listed in Table \ref{t:agetable}.

\subsection{Stellar Ages}
\label{s:age}

Our approach to age determination is to consider all the indicators available to us to better understand
the properties of each object and derive stellar age.
The lesson we got from the case of HIP 6177 \citep{2013A&A...554A..21Z} is that significant discrepancies between age
indicators may hide an astrophysical situation different from a normal young star 
(the expected typical target in our survey) and therefore require an in-depth investigation.

\begin{figure*}[h]
\includegraphics[width=18cm]{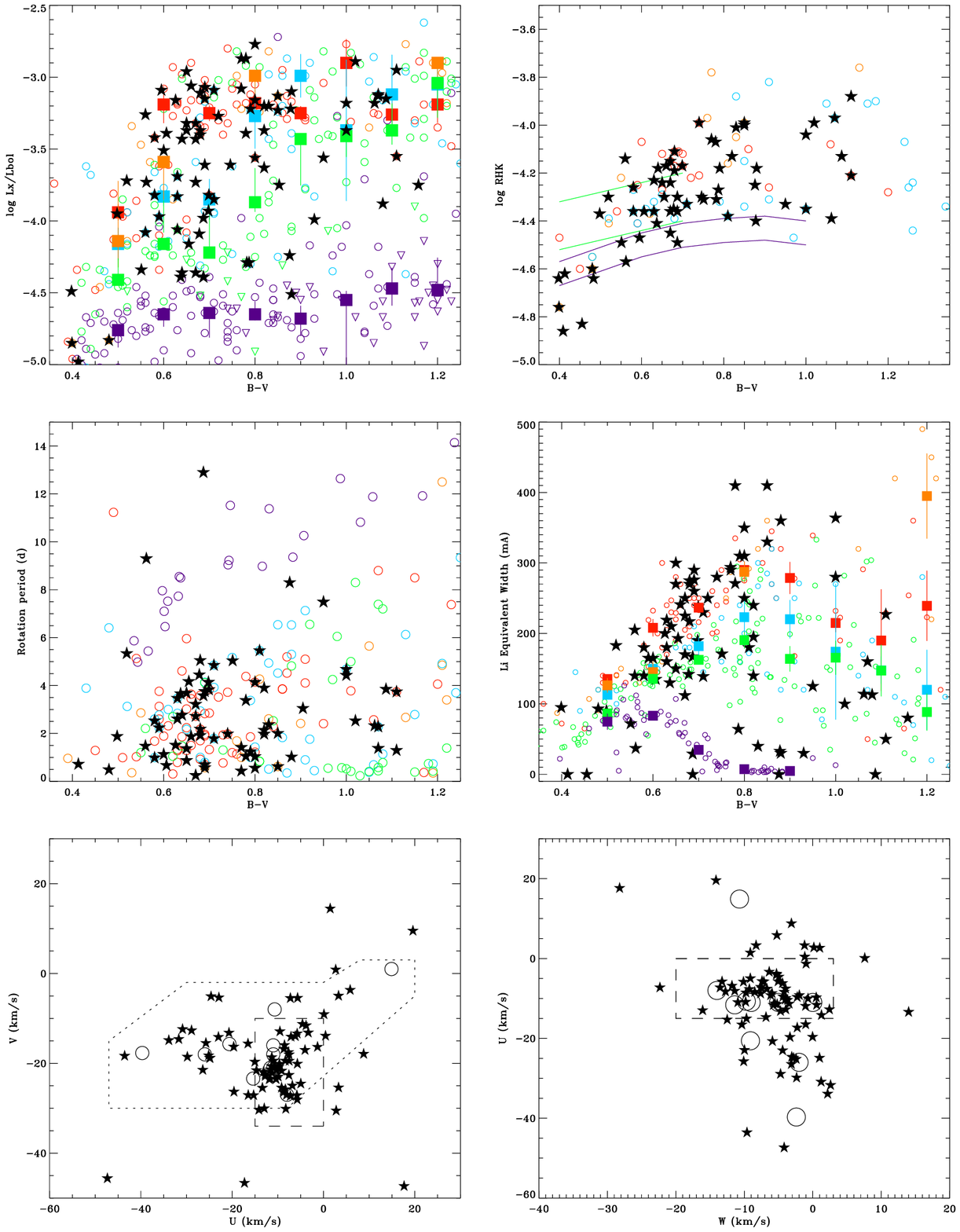}
\caption{Age indicators for the stars in the sample. From top-left: $R_{X}=\log L_{X}/L_{bol}$ vs B-V; $\log R_{HK}$ vs B-V;
Rotation period vs B-V; Lithium equivalent width vs B-V; U vs V space velocities, U vs W space velocities.
In all the panels, the black stars represent the targets. In the $R_{X}$, $\log R_{HK}$, rotation period and Lithium plots, orange circles
represent $\beta$ Pic MG; red circles: Tucana-Columba-Carina MG; Light blue circles: AB Dor MG; Green circles: Pleiades OC;
Purple circles: Hyades OC. In the $R_{X}$ vs B-V and EW Li vs B-V panels, the filled squares represent the median value of $R_{X}$ 
and EW Li for the
corresponding color bin. In the $\log R_{HK}$ plot, green  and purple lines represent the locus populated by Pleiades and Hyades
stars respectivey.
In the latter two panels, the positions of some young moving groups are marked as open circles.
The dashed-line box in both panels is the locus of the Nearby Young Population as defined in
\cite{2004ARA&A..42..685Z}. The dotted contours in the U vs V plot show the locus of kinematically young stars proposed
by \cite{2001MNRAS.328...45M}.}
\label{f:ageindicators}
\end{figure*}

In the previous sections we provided several parameters (Lithium EW, Ca II H\&K emission, X-ray emission, 
rotation period) which are known to depend on stellar age.
The stellar age from Lithium  was estimated for each target from a comparison of measured Lithium EW with that of confirmed
members of young moving groups and clusters of similar color (Fig.~\ref{f:ageindicators}).
For Ca II H\&K emission, X-ray emission, and rotation period, we adopted the age calibration by
\citet{2008ApJ...687.1264M}\footnote{There are some differences between our adopted ages for groups and clusters and those
on which the \citet{2008ApJ...687.1264M} calibration is based, but they are mostly confined to the youngest groups
($\beta$ Pic and Upper Scorpius), then in an age range for which our age estimate is primarity based on other indicators.}.
Our target stars have colors extending outside some of the validity limits of these calibrations
(B-V between 0.5 to 0.9 for $\log R_{HK}$ and X-ray emission).
As shown in Fig.~\ref{f:ageindicators}, the results from known group members support the use of these
relations for most of our target stars, even when they are slightly outside the formal boundaries, leaning toward the red side
(from B-V=0.9 up to B-V=1.2 -- 1.3).
On the blue side the situation is different, as both $\log R_{HK}$ and $R_{X}$ becomes smaller when increasing
temperature at fixed age, due the vanishing of the stellar magnetic activity when the external convective zone of the star
become too thin. 
Therefore, the extrapolation of the \citet{2008ApJ...687.1264M} calibrations would provide
a significant overestimate of the stellar age\footnote{Actually; from Fig.~\ref{f:ageindicators} it results that $R_{X}$
is smaller for stars with B-V $\sim$ 0.5 with respect coeval stars with redder colors, indicating that the
color dependence starts to become important already in the range of \citet{2008ApJ...687.1264M} calibration}. 
The results of the age calibration are included in Table 12 as upper limits.

The measurement of stellar age from rotation period (gyrochronology) is becoming increasingly popular,
as it is not vulnerable to the time variability that characterizes the choromospheric and coronal emissions.
However, it should be noted that a group of coeval young stars usually
distributes along two sequences: that of very fast rotators 
\citep[{\it `C`} sequence following the nomenclature by ][]{2007ApJ...669.1167B} and that of slower rotators, with rotation period
increasing with color ({\it 'I'} sequence; Fig.~\ref{f:ageindicators}).
The gyrochronology calibration applies only to stars on the {\it 'I'} sequence; for stars on 
the {\it `C`} its use is not appropriate and only supports a young age for the star.

Age from isochrone fitting has been derived in Sect.~\ref{s:isoc}. As discussed above, in most cases
isochrone fitting does not add tighter constraints on stellar age when compared to other methods, since
the star position on the CMD is very close to the main sequence.
For stars that are off of the main sequence, the use of other age indicators (e.g. lithium) allows us to
conclusively discriminate between case of young pre-main sequence stars and evolved post-main sequence
stars, and to assign corresponding ages.
Position on the CMD is also affected by the presence of companions. Having sensitivity with NaCo imaging
down to a few-AU separation  for stellar companions, and having checked for the presence of blended components in 
high resolution spectra for most of our target stars, we think we have identified most cases of binary companions
contributing significantly to the integrated flux of the system.  However, the possibility of a few remaining 
cases can not be excluded.

Finally, as discussed in Sect.~\ref{s:groups}, kinematic parameters alone can not
be considered as conclusive proof of a given stellar age value, as numerous old stars
occupy the same kinematic space of young stars.
Final assignment of membership to groups is based on both the kinematic analysis and the
consistency of the age indicators with the age of the group.
Ambiguous cases arise for stars where the other age indicators yield poor constraints, such as
early F stars. 
The existence of space velocities firmly outside the kinematic space on which confirmed young stars are found (see for example 
the \citet{2001MNRAS.328...45M} boundaries, also shown in Fig.~\ref{f:ageindicators}) is considered a strong indication for an age older than about 500 Myr.

The spread in rotation rates for coeval young stars \citep[see e.g.~][]{2010A&A...520A..15M} and the 
saturation of X-ray emission at fast rotation periods \citep{2003A&A...397..147P} make the dependence 
of these indicators shallow for young ages (ages smaller than the Pleiades), as discussed in e.g. \citet{soderblom13}.
As an example,  $\beta$ Pic, Tucana-Columba-Carina and
AB Dor members overlap on the $log L_{X}/L_{bol}$ vs B-V diagram in Fig.~\ref{f:ageindicators}.
Therefore, the exact numerical values resulting from age calibrations can not be taken at face value, 
but the consistency of activity and rotation is a strong indication of youth, unless fast rotation is
due to alternative causes (see below).
For such candidate young stars, lithium EW, kinematics and in a few cases isochrone fitting are more robust 
indicators of the stellar age.

Special care is required for close spectroscopic binaries. Tidal locking represents a viable alternative
to young age for explaining the fast rotational velocity and high levels of chromospheric and 
coronal activity\footnote{It should 
be noted that the two explanations are not mutually exclusive. There are cases of close tidally locked 
binaries that are found to be young based on lithium EW, membership to groups, or presence 
of disks (e.g. V4046 Sgr).}.
In these cases, indicators such as rotation period and coronal and chromospehric emission are not 
considered for age determination, and we rely only on indicators not directly linked to rotation, 
such as lithium, kinematics and isochrone fitting.
For the spectroscopic binaries in our sample, the orbital period is typically not available, as the binarity
was identified with just a handful of RV measurements, or even one single spectrum in the case of SB2.
Therefore, we can not conclusively infer the tidal locking nature of active binary stars like 
TYC 7188-575-1 and TYC 5206-0915-1.
Furthermore, the lack of an RV orbital solution prevents the determination of a system RV, adding 
uncertainties on the space velocities and effectively making them useless in some cases.
Lithium EW is available in most cases, and age estimates for candidate tidally-locked binaries 
are primarily based on this diagnostic. 
However, it is possible that lithium abundance is also affected in close binaries, as it is often
observed in RS CVn systems \citep{1992A&A...253..185P}.

By applying these techniques, we derived ages for our target stars.
The ages from individual indicators and the final adopted values are given in Table \ref{t:agetable}.
Most of them are confirmed to be young.
Besides the tidally-locked systems and the star rejuvenated by the progenitor of its WD companion 
\citep[HIP 6177,][]{2013A&A...554A..21Z},
other cases of old stars spuriously included in our sample are represented by the Li-rich giant
HD 99409, by stars for which our determinations of age indicators are in contrast with
the literature value (HIP 13008; HIP 114046), and by old interlopers in the kinematic space of
young stars (HIP 71908, formerly classified as an Her-Lyr member).

Most of our target stars have not been previously observed in deep imaging 
and in several cases their ages have not been studied in detail.
However, some of them were included independently in NICI imaging surveys \citep{2013ApJ...773..179W,2013ApJ...777..160B}.
For the seven stars in common, we obtained identical ages for 
one member of AB Dor MG (HIP~14684) and one of Her-Lyr (HD 139664).
HIP~99273 and TYC~8728-2262-1 were assigned to the $\beta$ Pic MG in both cases, but 
there is a slight difference in the adopted group age.
For HIP~11360, the star is classified as a $\beta$ Pic member in NICI papers, while we adopt
membership in the Tucana association, based on our revised system RV.
For HD~61005, we adopt membership in Argus (40 Myr) while \cite{2013ApJ...773..179W} adopt an age of 100 Myr
without assigning membership to any known group.
Finally, HIP 71908 is included in the Her-Lyr membership in \cite{2013ApJ...777..160B}, while we claim
it is a significantly older star with similar kinematics.
While we are convinced of our adopted choices, this comparison is a further indication of the
challenges represented by the age determination of young stars.

\section{Discussion and conclusions}
\label{s:conclusion}

We have performed a characterization of the targets of the NaCo-LP to probe the occurrence 
of exoplanets and brown dwarfs in wide orbits.
To this aim we acquired spectroscopic data using FEROS, HARPS and CORALIE and retrieved 
additional spectroscopic data from public archives.
From these spectra, we determined radial and projected rotational velocities, lithium equivalent widths, 
effective temperatures, and chromospheric emission of the targets.
We also analyzed photometric time series available from ASAS and Super-WASP archives to
derive rotation periods and investigate the stellar variability phenomena.
These data were used to investigate basic properties of the targets of our deep-imaging 
planet search program.  These basic properties included ages, distances, masses, and multiplicity.
Such information is needed for a proper interpretation of the direct imaging results, which
are presented in the companion paper by Chauvin et al.~(2013).

\begin{figure*}[h]
\includegraphics[width=18cm]{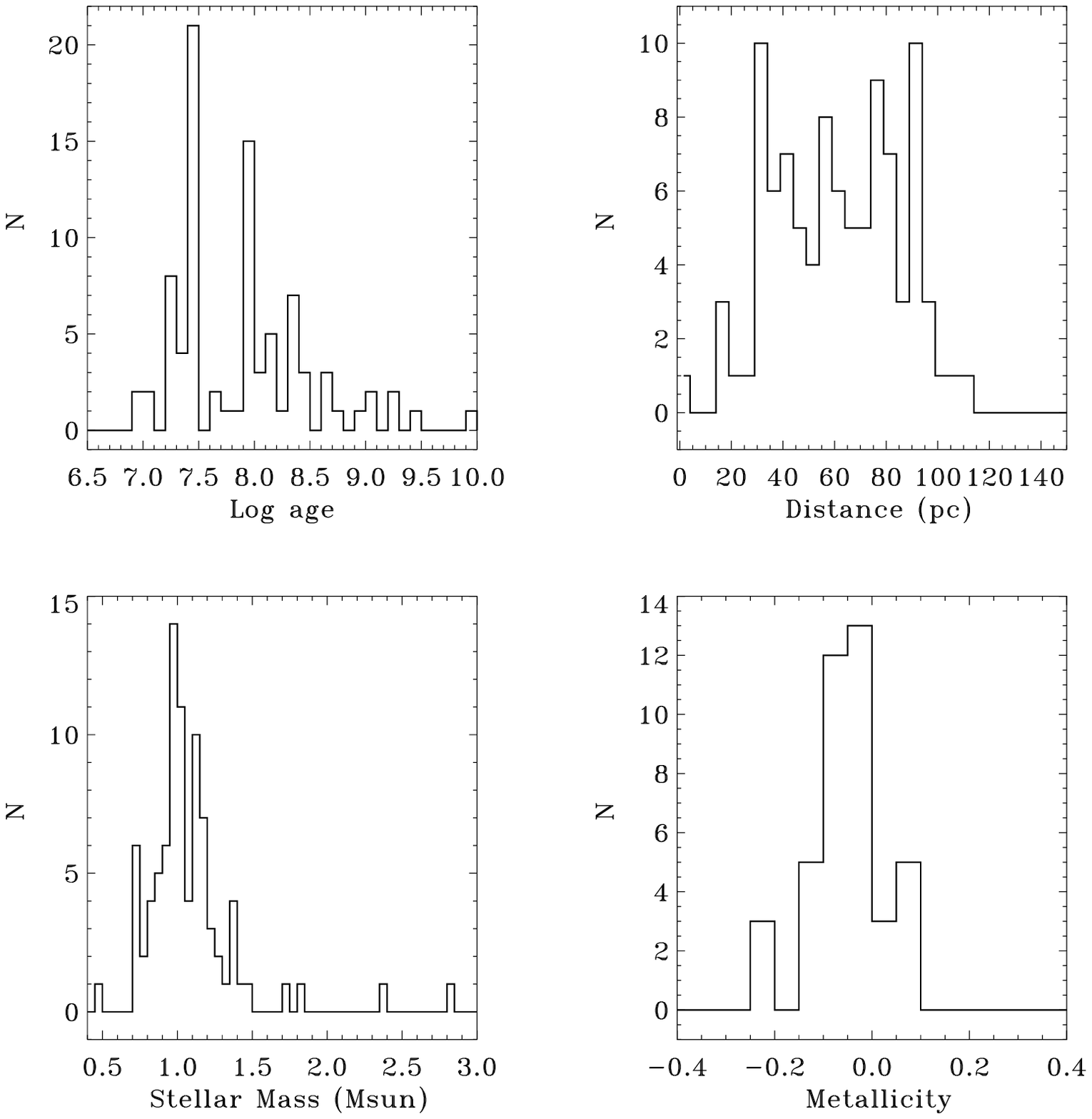}
\caption{Distribution of stellar age, distance, mass, and metallicity for the stars in 
the NaCo-LP sample.}
\label{f:histsample}
\end{figure*}

\begin{figure}[h]
\includegraphics[width=8cm]{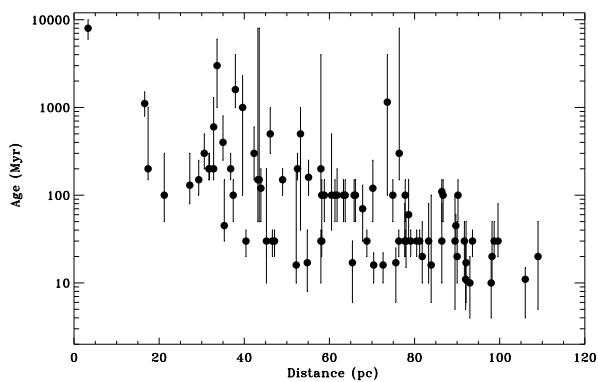}
\caption{Age vs distance for the stars in the NaCo-LP sample}
\label{f:agedist}
\end{figure}

Histograms showing the distribution of age, distance, mass and metallicity are in Fig.~\ref{f:histsample}.
The irregular behavior of the age distribution is due to members of MGs, which cluster at the same age
in the histogram.
The median age of the targets is 100 Myr.
Few targets have estimated ages below 20 Myr and even fewer have ages older than 500 Myr, representing
old interlopers for which our original selection criteria failed to remove. 
The presence of older interlopers is due to a variety of causes related to age estimation, as discussed in Sect.~\ref{s:age}.
The median distance of the stars in our sample is 63.7 pc. 
Considering that the  region closest to the star is saturated or hidden by the coronagraph in our
imaging observations (Chauvin et al.~2013), our survey is then typically sensitive to projected separation down to
about 20 AU.
Stellar age and distance from the Sun are correlated, as shown in Fig.~\ref{f:agedist}. 
This correlation is the result of the low space density of very young stars in the solar neighborhood,
and of several nearby young stars being previously observed in other direct imaging surveys.

The average metallicity of stars in the NaCo-LP sample is close to solar (median value [Fe/H]=--0.03), 
with a dispersion that is significantly smaller than that of old nearby stars studied in RV surveys, 
which include both moderately metal-poor stars and metal-rich stars 
\citep{2005ApJ...622.1102F,2008A&A...487..373S}. Furthermore, the stars 
with the lowest and highest metallicity in our sample are typically among the oldest ones, thus carrying limited 
weight in terms of detection limits of planetary companions. Therefore, the exploration of any correlation between
the frequency of giant planets in wide orbits and metallicity appears to be challenging for the direct
imaging technique.

The median stellar mass is $1.01~M_{\odot}$, with  94\% of the stars having masses 
between 0.70 and 1.50 $M_{\odot}$. Therefore, our sample is tuned to the study of the
occurrence of planets in wide orbit around solar-type stars and complements the results
of RV surveys, which are mostly focused in the same range of stellar masses.
Young, low mass stars are virtually excluded by our adopted magnitude limits, while
few early type stars with confirmed young ages were known at the time of our sample selection.
However, dedicated imaging surveys devoted to the search for planets around early type stars
were performed in recent years \citep{2011ApJ...736...89J,2012A&A...544A...9V} while other
authors focused their attention on low mass stars \citep{2012A&A...539A..72D}.
Appropriate merging of these samples and of those of other surveys which are mostly focused
on solar-type stars \citep[e.g.][]{2007ApJ...670.1367L} will allow a preliminary evaluation of the frequency
of planets in wide orbits. We will address this issue in the forthcoming paper by Vigan et al.~(2014).
The soon-to-come new generation instruments such as SPHERE and GPI will allow an extension of these results
to lower planetary masses and closer separations to the star, thus providing, when coupled with RV results,
a complete view of the occurrence of giant planets over a wide range of separations.

\begin{acknowledgements}
Based on observations collected at La Silla and Paranal Observatory, ESO (Chile):
Programs 
70.D-0081A, 072.D-0021, 072.C-0393, 074.D-0016, 074.A-9002, 074.A-9020,
076.C-0578, 076.A-9013, 077.C-0138, 077.D-0525,
081.A-9005, 082.A-9007, 083.A-9003, 083.A-9011, 083.A-9013,
084.A-9004, 084.A-9011 (FEROS),
072.C-0488, 074.C-0037, 075.C-0689, 076.C-0010, 077.C-0012, 077.D-0085, 077.C-0295, 079.C-0046, 079.C-0170, 
080.D-0151, 080.C-0712, 080.C-0664, 081.D-0008, 081.C-0779, 082.C-0390, 184.C-0815 (HARPS), 
184.C-0157 (NaCo).
Based on data obtained from the ESO Science Archive Facility.
This research has made use of the SIMBAD database and Vizier services, operated at CDS, Strasbourg, France 
and of the Washington Double Star Catalog maintained at the U.S. Naval Observatory. 
We have used data from the WASP public archive in this research. The WASP consortium comprises of the 
University of Cambridge, Keele University, University of Leicester, The Open University, The Queen's University Belfast, 
St. Andrews University and the Isaac Newton Group. Funding for WASP comes from the consortium universities and 
from the UK's Science and Technology Facilities Council.\\
We warmly thank the anonymous referee for helpful comments.
We thank the SACY team for providing reduced spectra of NaCo-LP targets from their archive.
We thank the SOPHIE Consortium for providing unpublished activity measurements.
We thank L.~Pastori for providing the CES spectra of HD16699A and B.
We thank B.~Mason for providing  astrometric data for individual objects collected in the Washington Double Star Catalog. \\
We acknowledge partial support from PRIN-INAF 2010 ``Planetary systems at young ages and the
interactions with their active host stars''.  JC was supported by the US National Science Foundation under
Award No. 1009203.

\end{acknowledgements}

\bibliography{nacolp}
\bibliographystyle{aa}

%
%
\addtocounter{table}{1}
%

%
\clearpage
%
\onecolumn

\include{tablong_ave}

\begin{table*}[h]
\caption{Measurements of spectroscopic parameters from CORALIE RV survey} 
\label{t:coralie}
\begin{center}       
\begin{tabular}{cccccccl} 
\hline
ID1 &      ID2       &  N   &      RV  &  vsini & $\log R_{HK}$  & Remarks \\
    &                &      &    km/s  &   km/s &               &  \\
\hline
 HIP 6177    &  HD 8049   &    &           &  3.52 & -4.252  & RV trend, Zurlo et al.~(2013)  \\
 HIP 14684   &  HD 19668  &  4 &  14.65    &  6.29 & -4.38   &   \\
 HIP 37923   &  HD 63608  &  5 &  17.73    &  2.67 & -4.372  &   \\
 HIP 79958   &  HD 146464 &  3 &   1.74    &       & -4.445  &   \\
 HIP 105384  &  HD 203019 & 10 &  24.43    &  5.86 & -4.356  &   \\
 HIP 105612  &  HD 202732 &  3 &   -1.7    &  3.28 & -4.476  &   \\
\hline
\end{tabular}
\end{center}
\end{table*}

\begin{table*}[h]
\caption{Measurements of spectroscopic parameters from dedicated CORALIE observations}
\label{t:coralie_new}
\begin{center}       
\begin{tabular}{lccccl} 
\hline
ID1            &  JD -2450000  &  RV             &  FWHM  &   S/N     & Remarks \\
               &               &   km/s          &  km/s  &           & \\
\hline
TYC 8128-1946-1 &  5927.6395   &  17.0$\pm$0.5   &   ...  &   94 & individual lines, $v \sin i \sim 50$ km/s \\ 
HIP 36312       &  5931.8242   &  16.62$\pm$0.07 &  40.1  &   43 & pipeline \\  
TYC 7796-2110-1 &  5931.8261   &    1.3$\pm$4.5  &   ...   &   15 & individual lines, $v \sin i \sim 90$ km/s \\
TYC 7188-0575-1 &  5931.8654   & -38.75$\pm$0.05 &  42.9  &   32 & manual recorrelation  \\ 
HIP 25436       &  6360.6450   &   21.8$\pm$3.8  &   ...   &   32 & individual lines, $v \sin i \sim 110$ km/s \\ 
HIP 25434       &  6360.6380   &   24.2$\pm$1.5  &   ...   &   17 & individual lines, $v \sin i \sim 50$ km/s \\ 
TYC 7188-0575-1 &  6357.7972   & -24.71$\pm$0.08 &  39.1  &   14 & manual recorrelation \\
HIP 35564       &  6357.7330   &   13.7$\pm$1.1  &   ...   &  108 & individual lines, $v \sin i \sim 60$ km/s \\ 
HIP 72399       &  6357.7867   &  30.31$\pm$0.05 &  15.1  &   17 & pipeline \\ 
                &  6357.8246   &  33.24$\pm$0.06 &  16.7  &   16 & pipeline \\ 
HIP 72400       &  6360.7397   &   1.35$\pm$0.02 &   8.1  &   23 & pipeline \\ 
HIP 13008       &  6509.9063   &    0.6$\pm$1.6  &   ...   &   27 & individual lines, $v \sin i \sim 80$ km/s \\ 
TYC 5606-0915-1 &  6508.6744   & 17.47$\pm$0.06  &  31.0  &   21 & pipeline \\ 
\hline
\end{tabular}
\end{center}
\end{table*}

\begin{table*}[h]
\caption{Measurements of spectroscopic parameters from HARPS spectra. The mean
measurement of individual spectra is listed when more than one spectrum is available.
For early type stars we list both the absolute RV obtained with the HARPS pipeline and the RV dispersion
of the differential RV obtained with the SAFIR code 
} 
\label{t:harps3}
\begin{center}       
{\scriptsize
\begin{tabular}{cccccccccccl} 
\hline
ID1             &  N & JD start  & JD end   & RV   & rms RV &  $v \sin i$ &  FWHM  & $\log R_{HK}$ & EW(Li)  &  $T_{\rm eff}$ & Remarks \\
                &    & -2450000  & -2450000 & (km/s) & (km/s)   &   (km/s) &   (km/s) &              &  (m\AA)   &  (K)    &  \\
\hline
HIP~6177        &  1 & 3263.   &    --     &  42.00 &   -- &   7.19 &  --      &         &    0 & 5045$\pm$17  & Zurlo et al.~(2013) \\  
HIP~11360       &  8 & 4439.68 &  4479.58  &   0.81 & 2.13 &        & $\sim$20 &         &   95 &              & pipeline \\  
                &    &         &           &        & 0.12 &        &          &         &      &              & SAFIR \\
HIP~14684       & 12 & 3311.79 &  3805.53  &  14.72 & 0.05 &        &   10.304 &         &  179 & 5451$\pm$7   &   \\  
HIP~32235       & 17 & 2948.87 &  3805.62  &  20.72 & 0.13 &  17.38 &   18.124 & -4.169  &  239 & 5493$\pm$11  &   \\  
HIP~36948       & 12 & 5343.45 &  5347.47  &  22.64 & 0.04 &        &  12.830  &         &  173 & 5493$\pm$15  &    \\
\hline
\end{tabular}
}
\end{center}
\end{table*}

\include{tables_nacolp}
\include{tablong_age}

\clearpage
\Online
\appendix

\twocolumn

\section{Literature compilation}
\label{app:lit}

The compilation of literature values for several stellar parameters (RV,
$v \sin i$, EW of Li 6707.8 \AA~ resonance line, $\log R_{HK}$, $T_{\rm eff}$)  is shown in
Table \ref{t:spec_lit}.

\begin{table*}[h]
\caption{Summary of measurements of spectroscopic parameters available in the literature} 
\label{t:spec_lit}
\begin{center}       
\scriptsize{
\begin{tabular}{lccccccccl} 
\hline
Star             & JD        &  RV         &  $v \sin i$   & EW Li & S      & $\log R_{HK}$ &  $T_{\rm eff}$  & Reference & Remarks \\
                 & -2450000  &  (km/s)       &  (km/s)    & (m\AA)  &        &              &   (K)    &           &         \\
\hline

TYC 5839-0596-1  &         &   20.0        & 36.5$\pm$3.7  &  100  &        &          &        &   SACY   & N=1, SB2, Halpha emission \\
TYC 5839-0596-1a & 5106.75 & +51.5$\pm$1.5 & 18.0$\pm$2.0  &       & 2.135  & -3.990   &        &   Des13  & FEROS, deblended \\
TYC 5839-0596-1b & 5106.75 & +21.3$\pm$1.5 & 17.0$\pm$2.0  &       &        &          &        &   Des13  & FEROS, deblended  \\
\hline
TYC 0603-0461-1  &         &  -12.9        &          &   160      &        &          &        &  SACY         &  N=1 Halpha emission \\
\hline
HIP 3924         &         &   10.7        &  28          &   138     &        &        &              &   Wic03       &  \\
                 &         &               & 33.5$\pm$1.9 & 132$\pm$5 &        &        &              &   Wei10       & \\
\hline

\end{tabular}
}
\end{center}
\end{table*}

\begin{table*}[h]
\caption{Summary of measurements of spectroscopic parameters available in the literature} 
\begin{center}       
\scriptsize{
\begin{tabular}{lccccccccl} 
\hline
Star             & JD        &  RV         &  $v \sin i$   & EW Li & S      & $\log R_{HK}$ &  $T_{\rm eff}$  & Reference & Remarks \\
                 & -2450000  &  (km/s)       &  (km/s)    & (m\AA)  &        &              &   (K)    &           &         \\
\hline
HIP 25434        & 2721.03    & 24.94$\pm$0.33 &   40  &            &  0.39  &  -4.30   &        &   Des06  &  \\  
                 & 2721.03    &                &       & 183        &        &          &        &   Des13  & FEROS, same spec as Des06 \\
                 & Jul-Aug 09 &  23.6$\pm$1.8  &       & 158$\pm$10 &        &          &        &   Moo13  &  \\
                 & 6360.64    &  24.2$\pm$1.5  &   50  &            &        &          &        &   Des13  & CORALIE \\
\hline
HIP 25436        & 2721.01    & 32.17$\pm$1.30  &  112   &       &   0.25 &  -4.57   &        &   Des06  &  \\  
                 & 2721.01    &                 &        &  107  &        &          &        &   Des13  & FEROS, same spec as Des06 \\
\hline

\end{tabular}
}
\end{center}
\end{table*}

\begin{table*}[h]
\caption{Summary of measurements of spectroscopic parameters available in the literature} 
\begin{center}       
\scriptsize{
\begin{tabular}{lccccccccl} 
\hline
Star             & JD        &  RV         &  $v \sin i$   & EW Li & S      & $\log R_{HK}$ &  $T_{\rm eff}$  & Reference & Remarks \\
                 & -2450000  &  (km/s)       &  (km/s)    & (m\AA)  &        &              &   (K)    &           &         \\
\hline
HIP 46634        &         & 27.3$\pm$0.2  &  4        &       &        &          &        &   Nor04  &  N=2 rms RV 0.1  km/s  comp. exchanged \\  
                 &         &                &          &  25   &        &          &        &   Cut99  &  \\
                 &         &                &          &       &        &          &        &   Str00  &  \\
                 &         & 27.1$\pm$0.4   &          &       &        &          &        &   Mon01  &  \\
                 &         & 27.07$\pm$0.04 &          &       &        &          &        &   Gal05  & multi-epoch, RV constant  \\
                 &         & 27.93$\pm$0.05 &  $<3$    &  33$\pm$6  &        &          &  5140$\pm$35   &   Str12  & N=96 DT=1129, single  \\
\end{tabular}
}
\end{center}
\end{table*}

\begin{table*}[h]
\caption{Summary of measurements of spectroscopic parameters available in the literature} 
\begin{center}       
\scriptsize{
\begin{tabular}{lccccccccl} 
\hline
Star             & JD        &  RV         &  $v \sin i$   & EW Li & S      & $\log R_{HK}$ &  $T_{\rm eff}$  & Reference & Remarks \\
                 & -2450000  &  (km/s)       &  (km/s)    & (m\AA)  &        &              &   (K)    &           &         \\
\hline
TYC 9010-1272-1  &         & 10.2          & 38$\pm$3.8 &  210  &        &          &             &    SACY   & N=2, rms RV 0.5 km/s, SB2?  \\ 
                 &         & 43.4$\pm$4.9  &    8       &       &        &          &             &    Nor04  & N=5 rms RV 11 km/s \\ 
                 & 3909.59 & 11.5$\pm$0.8  &            & 200   &  0.534 &  -4.137  & 5324$\pm$92 &    Des13  & FEROS, SB2 \\
\hline
HIP 70351        &         & -8.3          &  16.9$\pm$1.2 &  170  &        &          &        &    SACY      & N=1 \\
                 & 4940.77 & -8.6$\pm$1.2  &  12.3$\pm$0.7 &  164  &  0.460 &  -4.301  & 5624$\pm$50  &    Des13     & FEROS  \\
\hline
$\alpha$ Cir     & 3423.84 &  6.2$\pm$0.2   & 14.5$\pm$1.8 &         &        &        &               & Des13 & FEROS \\
\end{tabular}
}
\end{center}
\end{table*}

\begin{table*}[h]
\caption{Summary of measurements of spectroscopic parameters available in the literature} 
\begin{center}       
\scriptsize{
\begin{tabular}{lccccccccl} 
\hline
Star             & JD        &  RV         &  $v \sin i$   & EW Li & S      & $\log R_{HK}$ &  $T_{\rm eff}$  & Reference & Remarks \\
                 & -2450000  &  (km/s)       &  (km/s)    & (m\AA)  &        &              &   (K)    &           &         \\
\hline
TYC 5736-0649-1  &         &    -7.0       &  206.0$\pm$20.6 & 250      &        &          &        &    SACY      & N=3, rms RV 3.5 km/s, SB? \\
                 &         &               &                 &          & 0.401  & -4.360   &        &   Des13  & FEROS \\
\hline
HD 189285        &           &                &  9.0         &  140       &         &         &  5630         &  DaS09 & \\
                 &           &                &  8.0$\pm$2.0 &            &         &         &  5800         &  DOr12 & \\
                 & 3906.84   &  -19.5$\pm$0.1 & 8.3$\pm$0.9  &  141       &  0.455  &  -4.325 &  5545$\pm$35  &  Des13 & FEROS \\
                 & 4782.53   &  -19.6$\pm$0.2 & 8.3$\pm$0.7  &  135       &  0.449  &  -4.331 &  5620$\pm$23  &  Des13 & FEROS \\
\hline
\end{tabular}
}
\end{center}
\end{table*}

\begin{table*}[h]
\caption{Summary of measurements of spectroscopic parameters available in the literature} 
\begin{center}       
\scriptsize{
\begin{tabular}{lccccccccl} 
\hline
Star             & JD        &  RV         &  $v \sin i$   & EW Li & S      & $\log R_{HK}$ &  $T_{\rm eff}$  & Reference & Remarks \\
                 & -2450000  &  (km/s)       &  (km/s)    & (m\AA)  &        &              &   (K)    &           &         \\
\hline
TYC 9338-2016-1  &         &   9.5         &  29.9$\pm$0.5  &   259        &        &          &        &   SACY   & N=9, rms RV 0.8 km/s \\
                 &         &   5.5$\pm$1.9 &                &              &        &          &   5973 & RAVE3  &   \\ 
                 &         &               &  31.8$\pm$1.9  &   259$\pm$4  &        &          &        &  Wei10       & \\
                 & 4778.58 &   7.3$\pm$1.7 &                &   221        & 0.568  & -4.167   &        &  Des13 & FEROS \\

\hline
TYC 9529-0340-1  &         &   9.2         &  73.9$\pm$1.4  &   276    &        &          &        &  SACY    & N=11 rms RV=2.0 km/s \\
                 &         &               &  74        &  276     &        &          &  5481      &  DaS09   &  \\
\hline
\end{tabular}
}
 \tablebib{
Des13:~this paper; 
SACY: \citet{2006A&A...460..695T}; 
Aar08: \citet{2008AJ....136.2483A}; 
Alc96: \citet{1996A&AS..119....7A};
Alc00: \citet{2000A&A...353..186A};
Amm12: \citet{2012A&A...542A.116A};
Baa97: \citet{1997A&A...323..429B};
BaB00: \citet{2000A&AS..142..217B};
Bia12: \citet{2012MNRAS.427.2905B};
Bub11: \citet{2011AJ....142..180B};
Cin07: \citet{2007A&A...469..309C};
Cut99: \citet{1999A&AS..138...87C};
Cut02: \citet{2002A&A...384..491C};
Cut03: \citet{2003A&A...397..987C};
DaS09: \citet{2009A&A...508..833D};
Deb08: \citet{2008ApJ...684L..41D};
Des06: \citet{2006A&A...454..553D};
DOr12: \citet{2012MNRAS.423.2789D};
Fav95: \citet{1995A&A...295..147F};
Feh74: \citet{1974A&AS...13..173F};
Gal05: \citet{2005ESASP.560..567G};
Gal06: \citet{2006Ap&SS.304...59G};
Gle05: \citet{2005yCat.3244....0G};
Gon06: \citet{2006AstL...32..759G};
Gra06: \citet{2006AJ....132..161G};
Gre99: \citet{1999A&AS..137..451G};
Gui09: \citet{2009A&A...504..829G};
Hen96: \citet{1996AJ....111..439H};
Jen06: \citet{2006MNRAS.372..163J};
Jen11: \citet{2011A&A...531A...8J};
Lag08: \citet{2009A&A...495..335L};
LoS06: \citet{2006ApJ...643.1160L};
Mam02: \citet{2002AJ....124.1670M};
Mam04: \citet{2004PhDT........20M}; 
Mas06: \citet{2006A&A...450..735M};
Men08: \citet{2008ApJ...689.1127M};
Mon01: \citet{2001MNRAS.328...45M}
Moo06: \citet{2006ApJ...644..525M};
Moo11: \citet{2011ApJS..193....4M};
Moo13: \citet{2013MNRAS.435.1376M};
Nid02: \citet{2002ApJS..141..503N};
Nor04: \citet{2004A&A...418..989N};
Pau05: \citet{2005A&A...444..941P};
Pau06: \citet{2006PASP..118..706P};
RAVE3: \citet{2011AJ....141..187S};
Rei03: \citet{2003MNRAS.342..837R};
Rei04: \citet{2004A&A...415..325R};
Rei06: \citet{2006A&A...446..267R};
Rhe07: \citet{2007ApJ...660.1556R};
RoP02: \citet{2002A&A...384..912R};
Roy02: \citet{2002A&A...381..105R};
Roy07: \citet{2007A&A...463..671R};
Saa90: \citet{1990A&A...235..291S};
Sch07: \citet{2007ApJ...662.1254S};
Sch09: \citet{2009A&A...493.1099S};
Son02: \citet{2002A&A...385..862S};
Son03: \citet{2003ApJ...599..342S};
Son12: \citet{2012AJ....144....8S};
Sou10: \citet{2010A&A...515A.111S};
Str00: \citet{2000A&AS..142..275S};
Str12: \citet{2012AN....333..663S};
Tor00: \citet{2000AJ....120.1410T};
ViA09: \citet{2009A&A...501..965V};
VSOP: \citet{2007A&A...470.1201D};
Wai05: \citet{2005PASA...22...29W};
Wai11: \citet{2011PASA...28..323W};
Wei10: \citet{2010A&A...517A..88W};
Whi07: \citet{2007AJ....133.2524W};
Wic99: \citet{1999MNRAS.307..909W};
Wic03: \citet{2003A&A...399..983W};
Woo05: \citet{2005MNRAS.356..963W};
Wri04: \citet{2004ApJS..152..261W};
XHIP: \citet{2012AstL...38..331A};
Zor12: \citet{2012A&A...537A.120Z};
Zuk06: \citet{2006ApJ...649L.115Z}.
}
\end{center}
\end{table*}

\section{Notes on individual objects}
\label{a:notes}

\begin{description}

\item {\bf TYC 5839-0596-1}:  (ASAS 001208$-$1550.5) 
The star is flagged as SB2 in SACY and we confirm this classification. 
\cite{2012AcA....62...67K} reports a photometric rotation period P=2.532d based on the 
ASAS photometry. We find two significant photometric periods, P=1.644d and P=2.53d, of comparable powers 
in almost all the seasonal  Lomb-Scargle and Clean periodograms. 
The longer period is more consistent with $v \sin i=$ 18 kms$^{-1}$ and with the 
radius of a K0IV PMS subgiant.  P=1.644d is an alias, specifically a beat period between 
the star's rotation period and the $\sim$ 1d data sampling.  This is a result of the rotation of the Earth and the fixed longitude of the observation site. 
The star has also a very large X-ray luminosity ($\log L_{X}/L_{bol}=-3.0$).
While we do not have the orbital period of the SB2, 
the similar $v \sin i$ of the two components suggests that this is a tidally locked system.
The star also shows a moderately large lithium abundance, compatible with an age similar to the Pleiades.
We are unable to conclude whether this is a true young binary or if the system
is a RS CVn variable with a relatively large lithium abundance \citep[see e.g.][]{1992A&A...253..185P}.
The photometric distance was obtained assuming equal magnitudes for the components.

\item {\bf TYC 0603-0461-1 = BD +07 85}:   (ASAS 003858+0829.0) 
New close visual binary from the NaCo-LP; $\Delta H=0.10$ mag at a projected separation of
74 mas, corresponding to 4.3 AU at the photometric distance, after correction for
binarity. 
It is kinematically consistent with the Nearby, Young population.
We find three photometric periods: P=3.60d, P=0.566d, and P=1.381d. 
Although they are all significant, their relative power changes from season to season. 
We have no $v \sin i$ measurements that together with the $R = 1.026~R_{\odot}$ stellar radius could 
allow us to fix an upper value to the rotation period.
P=1.381d is the most significant period also revealed by the Clean analysis, and may
be the stellar rotation period. However, the other periods cannot be ruled out. 
Considering the binarity and the small magnitude difference between the components,
the multiple periodicities reported in the photometric analysis might be due
to the different components. 

\item {\bf HIP 3924 = HD 4944}:  (ASAS  005024$-$6404.1; 1SWASP+J005024.31-640404.0) 
Our FEROS spectra indicate the presence of an additional component, which is more evident in the red part of the
spectrum. 
Binarity is further supported by the RV difference of
about 9 km/s between our RV and that measured by \cite{2003A&A...399..983W}.
Therefore the star is classified as SB2.
We measured an RV difference of about 18 km/s between the components, with a line depth for the secondary that is
about 30\% that of the primary, at 6450\AA. 
Our projected rotational velocity and Li EW are significantly different from those reported in the
literature by \cite{2003A&A...399..983W} and \cite{2010A&A...517A..88W}.
The first discrepancy might be due to a different RV separation between the components at the time of observation.
For Li EW, our FEROS spectrum indicates that the two components have a similar lithium line depth.
We find various significant photometric periodicities. 
Only one, P=9.3d, is consistent with  v$\sin{i}=$ 8 kms$^{-1}$ and the stellar radius.  It is found in the 
complete ASAS series as well as in one segment. As the primary is classified as F7, it is possible that the photometric
period belongs to the secondary.
No period is derived from the Hipparcos photometric data.
Lithium and $\log R_{HK}$  indicators are consistent with an age comparable to the Hyades, while X-ray emission yields 
younger ages than the other indicators.
These discrepancies may be linked to the blending of the components.
The target's kinematics are only consistent with IC2391 and with
the Nearby, Young population. 
We adopt an age of 500 Myr, with a lower limit at the age of the IC2391 MG.

\item {\bf HIP 6177 = HD 8049}: Moderately old star rejuvenated by stellar wind from the progenitor of its WD companion.
The properties of the K dwarf and the WD companion and a description of the most probable evolution of the system
are presented in \citet{2013A&A...554A..21Z}. 

\item {\bf HIP 8038}:  New close binary from the NaCo-LP ($\Delta H=2.5$ mag at 0.44 arcsec).
There is also a wide companion, {\bf HIP 8039}, which is 1.9 mag brighter and at a projected separation of 14.97 arcsec
(this star was not observed as part of the NaCo-LP).
All the components are included in the aperture photometry of ASAS and SuperWASP (referred to as ASAS 014314$-$2136.7 and 1SWASP+J014314.17-213656.5, respectively).
Each season exhibits its own most significant period which differs from season to season in both ASAS and SuperWASP data. 
The most likely period in SuperWASP seems to be P=1.668d, but with a scattered phase light curve. 
No clear periodicity comes out from the Clean analysis or from the Hipparcos photometry.
As the wide companion HIP 8039 is 1.9 mag brighter, the tentative rotation period most likely belongs to the brighter component.
The target's  kinematics and age
are only  consistent with the  Young, Nearby population.

\item {\bf HIP 10602  = HD 14228 = $\phi$ Eri } (ASAS 021631$-$5130.7)
Proposed in \citet{2008hsf2.book..757T} and  \citet{2013ApJ...762...88M}  as   a  high
probability member of Tuc/Hor.
Our analysis confirms that it has kinematics consistent,  at the $\le 1.0$ sigma level,  with the Tuc/Hor association
\citet{2014ApJ...783..121G} tools also yield a very high membership probability.
Torres et al. investigated color-magnitude diagram placement and
found  it  to be  consistent  with  Tuc/Hor  membership. 
The star is quoted as binary in \cite{2004ARA&A..42..685Z} but the identification and properties
of the companion are not provided.
The candidate at about 90 arcsec  ({\bf \object{$\phi$ Eri B} = CPD-52 284}) is not physically associated, as reported
in the Washington Double Star (WDS) catalog \citep{2013yCat....102026M}. 
This early-type star exhibits non-periodic photometric variability.

\item {\bf HIP 11360  = HD 15115} (ASAS 022616+0617.6).
This F-type star's membership to the $\beta$ Pic MG was first proposed by \cite{2006ApJ...644..525M}.
\cite{2008hsf2.book..757T} concluded a possible membership (prob. 60\%).
\citet{2013ApJ...762...88M}  instead propose that the star is  a bona  fide member  of the
Columba  association.
These estimates are based on the only RV measurement available in the literature
previous to our work \citep[8.8$\pm$3.0 km/s;][]{2006ApJ...644..525M}. Our RV determination (1.7$\pm$1.0 km/s), which includes 
8 HARPS and 2 FEROS spectra, has better accuracy and concludes a value quite different from the previously published one.
Adopting our RV value,  \citet{2013ApJ...762...88M} and \citet{2014ApJ...783..121G} tools support a high membership
probability (98\%) for the  Tuc/Hor association.
The discrepancy between \cite{2006ApJ...644..525M} RV and our value is marginally significant
(about 2 sigma). It might be due to the presence of a close companion.  However, the high-precision
differential RVs obtained with the SAFIR code applied to HARPS data show a dispersion of only 120 m/s over 40 days,
with intra-night RV and line bisector variations suggestive of pulsation occurrences. 
The age determination from indirect indicators is inconclusive,
as expected for an F star. Both the young ages of the Tucana association and older ages, up to about 
200 Myr, are compatible with the data.
Verifying  the  true  membership  will  require
further investigation, including confirmation of the adopted RV value with
longer baseline observations.   
The star exhibits non-periodic photometric variability.
HIP 11360 hosts a debris disk that is unique 
for its large asymmetry, being detected out to 315~AU on the east side
and more than 550 AU on the west side \citep{2007ApJ...661L..85K}. 
\citet{2012ApJ...752...57R} showed that the geometry is wavelength-dependent, and found indications
of a gap in the disk at about 1 arcsec. 

\item {\bf TYC 8484-1507-1 = HD 16699B}
This star is a new close visual binary from the NaCo-LP observations (sep. 0.06 arcsec = 3.6 AU, $\Delta H$=0.23 mag).
SACY lists it as a possible SB. 
There is an additional component ({\bf HIP 12326 = HD 16699}) at 8.7 arcsec, corresponding to a projected separation of 530 AU.
According to \citet{1999A&AS..138...87C}, HD 16699 shows a composite spectrum, 
with one system of sharp lines and another system of broad lines superimposed. 
Since the RV is constant, this was interpreted as being due
to a companion at moderately large separation.
This would indicate a quadruple system. However, we note that the $v \sin i$ values
of the two components in \cite{1999A&AS..138...87C} are roughly exchanged with respect to
other studies in the literature (Table \ref{t:spec_lit}).
An analysis of the CES spectra used in  \citet{1999A&AS..138...87C} 
indicates that the components were indeed exchanged in the original study. 
In light of this fact, the new companion seen in the NaCo-LP is likely the same one appearing in the composite spectra
by \citet{1999A&AS..138...87C}. The separation of a few AU is compatible with the lack of a measurable RV variation,
given the short time baseline of the observations.
Indeed, our FEROS spectrum of the primary does shows no indication of additional components, while
a check of the spectra in the \citet{2006A&A...460..695T} database  
confirms the composite nature of the secondary's spectrum. 
The two components of TYC 8484-1507-1 have similar magnitudes but very different $v \sin i$ values, pointing 
to an intrinsically large difference between their rotational velocities, or to a strong misalignment of their
rotation axes, which would be unusual for such a close binary \citep{1994AJ....107..306H}.
The two stars are indistinguishable by the ASAS photometry  (ASAS 023844$-$5257.0). 
The blended light curve shows variability, but exhibits no significant periodicity.
The  target's  kinematics
match  only the  IC2391  open cluster, with a correspondence of $<$2.0 sigma.
The target's sky position is inconsistent  with confirmed
IC2391 members, which all lie around RA=8h.  However, the system could be
an outlier of the kinematic  association of young stars that \citet{2008hsf2.book..757T}  
proposes  to  comprise  IC2391  and  Argus.
The age indicators are somewhat puzzling. 
The secondary has a lithium line above that of Pleiades stars with similar colors 
(which would be consistent with the possible association with Argus)
while that of the primary is below. Ca II H\&K emission is also much stronger in the secondary (as was seen in SACY
spectra). The system is not resolved in the ROSAT All Sky catalog, but XMM observations clearly indicate
that the X-ray emission is concentrated on TYC 8484-1507-1 (the close visual pair
seen with NaCo).
These differences are not easy to reconcile.
The composite nature of TYC 8484-1507-1 might somewhat alter some of the parameters.
The very young isochrone age derived by \citet{2011ApJS..193....4M} (12 Myr) for TYC 8484-1507-1  is due
to this. After correcting the photometry for the flux of the new visual component, the star lies close
to the main sequence. 
Considering these uncertainties, we adopt an age of 100 Myr, with a lower limit of 40 Myr and an upper limit of 
500 Myr. 
The system has an additional very wide (216 arcsec = 13000~AU projected separation) common proper motion star,
{\bf HIP 12361 = HD 16743}. \citet{2011ApJS..193....4M} measured for this star a similar RV to those of 
HD 16699 A and B, and noted a fully
consistent trigonometric parallax. It is then likely that the stars form a quadruple system.
As expected for an early F star, HIP 12361 provides little additional contribution to the determination of the
age of the system. Isochrone age analyses yield an upper limit to stellar age of about 1.7 Gyr.
Finally, HIP~12361 hosts a debris disk \citep{2011ApJS..193....4M}.

\item {\bf HIP 12394 = HD 16978 = $\epsilon$  Hyi}  (ASAS 023932$-$6816.1) 
Early type star, proposed  as  member of  Tuc/Hor by \citet{2008hsf2.book..757T}
and \citet{2013ApJ...762...88M}. Our kinematics
analysis shows that  its motion does indeed match ($<1.0$  sigma) those  of the Tuc/Hor
group   in U, V, and  W velocities and BANYAN II on-line tool further supports membership.  The  star  meets  the  age
requirement of Torres et al.
A small X-ray luminosity is derived from ROSAT.
The star exhibits non-periodic photometric variability.

\item {\bf HIP 13008 = HD 17438 = HR 827}
Fast rotating mid F star. It is a possible binary, 
based on the Hipparcos acceleration and the difference between the Hipparcos and the historical
proper motion \citep{2005AJ....129.2420M}.
The fast rotation does not allow an accurate RV measurement,  but the existent RVs and line profile analysis present 
no compelling evidence for binarity.
The star was originally flagged as young because of the high level of chromospheric activity reported by \citet{2006AJ....132..161G}.
However, our own measurement indicates a significantly lower value ($\log R_{HK}=-4.64$).
Similar discrepancies between \citet{2006AJ....132..161G} and other sources are observed for other
F-type stars in our sample (HIP 35564, HIP 78747). These discrepancies might be due to the larger width of the
spectral region used by \citet{2006AJ....132..161G} to derive the S-Index (3~\AA~vs the usual 1~\AA). This
may induce spurious effects for F-type stars due to their shallow Ca II H and K absorptions. 
HIP 13008 exhibits non periodic photometric variability.
A slightly sub-solar metallicity was reported from Str\"omgren photometry \citep{2011A&A...530A.138C}. 
The isochrone age is $1.288\pm0.912$ Gyr.
The target's kinematics are  consistent with the Young, Nearby population.
We adopt the isochronal age, which is consistent with our activity measurement and kinematics.

\item {\bf HIP 14684 = HD 19668 = IS Eri}  (ASAS 030942$-$0934.8) 
\citet{2013ApJ...762...88M}  propose this star to be a bona fide  member  of
the AB Dor MG.
The target's kinematics  match the AB Dor group  within the 2.0 sigma
threshold, the BANYAN II on-line tool yields high membership probability, and age indicators support this assignment.
The rotation period P=5.46d was discovered by \cite{2010A&A...520A..15M} 
and subsequently confirmed by \cite{2011ApJ...743...48W}.
There is no X-ray counterpart in the ROSAT all sky catalog. The closest source in the ROSAT Faint Source
Catalog has an integration time significantly lower than the typical ones. This may be an indication of some
instrumental issues that caused the non-validation of a source,
as found for HD 61005 in \cite{2011A&A...529A..54D}. 
Therefore, we do not list in Table 11 any upper limit on X-ray luminosity.
The effective temperature resulting from the spectral type quoted in Simabd and Hipparcos catalog (G0)
is more than 500 K warmer than our photometric and spectroscopic ones. This is likely due to a wrong spectral
classification. We adopt the revised classification (G8/K0V) by \citet{2006PASP..118.1690M}.

\item {\bf TYC 8060-1673-1 = CD-46 1064}  (ASAS 033049$-$4555.9) 
Proposed  as  a  100\%  probability  member of  Tuc/Hor  by  \citet{2008hsf2.book..757T}.   
Our kinematic  analysis shows that  it matches  both the
Tuc/Hor and TW Hydrae associations at  the 1.0 sigma level in U,V, and W.  
The  star's sky position  and XYZ  galactic distances  are inconsistent
with other known members of TWA.  All the age indicators are consistent with an age younger than 100 Myr.
 \citet{2013ApJ...762...88M} and \citet{2014ApJ...783..121G} Bayesian analysis tools 
predict a very high probability of  Tuc/Hor membership at a distance of
43.5 pc. This predicted distance for membership is consistent with the
photometric distance of 40.4  pc.  
The rotation period P=3.74d was discovered by \cite{2010A&A...520A..15M} from ASAS data, and 
subsequently confirmed by SuperWASP data \citep{2011A&A...532A..10M} and by \cite{2012AcA....62...67K}. 

\item {\bf HIP 19775 = HD 26980} (ASAS 041423$-$3819.0) 
Proposed   as   a  bona fide member of Columba by  \citet{2008hsf2.book..757T} and
 \citet{2013ApJ...762...88M}.  The kinematic
analysis  shows  that  its  UVW motion matches Columba  at  the  $<2.0$  sigma
level and the BANYAN II on-line tool yields high membership probability. Also, the  ages from various youth indicators  are
consistent with the  age of Columba.  
\cite{2010A&A...520A..15M} reported a rotation period P=2.53d based on ASAS data. 
From the analysis of SuperWASP data,  \citet{2011A&A...532A..10M} found P=1.684d to be the most
significant periodicity.  Considering the $v \sin i=$ 13.9 kms$^{-1}$ and the stellar radius 
R = 1.015~R$_{\odot}$, we infer rotation axis 
inclination values of $i$=43$^{\circ}$ and $i$=27$^{\circ}$, respectively.
The observed maximum light curve amplitude up to $\Delta$V=0.07 mag is compatible with  both inclinations, 
although it is most consistent witih the larger $i$=43$^{\circ}$ value.  This suggests P = 2.53d as the more 
likely rotation period.

\item {\bf HIP 23316 = HD 32372} (ASAS 050052$-$4101.1; 1SWASP+J050051.86-410106.7) 
Proposed   as  a high-probability member of Columba by \citet{2008hsf2.book..757T}
and \citet{2013ApJ...762...88M}.  Our kinematic
analysis shows that  its UVW motion matches Columba at  the $<2.0$ sigma level,
reinforcing the membership assignment, which is also confirmed by the \citet{2014ApJ...783..121G} on-line tool.   
Also, the ages from various youth indicators   are consistent with the age of Columba.
The most significant period we find in the SuperWASP data is P=2.183d. 
Another significant period is P=1.8d, which is the 1-d beat period of the P=2.183d rotation period.
The ASAS data analysis does not provide any conclusion on the rotation period. 
The highest peak of the 
Hipparcos data periodogram is at P=17.8d, which coincides with the maximum of the window function.

\item {\bf HD 32981 = BD -16 1042 = TYC 5901-1109-1}
Proposed   as   a  100\%
probability member of  AB Dor by \cite{2008hsf2.book..757T}.  The kinematic
analysis shows  that its UVW motion matches  AB Dor at the  $<1.0$ sigma level. 
The  age of  the  star from
various youth  indicators is also  broadly consistent with the  age of
the AB  Dor group.  \citet{2013ApJ...762...88M} Bayesian  analysis tools 
predict a 97\% probability of AB Dor membership at a distance of 79
pc \citet{2014ApJ...783..121G} return a very low membership probability, as most of the proposed AB Dor members. 
The predicted distance  for membership is roughly consistent with
the  photometric  distance  of  86.7  pc.  
\cite{2010A&A...520A..15M} found a rotation period P=0.985d.

\item {\bf BD$-$09 1108 }: (ASAS 051537$-$0930.8) 
Proposed as  an  85\%
probability member of Tuc/Hor by  \cite{2008hsf2.book..757T}.  The kinematic
analysis shows that  its UVW motion matches Tuc/Hor at  the $<$2.0 sigma level,
reinforcing  the membership  assignment. The  age of  the  star from
various  youth indicators  is  also  consistent with  the  age of  the
Tuc/Hor association. \citet{2013ApJ...762...88M} Bayesian analysis tools 
predict   a 17\%  probability   of  Tuc/Hor  membership   and  a 78\%
probability of Columba membership at  distances of 80.5 pc and 74.0 pc
respectively.  These  predicted  distances  for  membership  are  both
smaller  than  the  photometric  distance  of 93.6  pc.  
\citet{2014ApJ...783..121G} on-line tool return instead a very low membership probability on both associations when 
including our photometric distance while a 54\% membership probability for Columba at a distance of 67 pc. 
\cite{2010A&A...520A..15M} found a rotation period P=2.72d, which was subsequently confirmed by \cite{2012AcA....62...67K}.

\item {\bf HIP 25434 = HD 274197} This is the secondary of a wide binary system, with the 
star {\bf HIP 25436 = HD 35996} being the primary. 
HIP 25436 is brighter by 0.8 mag in V band. The two components have a separation of about 
12 arcsec.
Using three epoch RV measurements 
\citep[][and our CORALIE observations]{2006A&A...454..553D,2013MNRAS.435.1376M}, we found that the primary is an SB.
There is also additional evidence that one of the components is itself a binary.
The slope of -0.070 deg/year for the position angle measured by Hipparcos
is not compatible with the much smaller long term trend seen in the visual
measurements (as derived by us from the individual measurements in WDS, kindly provided 
by Dr. B. Mason). Furthermore, there are significant differences between
the short term (3.25 yr baseline) proper motion measured by 
Hipparcos \citep{2007A&A...474..653V} and the long term proper motion
which includes old photographic plates \citep[e.g. Tycho2,][]{2000A&A...355L..27H}.
This supports an orbital period of a few years for the SB.
In the ASAS photometry, the two components are indistinguishable from HIP 25434  
(ASAS 052623$-$4322.6;.1SWASP+J052622.97-432236.2)
We find a significant period P=5.34d in only one ASAS season and P=6.6d in the SuperWASP data 
(only 59 measurements available). These  periods are 'uncertain' and the component to which they belong is
ambiguous. Considering the early spectral type of the primary, they are more likely due to photometric
variations of the secondary, HIP~25434. However, they are both inconsistent with secondary's stellar radius and $v \sin i$,
and would imply an age much older than that derived from other methods.
\cite{2013MNRAS.435.1376M} propose membership to the Columba association.
The space velocities and association to groups depend sensitively on the adopted proper motion.
Using the Hipparcos proper motion, the kinematics are only consistent with the Nearby Young Population.
Adopting the Tycho2 proper motion results in a high membership probability to the Columba association
using  \citet{2013ApJ...762...88M} and \citet{2014ApJ...783..121G} Bayesian analysis tools.
Considering that the Hipparcos astrometry is likely biased by the astrometric motion of the additional component
orbiting HIP 25436, the Tycho2 proper motion should be preferred.
The age indicators are consistent with Columba membership.

\item {\bf TYC 9162-0698-1 = HD 269620}  (ASAS 052927$-$6852.1)
The star's kinematics match Tuc/Hor, Columba and Carina at $<$2.0 sigma.
The estimated age of the star is also broadly consistent with the ages
of these groups, as  all age indicators are within the distribution of known members.
\citet{2013ApJ...762...88M} Bayesian analysis predicts an 85\%
probability of  being a member  of Columba at  a distance of  93.5 pc.
This is  close to the photometric  distance estimate of  98.7 pc.  
Membership to Columba association was also recently proposed by \cite{2013MNRAS.435.1376M}.
On the other hand, \citet{2014ApJ...783..121G} BANYAN II web tool does not confirm this
assignemnt, suggesting a possible membership to Carina (15\%) or the field.
The star is projected in front of the LMC. 
\cite{2012AcA....62...67K} finds a photometric rotation period P=2.613. We also find the same 
period P=2.61d either from Lomb-Scargle or Clean analysis. 

\item {\bf TYC 5346-0132-1 = BD-8 1195}: (ASAS 053835$-$0856.7) 
Proposed  as  a  member  of  Columba  by  \citet{2008hsf2.book..757T}.   
At the  $<2.0$  sigma  level,  its kinematics  match  TWA,
Columba, and $\eta$ Cha.  However, the target's distance and position in the sky
are inconsistent  with  TWA or $\eta$ Cha membership. Also, the
age indicators are  more consistent with  Columba than the
other  groups.  Thus,  the analysis  reinforces the  Columba membership
assignment.   \citet{2013ApJ...762...88M}  tools  
predict 100\% probability of  Columba membership at a distance of
80.5 pc. This predicted distance for membership is nearly equal to
our photometric distance of 81.2  pc. \citet{2014ApJ...783..121G} on-line tool yields
a probability of 53.6\%. 
ROSAT shows no X-ray counterparts within the adopted matching radius of 30 arcsec, but the bright 
X-ray source \object{1RXS J053835.1-085639}
lies just above the threshold (angular distance of 32.5 arcsec from the optical source).
The position of the X-ray source XMMSL1 J053834.6-085645 is 6 arscec apart from TYC 5346-0132-1,
suggesting that \object{1RXS J053835.1-085639} is associated to TYC 5346-0132-1.
The photometric period P=1.983d is found by both the Lomb-Scargle and the Clean analysis at high 
confidence level and in good agreement with the earlier determination by \cite{2012AcA....62...67K}.

\item {\bf  HIP 30261 = HD 44748}:  (ASAS 062157$-$3430.7; 1SWASP+J062157.25-343043.7) 
This target's kinematics do
not match, at the $<$2.0
sigma level, any of the  investigated groups and associations.  The kinematics are  also not consistent with the Nearby,
Young population.  The star is estimated to be quite young (60 Myr) based on Lithium  and estimated to be slightly older than the Pleiades
based on other indicators.
With either a Scargle-Lomb or Clean analysis, the photometric period P=4.16d is found in almost all the ASAS seasons and 
all the SuperWASP data. The same rotation period was also found by \cite{2012AcA....62...67K}. 

\item {\bf TYC 7617-0549-1 = CD-40 2458}:  (ASAS 062607$-$4102.9) 
Proposed  as  a  70\%  probability  member  of  Columba  by  \citet{2008hsf2.book..757T}.  
At the $<$1.0 sigma level, its kinematics match both Columba
and Carina.  However, the  target's X galactic distance is inconsistent
with other Carina members, which have a relatively tight distribution.
The  age of  the  star is  broadly consistent  with the  Columba
association.   Thus,  the analysis  reinforces  the Columba  membership
assignment. \citet{2013ApJ...762...88M} and \citet{2014ApJ...783..121G} Bayesian analysis tools 
predict 100\% probability  of Columba membership at a  distance of 77.5
pc.  This  predicted distance  for membership is  nearly equal  to our
photometric distance of 77.8  pc.  
\cite{2010A&A...520A..15M} found a rotation period P=4.13d in ASAS data and P=4.14d 
in SuperWASP data \citep{2011A&A...532A..10M}.  This was subsequently confirmed by \cite{2012AcA....62...67K}. 

\item {\bf TYC 9181-0466-1 = HD 47875}:  (ASAS 063441-6953.1) 
New close visual binary from our program ($\Delta H = 1.5$ mag at 1.89 arcsec).
All indicators are consistent with an age younger than 100 Myr, but there is no previously published association with MGs.
The target's  kinematics match  Tuc/Hor and  TWA at  the $<2.0$  sigma level.
However, its  position on the sky  and estimated age reject it as  a TWA member.
\citet{2013ApJ...762...88M} Bayesian analysis  predicts a 97\% chance  of it being a
Tuc/Hor  member  at   65.5  pc, not so different from our photometric distance
of 77.7 pc (including a correction for binarity).   Thus, the  star  might  be  a  previously
unrecognized member of Tuc/Hor. However the updated BANYAN II web tool does not support 
the Tuc-Hor membership.
With both period 
search methods, we find P=2.77d to be the most significant period in most seasons. A second period P=1.56d has comparable power in several seasons.
 Considering v$\sin{i}=$ 13.5 kms$^{-1}$ and stellar radius R = 1.075~R$_{\odot}$,
we infer rotation axis inclination values of $i$=43$^{\circ}$ and $i$=22$^{\circ}$, respectively.
The observed maximum light curve amplitude up to $\Delta$V=0.04 mag is more
compatible with the longer period P=2.77d.

\item {\bf HIP 32235 = HD 49855}:  (ASAS 064346$-$7158.6)	
Proposed   as  a bona fide member  of Carina by \citet{2008hsf2.book..757T} and
\citet{2013ApJ...762...88M}. Its kinematics match 
both Columba and Carina at  the $<2.0$
sigma level and the membership to Carina was fully supported by \citet{2014ApJ...783..121G} on-line tool.  The target's
YZ  galactic  distances differ  slightly  from  what  is expected  for
Columba.  The age  is  consistent with  both Columba  and Carina.
Thus, the analysis reinforces the Carina membership assignment. 
\cite{2010A&A...520A..15M} found a rotation period P=3.83d in ASAS data, subsequently confirmed by 
\cite{2012AcA....62...67K}.

\item {\bf HIP 35564 = HD 57852 = HR 2813}: (ASAS 072021$-$5218.7)
Fast-rotating early type star, proposed as  a member  of
Carina-Near  moving group by  \citet{2006ApJ...649L.115Z}.   The target's
kinematics match  the Carina-Near  group at the  $<1.0$ sigma  level and
show no indication of matching  any other group.  The age estimates are
also consistent with the age of this group.  Thus, the Zuckerman
et  al. membership  assignment is  confirmed. 
The RV measurements show a scatter of a few km/s when combining data from several sources, 
indicating it is an SB.
The CCF shows a broad profile with no indication of additional components.
The star also exhibits non-periodic photometric variability.
It has a moderately wide (9 arcsec projected separation) companion {\bf HD 57853 = HR 2814}, a late F star which is
a spectroscopic triple \citep[see ][]{1990A&A...235..291S,2006A&A...454..553D}. 
The total mass of the close triple was estimated to be about $2.4~M_{\odot}$ 
\citep{1990A&A...235..291S}. The system is therefore a hierarchical quintuple.

\item {\bf TYC 8128-1946-1 = CD-48\,2972 } (ASAS 072823$-$4908.4)
Proposed  as  a  100\%  probability   member  of  Argus  by  \citet{2008hsf2.book..757T}.   
At the $<$1.0 sigma  level, its kinematics  match both the Argus
association and IC2391 cluster.  
The age diagnostics support a young age, compatible with Argus membership.
\citet{2013ApJ...762...88M} Bayesian  analysis  tools, with our adopted
parameters as input, predict 100\% probability
of Argus membership at a distance of 76.5 pc. 
On the other hand, membership is rejected by \citet{2014ApJ...783..121G} on-line tool.
The FEROS RV by SACY and our measurement made with CORALIE differ
from one another by 3.1 km/s. Considering the large $v \sin i$ of the star, we consider it a possible SB.
Additional measurements are needed for confirmation of binarity.
 \cite{2011A&A...532A..10M} report a rotation period P=1.038d, subsequently confirmed by 
\cite{2012AcA....62...67K}.
{\bf HIP 36312 = HD 59659} is an F7 star at 81 arcsec from  TYC\,8128-1946-1.
According to \cite{2008hsf2.book..757T}, the proper motions and distances are compatible 
with a physical association, but no spectroscopy was available to them to confirm this.
The RV measured by us on two epochs is consistent with that of  TYC\,8128-1946-1; the trigonometric distance of 
HIP 36312 is compatible with the photometric and kinematic ones for  TYC\,8128-1946-1.
The star was not detected in X-ray but it is moderately active and lithium-rich.
Taking into account these indicators' reduced age sensitivity for F stars, we consider the age
to be compatible with that of TYC\,8128-1946-1.
Therefore, we classified the two objects as physically associated, adopting the trigonometric distance
of HIP 36312 for TYC\,8128-1946-1. We also added 
HIP 36312 to the list of probable members of the Argus association.

{\bf HIP 36414 = HD 59704 }(ASAS 072931$-$3807.4)  
Proposed as  a member  of
Carina-Near  moving group by \citet{2006ApJ...649L.115Z}.   
However, the target's
kinematics do not match  the Carina-Near  group at the  $<2.0$ sigma  level. 
On the other hand, the age of the star
is consistent with  200 Myr or  greater.  
According to \citet{2011AJ....141...52T} and \citet{2004A&A...418..989N}, the star is an 
SB without an orbital solution ($\sigma_{RV}=1.2$ km/s). Our somewhat discrepant FEROS RV  \citep[$\Delta$ RV=2.6 km/s with
respect to][]{2004A&A...418..989N} further supports 
the presence of RV variability. 
We therefore classify this star as a likely SB, though additional measurements would be required
for orbit determination. The system RV is also  needed for a better assesment of MG membership.
We also found non-periodic photometric variability.
The star {\bf \object{CCDM J07295-3807B}}, at about 12 arcsec separation, is not physically associated \citep{2011AJ....141...52T}.

{\bf HIP 36948 = HD 61005}
Proposed as  a member  of
Argus  by  \citet{2011A&A...529A..54D}.
 At  the  $<1.0$ sigma  level  its
kinematics match only the Argus association. The ages derived for this
star are  all  $\sim 100$ Myr, slightly older  than the  $\sim 40$ yr age  of the
group, but compatible with the Argus membership assignment.  
\citet{2013ApJ...762...88M} also  propose this star  as a bona fide  member of
Argus and this is further supported by \citet{2014ApJ...783..121G} on-line tool.
In the Spitzer FEPS sample, this star exhibited the largest $24\mu$m excess with respect to the photosphere 
 \citep{2008ApJ...673L.181M}.
The star's disk was first resolved with HST \citep{2007ApJ...671L.165H}, indicating the presence of two components: 
an external one shaped by interaction with the interstellar medium, and an internal ring or disk.
Observations in the H-band performed as part of our NaCo-LP  
\citep{2010A&A...524L...1B} enabled the characterization of the inner ring, showing
a significant offset of its center with respect to the star and also showing a rather sharp inner rim. 
These properties might be due to the perturbation by a planetary companion, 
undetected in the available observations.

{\bf HIP 37563 = HD 62850}
Proposed as  a member  of
Carina-Near  moving group by \citet{2006ApJ...649L.115Z}.   The target's
kinematics match  the Carina-Near  group at the  $<1.0$ sigma  level and
show no indication  of matching any other group.   Thus, the Zuckerman
et  al. membership  assignment is  confirmed. Our Lomb-Scargle and Clean analyses of the ASAS (ASAS 074236-5917.8) timeseries 
revealed two periods, P = 3.65d and P = 1.37d, in 7 out of 8 seasons as well as in the complete series. 
Considering the $v \sin i =$ 14 kms$^{-1}$ and the stellar radius R = 1.055~R$_{\odot}$,
we infer rotation axis inclination values of $i$=75$^{\circ}$ and $i$=21$^{\circ}$, respectively.
An observed maximum light curve amplitude never exceeding $\Delta$V=0.02 mag is observed.
The photometric amplitude and longer period are fully compatible with the activity level
and age of the star. The shorter period would make the star an outlier in
Fig.~\ref{f:rossby}-\ref{f:prot_calc}. We therefore adopt P = 3.65d as the rotation period.

{\bf HIP 37923 = HD 63608}
Together with its wide
(23.1 arcsec projected separation) companion {\bf HIP 37918 = HD 63581}, HIP 37923 was proposed as  a member  of the
Carina-Near moving group by \citet{2006ApJ...649L.115Z} and as a member of
the IC2391 cluster by \citet{2012AJ....143....2N}.
The target's kinematics
match  the Carina-Near  group  at the  $<$1.0  sigma level  in  V and  W
velocities, but match at 3.2 sigma in the U velocity; HIP 37918
matches at 2.1  sigma in the U velocity.  
However, the age of the components is consistent with
that of  the Carina-Near group.
The kinematics do not match well with IC2391.
Thus,  the  membership assignment  to
Carina-Near is still likely, although  not as robust as other proposed
members.  
The difference in Li EW between the components is striking, considering the small difference in temperature.

{\bf TYC 8927-3620-1 = HD 77307} (ASAS 085849$-$6115.3)
Fast rotating star with large lithium abundance, indicating a young age of about 20 Myr.
The star was resolved as a very close visual binary (sep 0.1 arcsec) in NaCo-LP ($\Delta H=0.50$ mag), with
an additional bright star at 4.9 arcsec.
The latter ({\bf \object{2MASS 08584794-6115161} = WDS 08588-6115B})  is also included in 2MASS and WDS catalogs,
being only 0.5 mag fainter that the central star.
There is also an additional object ({\bf \object{2MASS 08584777-6115335} = WDS 08588-6115C}) listed in WDS at 19 arcsec.
It has a different proper motion from TYC 8927-3620-1 in the UCAC4 catalog and therefore
is not physically associated.
A subgiant luminosity class is reported in SACY.
The target's  kinematics are only consistent with  the Young, Nearby population.
Using both period search methods, we find P=0.874d in all ASAS seasons.
The same period is also found by \cite{2010OEJV..128....1B} and \cite{2012AcA....62...67K} from ASAS data.

{\bf HIP 46634 = HD 82159B} (ASAS 093035+1036.0)
Member of a wide pair whose components have similar brightness ($\Delta V=0.01$ mag).
We could not identify the rotation period, as the most significant frequency changed from season 
to season. No significant period comes out from the Hipparcos data analysis.
\cite{2012AN....333..663S} classified it as a single star based on extensive RV monitoring.
\cite{2004A&A...418..989N}  instead flagged it as being RV variable  (rms RV 36 km/s, N=2). Taking 
also the measured $v \sin i$ into account, we speculate that the spectroscopic measurements of the two components were
exchanged in \cite{2004A&A...418..989N}. The companion, {\bf HIP 46637 = GS Leo = HD 82159A}, 
is in fact a close SB \citep[period=3.86 days][]{2012AN....333..663S}. 
\cite{2012AN....333..663S} report a HIP 46637 rotation period of 3.05 days.
Its age indicators, along with the integrated X-ray luminosity, are altered by the fast rotation 
induced by the close companion.  These indicators are therefore not considered for age dating of the system. 
We note the large difference in lithium EWs in spite
of the very similar temperatures of the stars.
The Lithium EW of HIP~46634 and $\log R_{HK}$ (the latter kindly provided by the SOPHIE Consortium)
indicate an age intermediate between that of the Pleiades and Hyades. 
At the $<2.0$
sigma level, the kinematics do not match any of the investigated groups and associations.  
Its kinematics are similar to that of the Hyades. However, the solar metallicity derived by 
\citet{2012AN....333..663S} argues against a physical link with this cluster.
The G5 spectral type of HIP 46634 corresponds to an effective temperature warmer than 400~K than the photometric one.
As the spectroscopic temperature by \cite{2012AN....333..663S} is slightly cooler than the photometric one we argue 
that the spectral classification needs to be revised.

{\bf HIP 47646 = HD 84199} (ASAS 094251$-$2257.9; 1SWASP+J094250.81-225755.7) is an F5V star 
that exhibits non-periodic photometric variability.
A low level of chromospheric activity and low lithium content 
are derived from our FEROS spectra ($\log R_{HK}=-4.83$).
The kinematic parameters are distant from the loci of investigated groups and associations. 
This is consistent with a moderately old age. 
Taking into account isochrone fitting and \citet{2011A&A...530A.138C} metallicity yields an age of 1.156$\pm$0.950 Gyr. 
However, the star exhibits common proper motion and astrometric agreement with the K2.5 star {\bf HIP 47645}
(projected separation of about 1200\,AU), which appears to be much younger, based on high chromospheric activity \citep{2006AJ....132..161G}.
The system's integrated X-ray emission ($\log L_{X}=29.25$) also supports the younger age. 
One way to reconcile all these observations would be to conclude that HIP 47645 is a tidally-locked SB, but 
no existing spectroscopic data confirms this hypothesis.

\item {\bf TWA 21 = HD\,298936 }
Formerly listed by \cite{2003ApJ...599..342S} as a candidate member of TWA, and classified  by \cite{2004ARA&A..42..685Z} and \cite{2005ApJ...634.1385M} as a
member, it was later 
rejected as a TWA member by \cite{2008hsf2.book..757T}, based on the separation
in the $Z$ direction, with respect to confirmed members. 
\cite{2009A&A...501..965V} included it as a possible member of the Carina association (age 30 Myr).
Our kinematic analysis, which includes the trigonometric parallax recently
measured by \citet{2013ApJ...762..118W}, shows that this star is
consistent  with both the  TWA and  Columba groups  at the  $<$2.0 sigma
level.   The match  to Columba  in UVW  space is  actually  within 1.0
sigma.  \cite{2013ApJ...762...88M} Bayesian analysis tools predict a  99\%  probability  
of  Columba membership and only ~1\%  probability of TWA
membership.  
On the other hand, BANYAN II web tool yields a very high membership probability
to Carina MG.
In contrast to these results, \citet{2013ApJ...762..118W} recently showed, using a kinematic
trace-back study,  that
the kinematics of TWA 21 are  consistent with the TWA group. The age from  isochrone fitting is intermediate between
the quoted ages for TWA and Columba. The most probable lithium age is around
15-20 Myr, with the sequence of Carina, Columba and Tucana representing an upper limit. 
Independently from any membership assignment, we adopt the isochrone age of 17 Myr.
\cite{2010A&A...520A..15M} found a rotation period P=4.43d in ASAS data.

\item {\bf TYC 7188-575-1} (ASAS 102204$-$3233.4; 1SWASP+J102204.47-323326.9)
The star is an SB, with a peak-to-valley RV difference of 83 km/s.
The four available RV measurements are spread out over year timeframes and do not allow an estimate of
orbital period, which might conclude the possibility of a tidally-locked system.
Our FEROS spectrum reveals a complex CCF profile, suggesting a possible SB2 system.
However, the comparison with CORALIE spectra shows that the spectral lines have comparable width,
while moving tens of km/s with respect to the FEROS spectrum.
Therefore, the system may be triple or, more likely, the spectral lines are deformed by 
the presence of starspots, given 
the exceptionally high activity level (Balmer lines down to H$\epsilon$ were seen in emission). 
The star has  the largest photometric variability (0.28 mag) in our sample.
 The X-ray emission is also very strong.
The Lomb-Scargle and Clean analysis of ASAS and SuperWASP data 
reveal three major periodicities: P=4.27d, P=1.297d, and P=0.81d, in order of decreasing power. 
The longest period is listed in ACVS. 
However, it is not consistent with the v$\sin{i}=$ 25 kms$^{-1}$ and the stellar radius of a K0V star. 
Therefore, we adopt P=1.297d as the rotation period.
The extremely high activity level might be due to the interaction between the two stellar components, 
especially if they have already left the main sequence (RS\,CVn-type variability). 
The presence of lithium (comparable 
to that of Pleiades stars of similar spectral type)
would also be consistent with an RS\,CVn system, as these stars often show a moderate 
lithium content \citep{1992A&A...253..185P}, although of still debated origin.
Adopting the mean RV from the available measurement and using the photometric distance, assuming a main sequence star,
yield kinematic parameters that are compatible with the Nearby, Young Population. However, center of mass RV
is needed for a better assessment. 

\item {\bf TYC 6069-1214-1 = BD-19 3018} (ASAS 102737$-$2027.2; 1SWASP+J102737.33-202710.6) 
Only SACY spectroscopic data available.
The most important photometric period is P=3.91d (P=4.14d in SuperWASP data). 
There exists also a second period, P=1.34d, in both ASAS and SuperWASP data, as reported by \cite{2012AcA....62...67K}.
However, the absence of known $v \sin i $ prevents us from distinguishing between the correct one and the alias. 
At $<2.0$ sigma, the  target's kinematics  match only  the Carina  association.  The star's age  from multiple youth indicators is 
intermediate between that of Carina  association and of AB Dor MG, making the Carina membership unlikely.

\item{ \bf TYC 7722-0207-1 = HD 296790} (ASAS 102204$-$3233.4; 1SWASP+J103243.90-444055.6)
Classified as a possible TWA member by \cite{2003MNRAS.342..837R}, but our analysis of kinematics and
age indicators does not support it.
The Lomb-Scargle and Clean analysis of the photometric time series reveal two periods, 
P=2.01d and P=1.975d, of comparable power in almost all seasons. Both may be real, arising from Surface Differential Rotation. 
These periods are also consistent with the v$\sin{i}=$ 27.2 kms$^{-1}$. 
A period P=4.29d is listed in ACVS and reported by \cite{2012AcA....62...67K}.  
Such a value is not consistent with v$\sin{i}$ and the stellar radius.

\item {\bf TYC 7743-1091-1 = HD 99409}: (ASAS 112558$-$4015.8; 1SWASP+J112558.28-401549.9) 
This is a late type star with a moderately strong lithium line, as measured by 
\cite{2003A&A...399..983W,2010A&A...517A..88W}.
Luminosity class II-III is quoted in the Michigan Spectral Survey Catalog \citep{1982mcts.book.....H}.
DDO photometry by \citet{1987AJ.....93..616N} also supports luminosity class III and excludes
a dwarf star.
Our analysis of a FEROS archive spectrum confirms the giant status of the star.
We derived $T_{\rm eff}=4931\pm59$ using ARES.  From this we concluded $\log n(Li) \sim 1.9$ dex.
The most important photometric period is P=39$\pm$7 d, based on both our methods of period search. 
The star is found to be a hard X-ray source (HR1=0.86$\pm$0.08). 
The RVs from \cite{2003A&A...399..983W} and the FEROS spectrum differ by 4.3~km/s, making it a suspected SB.
HD 99409  is therefore not a bona fide young star, but rather should be added to the census of 
lithium-rich giants.
The moderately large $v \sin i$ (17 km/s), fast rotation (39 days), and strong X-ray emission are remarkable
for a giant star.
Similar features have been reported for other lithium-rich giants, 
and indeed the frequency of lithium-rich stars is much larger for fast-rotating giants than for
normal giants \citep{2002AJ....123.2703D}.
These features might be linked to planet engulfment phenomena 
\citep{1999MNRAS.308.1133S,2010ApJ...723L.103C,2012ApJ...757..109C},
but alternative mechanisms have also been proposed \citep{2012ASPC..464...51D}.
The kinematic parameters are inconsistent with all of the groups investigated.

\item {\bf HIP 58240 = HD 103742} (ASAS 115642$-$3216.1; 1SWASP+J115642.30-321605.3)
Together
with its wide companion {\bf HIP 58241 = HD 103743} (ASAS 115644$-$3216.0; 1SWASP+J115643.77-321602.7), 
HIP 58240 was proposed by \cite{2006ApJ...649L.115Z} to be  a member  of the
Carina-Near  moving group.   
The kinematics of HIP 52840 match the  TWA, Argus, and Carina-Near moving  group at the
$<$2.0 sigma  level.  The age  diagnostics measured for this  target are
inconsistent   with  the  expected   ages  for   the  TWA   and  Argus
associations.    
The kinematics of HIP 58241 match   Argus,the   Carina-Near   
moving  group,   the   Hercules-Lyra
association, and IC2391 at the $<$2.0  sigma level. 
Considering both components and the age diagnostics, the membership to
Carina-Near is likely.
For both stars, we could  derive the rotation period neither in ASAS nor in 
SuperWASP data, each season showing its own significant periodicity. This may be
linked to the blending of the components in ASAS and SuperWASP images.
HIP 58241 is a nearly perfect Solar analog (except for multiplicity) 
at an age of 200\,Myr \citep{2012MNRAS.423.2789D}. This component was not observed in the NaCo-LP.
\cite{2010AJ....140..510T} and \cite{2011AJ....141...52T} identified an additional source at 
about 4 arcsec from HIP 52841. Its physical association to HIP~52841 requires confirmation.
WDS lists an additional component, {\bf \object{CD-31 9365\,C}}, that is an unrelated F2 background star 
\citep{2006A&A...460..695T}.

\item {\bf TYC 9231-1566-1  = HD 105923} (ASAS 121138$-$7110.6)
New close visual binary ($\Delta H = 3.02$ mag at 1.98 arcsec). 
Proposed by \cite{2008hsf2.book..757T} as an 80\%
probability member of the $\epsilon$ Cha association, with
a kinematic distance of 115 pc, and by \cite{2002AJ....124.1670M} as a high probability member of 
LCC, with a kinematic distance of 98.5 pc.
\cite{2002AJ....124.1670M} estimated $A_V=0.48\pm0.08$, while
\citet{2011A&A...532A..10M} derived a significantly lower value (E(B-V)=0.03, $A_V=0.1$).
This discrepancy originates from the two different spectral classifications in \cite{2002AJ....124.1670M}
and \citet{2006A&A...460..695T} (G3IV vs G8V respectively). 
Our spectroscopic effective temperature ($5530\pm40$~K) supports an intermediate value, 
E(B-V)=0.08.
Adopting this latter value, we infer a photometric distance 
of 98 pc for an age of 10 Myr. 
The age indicators clearly point to a very young age, consistent with those
of $\epsilon$ Cha and LCC. An age younger than $\beta$ Pic MG is likely, thus favouring the $\epsilon$ Cha membership.
However, considering the uncertainty in the distance and reddening, and the influence
of the newly discovered binary companion, we consider 20 Myr to be an upper limit for the stellar
age.
\citet{2011A&A...532A..10M} report a rotation period of P=5.05d, which they classified as 
'confirmed'.  This was subsequently also confirmed by \cite{2012AcA....62...67K}. 
The star has no significant RV variations.

\item {\bf TYC 8979-1683-1 = CD-62 657 } (ASAS 122825$-$6321.0):
Classified as an LCC member by \cite{2012AJ....144....8S} 
with a photometric distance of 73 pc.
We derived a similar photometric distance of 75.6 pc, assuming no reddening.
However, the discrepancy between the photometric (5306~K) and the spectroscopic 
(5650$\pm$145~K) effective temperature, along with the spectral type (G7V), may indicate the 
presence of some amount of interstellar or circumstellar absorption ($E(B-V) \le 0.10$ mag).
This may increase the distance estimate up to about 80 pc, and would imply an even younger
age from Li EW.
Our analysis indicates that 
the target's kinematics match TWA, US and Corana Australis  at $<2.0$
sigma.   The  star's  age  from  multiple  youth  indicators  is  also
consistent with these young  ages.  The target's declination is further
south than any known member of TWA, implying that a membership is unlikely.  The
star may belong to the outskirts of UCL or LCC, in the direction of the Sun, having a
photometric  distance which is  closer  to the mean distances of the
ScoCen subgroups.  
The most important period in the photometric time series is P=3.35d,  which is the same as the one found by \cite{2012AcA....62...67K}.
However, it is inconsistent with $v \sin i=$ 31.9 kms$^{-1}$ and the stellar radius. 
Using both period search methods, we also find, in all the seasons and in the complete series, 
two other important periods of comparable power: P=1.42d and P=0.76d. 
 Considering the stellar radius R = 1.035~R$_{\odot}$,
we infer rotation axis inclination values of $i$=60$^{\circ}$ and $i$=37$^{\circ}$, respectively.
An observed maximum light curve amplitude up to $\Delta$V=0.11 mag is observed,
 suggesting that the star's rotation period is likely P = 1.42d.

\item {\bf TYC 8989-0583-1 = HD 112245} (ASAS 125609$-$6127.4) 
New close visual binary discovered in our program ($\Delta H=2.6$ mag).
Classified as an LCC member in \cite{2012AJ....144....8S},
with a photometric distance of 68 pc (close to our adopted value of 65 pc).
The target's kinematics  match Tuc/Hor, TWA, and UCL 
at  $<2.0$  sigma.    The  star's  age  from  multiple  youth
indicators is consistent  with it being very young, compatible with the ages of Tuc/Hor, but
likely younger.  The target's declination is further south
than  any known member  of TWA,  implying that it is likely  not a  member of this association.  The
target's  photometric  distance is  also  only  half  of the  distance
expected for membership in one of the ScoCen subgroups.  However, it could be part of the 
population of closer stars linked to these groups, as studied by \cite{2012AJ....144....8S}.
We do not have spectra for the derivation of spectroscopic temperature. From 
spectral type (K0Ve) and photometric temperature (4929~K), a small amount
of absorption can not be excluded.
The Lomb-Scargle analysis of the photometric timeseries indicated P=4.89d, P=1.25d, and P=0.83d as the most 
important periods. However, the two longer ones are inconsistent with $v \sin i=$ 47 kms$^{-1}$ and the stellar radius R = 1.095~R$_{\odot}$. 
The Clean analysis shows P=1.25d and P=0.556d to be the most important. We adopt P=0.556d as the rotation period, 
which is also reported  
by \citet{2010OEJV..128....1B}.

\item {\bf TYC 9245-0617-1 = CD-69\,1055} (ASAS 125826$-$7028.8) 
The star is a WTTS, as indicated by its strong lithium, 
filled-in H$\alpha$ line, strong X-ray emission and near-IR excess.
\cite{2008hsf2.book..757T} proposed it as  an 80\%  probability member of  the $\epsilon$ Cha  association, 
while \cite{2002AJ....124.1670M} and \cite{2012AJ....144....8S} proposed it to be a member of LCC.
Kinematic distances of 101 pc and 85 pc were derived by \cite{2008hsf2.book..757T} and \cite{2002AJ....124.1670M},
respectively, while \cite{2012AJ....144....8S} 
obtained a photometric distance of 75 pc. 
\cite{2002AJ....124.1670M} estimated $A_V=0.46\pm0.12$ while \citet{2011A&A...532A..10M}
derived a slightly lower value (E(B-V)=0.11, $A_V=0.34$).
Our spectroscopic and photometric temperatures are equal for E(B-V)=0.09.
Adopting this latter value, we infer a photometric distance 
of 93 pc for an age of 10 Myr, which is supported by the Li EW of 410 m\AA.
This is consistent with the \cite{2008hsf2.book..757T} kinematic
estimate and, together with our kinematic analysis for this distance, supports the 
association with $\epsilon$ Cha.
\citet{2011A&A...532A..10M} and \cite{2012AcA....62...67K} report a rotation period of P=2.007 and P=1.989d, 
respectively. From the RV data, there is an indication of RV variability on timescales 
of a few years and amplitude of a few km/s. Considering the young age and the probable presence
of circumstellar material, this might be due to variable obscuration from the
circumstellar disk \citep[see e.g.][for the case of T Cha]{2009A&A...501.1013S}
as well as the presence of a companion. 
We therefore consider the star as a possible SB, but we keep it in the sample for the
statistical analysis.

\item {\bf HIP 63862 = HD 113553}: (ASAS 130517$-$5051.4)
This target's kinematics do
not match any of the  groups and associations investigated at the $<2.0$
sigma level.  Its kinematics are  also not consistent with the Nearby,
Young population.  The star is estimated to be slightly older  than ~100 Myr.
The most important period is P=4.5d, found in almost all seasons, which confirms the period P=4.456d 
found by \cite{2012AcA....62...67K}.

\item {\bf TYC 7796-2110-1 = CD-41 7947}  (ASAS 133432$-$4209.5; 1SWASP+J133431.89-420930.7)
was reported as a suspected SB2 in the SACY survey.
The photospheric lines appear very broad in our FEROS spectrum; the target might be indeed an unresolved SB2 with 
fast rotating (and very active) components. 
The CORALIE spectrum also shows a complex profile,  but we can not confirm
the SB2 status from these data (similar line width). 3-epoch RVs show large scatter, but considering
the large measurement errors, this is not conclusive for a binary classification.
The star was classified as an LCC member in \cite{2012AJ....144....8S}, 
with a photometric distance of 93 pc, very similar to our adopted value of 92 pc.  
The rotation period P=1.022d is found in both ASAS and 
SuperWASP data by both period search methods. 
The target's  kinematics and age are consistent with the Young,
Nearby population at the $<$ 2 sigma level, and with UCL and LCC at the $<$ 3 sigma level.
We noted a large difference in Li EW between SACY and \cite{2012AJ....144....8S} (400 vs 315 m\AA, respectively).
Age indicators are consistent with LCC, but the very fast rotation and possible binarity make the
parameters uncertain. 

\item {\bf TYC 9010-1272-1 = HD 124831}: (ASAS 141805$-$6219.0) 
New close visual binary (projected separation 0.27 arcsec = 25 AU, $\Delta H$ = 1.01 mag).
The star is also listed as a possible SB2 in SACY.  The analysis of our FEROS
spectrum shows a strongly asymmetric CCF, which is likely due to an additional
component. 
Significant RV variations are reported in the literature.
Considering the RV difference between the components, $\sim$ 20 km/s in our FEROS spectrum, 
and the line depth ratio of about 0.7, it is plausible that the newly discovered visual companion is not the main object
responsible for the RV variations and spectral contamination. The system might therefore be triple.
The period P=1.477d is the most important period found by either Lomb-Scargle or 
Clean analysis. This confirms the earlier determination by \cite{2012AcA....62...67K} and is consistent with $v \sin i $= 38 kms$^{-1}$ and the stellar radius. 
The target kinematics  match several groups subgroups at the $<2.0$ sigma  level.  
Considering the star age, membership to Tuc/Hor or COL, seems plausible, but improved understanding
of the multiplicity and of the system RV are needed.

\item {\bf HIP 70351 = HD 125485}: (ASAS 142339$-$7248.5) 
Young star with an age of about 100 Myr, consistently derived from various indicators.
The most important period in the photometric time series is P=3.58d, 
but also the period P=1.385d is present in both 
Lomb-Scargle and Clean periodograms. No period is found from Hipparcos data.
The target's kinematics do
not match any of the  groups and associations investigated at the $<2.0$
sigma level.  Its kinematics are  also not consistent with the Nearby,
Young population.  

\item {\bf HIP 71908 = GJ 560A = $\alpha$ Cir}. (ASAS 144230$-$6458.5) 
roAp star. Isochrone fitting, carried out using the effective temperature 
by \cite{2008MNRAS.386.2039B}, yields an age of $1.10\pm0.17$ Gyr.
\citet{2009MNRAS.396.1189B} report a rotation period P=4.479d for the primary. 
No period is derived from ASAS nor from Hipparcos data.
The wide companion {\bf GJ 560B = SAO 252852} is included by  \cite{2006ApJ...643.1160L} in the list of  Her-Lyr 
members.  HIP 71908 is also proposed by \cite{2012AJ....143....2N}  as a
member of $\beta$ Pic MG.  The kinematics of the 
star  are  consistent  with  the  $\epsilon$ Cha  moving  group  and  Her-Lyr
association at the  $<$1.0 sigma level. BANYAN II web tool yields a very high membership probability
to $\beta$ Pic MG. However, the  age measured by
comparing isochrones in the HR diagram indicates that the primary has an age around 1 Gyr, making it
much older than  either of these groups.
There are also no indications of young age for the secondary, as the
system is not detected with ROSAT, the level of chromospheric activity 
is low and lithium is not detected in the spectrum.
These age estimates rule
out kinematic membership in either of these young associations, and the
star is therefore likely an older field dwarf interloping
in the  kinematic space where young stars are concentrated.   
We adopt the isochrone age of the primary.

\item {\bf HIP 71933 = HD 129181} (ASAS 144244$-$4848.0)
RV  measurements from SACY, \cite{2004A&A...418..989N}, \cite{2011PASA...28..323W}
and our FEROS spectrum  show a rather large scatter (4 km/s), suggesting binarity.
Indication of binarity comes also from the astrometric acceleration detected
by Hipparcos.
The small $v \sin i $ quoted by \cite{2004A&A...418..989N} (2 km/s) is highly discrepant with
the fast rotation reported by \cite{2011PASA...28..323W} and with our own spectrum (70 km/s). The \cite{2004A&A...418..989N}
RV measurement should therefore be taken with caution.
The broad photospheric lines make the RV measurement rather uncertain, and 
an SB2 nature cannot be ruled out. However, the deformation of the line profile 
might also be due to spots. This issue was also discussed by \cite{2011PASA...28..323W}, who
favour the star spot interpretation. We therefore consider this star as only a possible binary.
Our analysis of photometric data reveals marginal evidence of a period P=0.67d in two ASAS seasons and 
in the Hipparcos data. 
This star is very active and is also found to be a relatively hard X-ray source (HR1=0.58$\pm$0.07).
The strong lithium EW is consistent with an age between about 30 and 150 Myr.
The star is clearly above the main sequence on the color-magnitude diagram, with an isochrone age of $16\pm6$ Myr.
This is marginally inconsistent with the Li estimate, but the fast rotation makes the Li EW
measurement uncertain.
The kinematics  are fully consistent with UCL when Tycho2 proper motion is adopted 
(1.9 sigma discrepancy with respect to VL07) and  \cite{2004A&A...418..989N} RV is excluded.
Isochrone age is also fully compatible with UCL but the distance is closer then that of most of its known members. 
We therefore conclude that this is a truly young star located in front
of UCL. It is uncertain whether it is physically related to UCL and/or LCC. It may be a binary with a period of a few years,
but this needs confirmation.

\item {\bf HIP 72399 = HD 130260} (ASAS 144810$-$3647.0; 1SWASP+J144809.66-364702.2)  
This is a very active K dwarf, but without detectable lithium.
Four RV determinations (one by SACY and the others ours from HARPS and CORALIE data) show that the star is
an SB1. The two CORALIE spectra were taken with a time separation of about one hour and have an RV
difference of about 3 km/s, indicating a short period for the binary orbit.
Our analysis of ASAS data confirms the P=3.897d period reported by \cite{2012AcA....62...67K}. 
We find the same period in the SuperWASP data as well.
The K5V star  {\bf HIP 72400 = HD130260B} at 10 arcsec  is a probable wide companion. 
It has similar proper motion and parallax, and slightly different RV (which can be understood as being due to
our poor sampling of the HIP 72399 binary orbit). Both components have rather large parallax errors.
The X-ray emission of the two components is not spatially resolved by ROSAT, but the XMM source reported
in \citet{2008A&A...480..611S} is much closer to the SB component, suggesting that it is the main X-ray emitter.
The lack of lithium in the primary yields an age lower
limit of about 300 Myr, which is not consistent with the estimates based on coronal and chromospheric
emission and rotation period. 
The kinematics of HIP 72400 are slightly outside the box of very young stars, but fully
compatible with a moderately young age. Its slow rotation, as derived from the CORALIE spectrum, argues against a 
star as active as the primary. All these facts support the hypothesis that the high activity and fast rotation 
of  HIP 72399 is due to tidal locking and not to a young age. However, we can not conclusively demonstrate this
without a measurement of the binary's orbital period.

\item {\bf TYC 7835-2569-1 = HD 137059} (ASAS 152517$-$3845.4;1SWASP+J152516.99-384526.0)
Close visual binary (1.05 arcsec separation, $\Delta V=0.3$ mag, $\Delta H=0.1$ mag) first discovered
by \citet{1996A&A...307..121B} and confirmed by NaCo-LP data.
The star was found to be SB2 from CORAVEL spectra. 
\citet{2007ApJ...658..480M} classified it as a member of the Lupus association (age 5-27 Myr),
with a kinematic distance of 89 pc, but the star is not included in the recent study of the Lupus
star-forming region by \cite{2013arXiv1309.7799G}.
Kinematic parameters are consistent with $\beta$ Pic, Cha-Near, CrA and Upper Sco. However, the unresolved binary 
may imply that the RV measurements used in these analyses are very uncertain, making the kinematic analysis difficult to interpret.
Furthermore, lithium EW and X-ray flux 
are below those of most members of the youngest associations, such as $\beta$ Pic and Tuc-Hor.  Rather, the fluxes are more
compatible with an age older than those quoted in \citet{2007ApJ...658..480M} for Lupus and the ages of
other groups mentioned above. The star is therefore most likely a moderately young star
without associations to known groups.
Our kinematic distance, using the AB Dor MG sequence as a reference,  and including binary correction,
is  70.2 pc. 
The SuperWASP data analysis gives a period P=3.69d. 
In contrast, the ASAS data analysis gives two periods, P=2.32d and P=1.75d. 
 Considering the stellar radius R = 1.009~R$_{\odot}$ and $v \sin i=$ 17\,km\,s$^{-1}$, 
we infer rotation axis inclination values of $i$=50$^{\circ}$ and $i$=36$^{\circ}$, respectively.
The observed maximum light curve amplitude never exceeds $\Delta$V=0.04 mag,
suggesting that both rotation periods are possible.
The present data do not
allow us to be sure about the correct rotation period.
The multiple photometric periods might belong to the different components, considering their similar brightnesses.

\item {\bf HIP 76829 = HD139664}: (ASAS 154111$-$4439.7) 
Proposed as a member of the
Her-Lyra association  by \cite{2006ApJ...643.1160L}. The kinematics
of the  star are consistent  with Her-Lyr at  at the  $<2.0$ sigma level.
Beside kinematics, additional indications of youth are provided by coronal and chromospheric emission
and fast rotation but the sensitivity of these diagnostics for mid-F stars is not so strong. 
No lithium was detected in our analysis of the HARPS spectra.
The age of the star resulting from X-ray and Ca
H\&K is compatible with Her-Lyr membership.
We find marginal evidence of a period P=0.719d, which is consistent with 
$v \sin i=$ 71.6\,km\,s$^{-1}$ and the stellar radius. However, the measured period is very uncertain.
HIP 76829 hosts a debris disk, which was spatially resolved by \citet{2006ApJ...637L..57K} using HST. The disk is oriented 
nearly edge-on. The emitting material is concentrated on a belt at a separation of about 83~AU.

\item {\bf TYC 6781-0415-1 = CD-24 12231}: (ASAS 154131$-$2520.6; 1SWASP+J154131.21-252036.4) 
Very young star with a lithium age younger than 10 Myr (from B-V color).
Classified  by \cite{2008AJ....136.2483A} as a probable member of US 
with a kinematic distance of 109$\pm$11 pc.
This star is found to be a hard X-ray (HR1=0.94$\pm$0.03; see Sect.~\ref{s:X-ray}) and radio \citep{2004A&A...423.1073M} source.
This is consistent with the very young age of US and the age derived from the lithium EW of the star.
\cite{2004A&A...423.1073M} quotes the star as RV constant over 4 consecutive nights,  but does not provide RV values.
Individual RVs from SACY and \cite{2008AJ....136.2483A} are also consistent with a single star.
We note a highly discrepant effective temperature from B-V and V-K colors (5373~K vs 4569~K); reddening can not 
account for this.
The temperature from B-V is consistent with the G9IVe and G8e spectral classifications from SACY and 
\cite{2009ApJ...705.1646C} respectively. 
The high level of activity implies high levels of photometric variability.  This would result in significant errors 
when combining magnitudes taken several years apart. However, from ASAS data, the variability in V band over 10 years is within 
0.18 mag, and thus unable to explain the discrepancy. 
Alternatively, significant K band excess could be due to a circumstellar disk. 
Using only the B-V color, since it is consistent with the spectral classification, we derive a photometric distance of 
89 pc using the $\beta$ Pic sequence (upper limit).  We derive 106 pc for
a 11 Myr age. Within the uncertainties, these distances are compatible  with the kinematic distance by \cite{2008AJ....136.2483A}.
The kinematic analysis made with the longer distance supports the association with US. 
The ASAS data analysis gives two likely periods, P=1.05d and P=20d, 
whereas the SuperWASP data analysis gives a period P=1.05d as the most significant, although periods  of 20d 
and longer are also present. The rotation period P=1.05d is compatible with the stellar radius 
and $v \sin i$. 
The rotation period P=1.05d is 'uncertain'.

\item {\bf TYC 6786-811-1  = HD 140722C}. (ASAS 154611$-$2804.4; 1SWASP+J154610.73-280423.3)
The star was first resolved into a close visual binary (0.11 arcsec separation) by \citet{2000A&A...356..541K}.
The companion was also detected in our NaCo observations.
This star is classified as a US-B (which corresponds to UCL) member \citep{2000A&A...356..541K,2007ApJ...662..413K}.
No significant period is found either in ASAS or in SuperWASP photometric data.
The tight triple system {\bf HIP 77235 = HD 140722 = HR 5856} at 50 arcsec has similar proper
motion and radial velocity. The discrepancy, slightly exceeding the quoted errors, may be due
to the multiplicity of both the wide components. 
HIP 77235 is a close triple.
AB components have a separation of 0.58 arscec with A further resolved into a closer 
pair with 0.20 arcsec projected separation. The integrated spectral type of the triple is F2IV.
The trigonometric distance of HIP 77235 is $71.1\pm 5.5$ pc, compatible with the 
photometric distance of TYC 6786-811-1 (which is 78.6 pc including the correction for binarity),  and further
supporting the physical association. 
However, the literature contains no indications of youth for the tight triple, and it was not detected
in X ray by ROSAT. This may be due to the early spectral type of the components, which should all be
F stars based on the magnitude differences with respect to the primary. In the end, we consider this target to be inconclusive.  
Age indicators (Li and X-ray emission) for HD 140722C support a young age (nominal ages of 60 Myr for
Li and 120 Myr for X-ray), but not as young as the UCL age (16 Myr). 
The derived distance  also seems to argue against a membership with the UCL group.
Using the Hipparcos distance of the wide companion, the UVW velocities of the star is not consistent with any group, 
within 2.0 sigma.  
Thus, our kinematic and age analysis make UCL membership unlikely. The star might be an outlier in the kinematic 
distribution of most nearby young stars.  However, an unresolved 
binarity could make the adopted RV measurements uncertain.

\item {\bf HIP 78747 = HD 143928 = HR 5975} (ASAS 160437$-$3751.8)
Mid F star, originally classified by \citet{2006AJ....132..161G} as young because of the high $\log R_{HK}$.
But much lower activity levels are reported in \citet{2009A&A...493.1099S} and also measured in our FEROS spectrum. 
The star was also not detected by ROSAT. The isochrone  age is $2.357\pm0.627$~Gyr, taking the subsolar
metallicity by \citet{2011A&A...530A.138C} into account.
No period is found either in ASAS or Hipparcos data.
The target is likely an old star, but with young star kinematics.
The companion listed in WDS and CCDM {\bf (\object{CD-37 10680B}}, V=13.0 at about 40 arcsec) is not physically associated.

\item {\bf TYC 6209-0769-1 = BD-19 4341}: (ASAS 161414$-$1939.6) 
Moderately young star with an age similar to the Pleiades.
We find two periods of comparable powers, P=2.14d and P=1.86d, using both 
a Lomb-Scargle and Clean analysis of the ASAS timeseries. The rotation period  P=2.14d is 'likely'. 
The target's kinematics are consistent with the Young, Nearby
population.

\item {\bf HIP 79958 = HD 146464}: (ASAS 161916$-$5530.3) 
This is a moderately young star with an age similar to the Pleiades.
The period P=2.325d that we find with both Lomb-Scargle and Clean was also reported by 
\citet{2002MNRAS.331...45K} from Hipparcos data and by \cite{2012AcA....62...67K} from ASAS data. 
The rotation period derived from $\log R_{HK}$ is much longer than the observed one (Fig.~\ref{f:prot_calc}).
As there are no additional plausible longer period periodicities in the photometric time-series, the discrepancy is
likely due to one of the following: (1) the chance observation of low-activity phases on the 3-epoch CORALIE monitoring, or (2) errors in
the transformation from S Index to  $\log R_{HK}$, as the color of the star is slightly outside
the validity range of the \cite{1984ApJ...279..763N} prescriptions.
At the $<2.0$
sigma level, the target's kinematics and
age are only consistent with  the Nearby, Young population .

\item {\bf HIP 80290 = HD 147491 = V972 Sco} (ASAS 162323$-$2622.3; 1SWASP+J162322.93-262216.3)
Close visual binary observed with NACO (projected separation 3.3 arcsec; $\Delta H$=1.9 mag).
A source at similar position close to the star is reported in the 2MASS catalog ({\bf \object{2MASS	16232272-2622168}}),
but with a magnitude much brighter than our measurement. 	
\cite{1989IBVS.3334....1Y} classified it as a $\delta$ Scu variable. 
This classification was revised by \cite{1995IBVS.4188....1H}; we confirm it is a BY Dra rotational variable.
We find a period P=1.513d in both ASAS, SuperWASP, 
and Hipparcos data, in agreement with the earlier determination by \cite{2012AcA....62...67K}. 
The star is projected close to the globular cluster M4.
The star's kinematics match
Corona-Australis at  the $<2.0$ sigma  level.  The estimated age  of the
star is  quite young and it  may have membership  consistent with CrA.
The star is  also kinematically equivalent  with the
Nearby, Young population.
The star {\bf WDS J16234-2622B  = 2MASS J16232242-2622230 }, at a projected
separation of about 10 arcsec, is likely not physically associated. 

\item {\bf HIP 80758 = HD 148440}: (ASAS 162920$-$3057.7; 1SWASP+J162920.16-305739.8) 
Young age (adopted value of 20 Myr) is derived consistently from various indicators, including isochrone fitting
(19$\pm$14 Myr).
The target's kinematics is only consistent with the Young, Nearby population.
We find P=3.38d from ASAS data and P=3.55d from SuperWASP data. 
No period is derived from Hipparcos data.

\item {\bf TYC 6818-1336-1 = HD 153439}: (ASAS 170043$-$2725.3) 
This star was classified as a possible member of the Upper Scorpius association
\citep{2008AJ....136.2483A}, with a kinematic distance of 130$\pm$12 pc.
There is a severe discrepancy on the star's spectral type: G0IV according to SACY
and F5 according to the Michigan spectral atlas. The latter is adopted by \cite{2008AJ....136.2483A}. 
This has a significant impact on the presence of reddening, and therefore on the
distance to the star.
Considering the better  agreement of photometric temperatures from B-V and V-K,
and the higher spectral resolution of SACY spectra, we adopt the un-reddened case.
The age of the system is also quite uncertain. The star is found to be a relatively hard X-ray source
(HR1=0.46$\pm$0.20), which supports a very young age. 
But Lithium EWs from SACY and \cite{2008AJ....136.2483A} differ by 30 m\AA~. Taking into account these
uncertainties, ages from 10 to 100 Myr are possible.
The most important photometric period is P=1.13d, which is found in only three seasons. 
The rotation period P=1.13d is 'uncertain' and does not contribute meaningful age constraints.
Adopting an age of 30 Myr, we derive a photometric distance of 89.5 pc.  But if the star is younger than 15 Myr, then the distance would be farther 
than 100 pc (105 pc using $\beta$ Pic sequence).
The star's kinematics are only consistent with the Young, nearby 
population and Corona-Australis association at the $<$2.0 sigma level.

\item {\bf TYC 6815-0874-1 = CD-25 11922}: (ASAS 170433$-$2534.3) 
Fast rotating star, flagged as a possible SB2 in SACY. 
There are two spectra from SACY; the line profile changed between the two epochs, 
but the line width is similar. We thus cannot claim conclusive inference that it is an SB2.
From the ASAS photometric data, we find a period P=1.373d using both Lomb-Scargle and Clean analyses.
In a comparison of spectral type vs photometric effective temperature, there are indications of some amounts of reddening
 ($E(B-V) \sim 0.1$ mag). Taking as reference the $\beta$ Pic sequence, we derive a  distance  of 109 pc. 
The possibility that the star is an SB2 adds uncertainty to the derivation of stellar properties
from photometry.
The target's  kinematics and age  are only consistent with the Young,
Nearby population.
The star position is at the edges of the US group. The UVW velocities all match those of the US group, but only at 
the $<$3.6 sigma level.
The nominal position of the X-ray source \object{RXS J170428.0-253520}, classified in SACY as the X-ray counterpart of TYC 6815-0874-1, is 93 arcsec apart from that of the optical source, but with a very large position error (41 arcsec). 
We also noted that the integration time in the ROSAT catalog is significantly smaller than the usual ones.
Furthermore, the X-ray source is hard (HR1=1.00$\pm$0.32), which is unusual for the 
kind of stars we are investigating.
Without additional X-ray observations, we are unable to conclude on the association of the
ROSAT source with our target star.

\item {\bf TYC 6815-0084-1 = CD-25 11942} (ASAS 170601$-$2520.5) 
is also a fast-rotating suspected SB2 in SACY (no clear signature of binarity in our spectrum).
TYC 6815-0084-1 is a possible member of Upper Scorpius association according to \cite{2008AJ....136.2483A}, with a kinematic
distance of 127 pc. Assuming a reddening E(B-V)=0.08 from the comparison of spectral
type and photometric temperatures and adopting as reference the sequence of $\beta$ Pic MG, we obtain a photometric distance
of 79 pc, which increases to 92 pc for a 11 Myr age. The lithium EW is compatible with a very young age (10-20 Myr) and then wth US
membership. 
The space velocities are also compatible with US, and matching improves increasing distance to 110-130 pc.
The sky position is only 0.7 deg outside the formal boundaries of US adopted in \citet{2008hsf2.book..235P}.
We find more photometric periods that persist in different seasons and are all close 
or shorter than 1 day, which is expected from the high v$\sin{i}=$ 57 kms$^{-1}$. However, it is difficult to say which 
is the correct one. One likely period is P=1.243d, but also P=1.03d and P=0.507d may be the rotation periods. 
A very similar period (1.246d) is also listed in \citet{2012AcA....62...67K}. 
The star has similar kinematic parameters to TYC 6815-0874-1, also in our sample.
The projected separation of the two stars is about 21 arcmin, too large for a bound system ($\ge 100000$~AU) but
considering the similar ages a common origin for the two stars can not be excluded.
The photometric distances are roughly consistent taking the uncertainties due to
reddening, age and possible binarity of both objects into account.

\item {\bf TYC 7362-0724-1 = HD 156097}: (ASAS 171658$-$3109.1) 
G5V star with large lithium content. Only one SACY measurement is available.
We find a photometric period of P=1.92d.
The star's kinematics match Corona-Australis at the $<2.0$ sigma level.  
The estimated age  of the star is  quite young (about 20 Myr) and it  could be consistent
with  CrA membership. The position on the sky is slightly outside the nominal edges of US and UCL;
membership in  the US or UCL groups can not be excluded.

\item {\bf TYC 8728-2262-1 = CD -54 7336}: (ASAS 172955$-$5415.8) 
Proposed by  \citet{2008hsf2.book..757T} as  a  90\% probability  member  of  $\beta$  Pic.   
At the $<2.0$ sigma  level, its kinematics  match $\beta$ Pic,
TWA, Chamaleon, and Corona Australis.  The  target's position on  the sky  is inconsistent
with  the TWA  known member  distribution.  Ages derived  from
various methods  are fully compatible with $\beta$ Pic MG. .
Thus, the analysis reinforces the $\beta$ Pic membership assignment.  \citet{2013ApJ...762...88M} 
Bayesian analysis tools predict  100\% 
probability of  $\beta$ Pic membership at a  distance of 71.5  pc. This
predicted distance  for membership is consistent  with the photometric
distance  of 70.4 pc. 
\citet{2014ApJ...783..121G} on-line tool yield only a 19\% membership probability.
\cite{2012AJ....144....8S}  classified the star instead as a US star  
with a photometric distance of 72 pc.
\citet{2010A&A...520A..15M} and \cite{2012AcA....62...67K}  report a rotation period P=1.819d. 
However, the secondary period P=0.620d that is also found in 5 out of 9 seasons may be more consistent with 
v$\sin{i} = 35.3$  kms$^{-1}$ and the stellar radius.
The rotation  period P=0.620d is 'uncertain'.

\item {\bf HIP 86672 = HD 160682 = V2384 Oph} (ASAS 174230$-$2844.9)
Our analysis does not reveal evidence of eclipses, although \citet{2005yCat..34460785M} listed it in their catalogue of eclipsing binaries. \citet{2005yCat..34460785M} give neither an orbital 
period  nor find the secondary minimum, which makes their classification questionable.
We derive the same period P=3.755d from either Lomb-Scargle or Clean analyses. This value is similar 
to the P=3.912d period found by  \cite{2012AcA....62...67K} 
The star's kinematics match
Corona-Australis at  the $<2.0$ sigma  level.  The estimated age  of the
star is young (30 Myr), but marginally  inconsistent with the CrA's younger proposed age.  

\item {\bf HIP 89829 = HD 168210} (ASAS 181952$-$2916.5)
Ultrafast rotating star ($v \sin i$ 114 km/s), proposed by \citet{2008hsf2.book..757T} 
and \citet{2013ApJ...762...88M}  as a high-probability member of  $\beta$ Pic MG 
while membership is rejected by BANYAN II.  
At the $<$2.0 sigma  level, its  kinematics  match  $\beta$  Pic.  
The large lithium EW and high levels of activity and photometric variability 
are consistent with the $\beta$ Pic MG membership assignment. 
Isochrone fitting further supports the pre-MS status (age of $15\pm3$ Myr).
The star resides in a field that is very crowded with fainter stars. 
We derive the same period P=0.571d from either Lomb-Scargle or Clean analyses. This period is
also reported by  \cite{2012AcA....62...67K} and in ACVS. 
No period is found in the Hipparcos data. 
According to \cite{2011PASA...28..323W}, the star is likely single, with line profile variations
most likely due to star spots.

\item {\bf HIP 93375}: (ASAS 190106$-$2842.8) 
Proposed by \citet{2008hsf2.book..757T}   as  a high-probability member  of AB Dor.  
The target's age from various youth indicators is about 100  Myr, in agreement with the range expected
for AB Dor members. Thus, the analysis reinforces the AB Dor membership
assignment.  \citet{2013ApJ...762...88M}  Bayesian  analysis  tools  predict a 
91\% probability  of AB Dor  membership at a
distance  of  63.5 pc.   This predicted distance for membership is
consistent  with  the  measured  distance   of  70.4  pc. On the other hand,
the membership is formally rejected by \citet{2014ApJ...783..121G} on-line tool. 
We find no rotation period either in the ASAS data or  in the Hipparcos data.
There is another star, {\bf WDS J19011-2843B = CD-28 15269B}, at 
11 arcsec projected separation. TYCHO2 and UCAC give no proper motion measurements for the possible secondary.
However, the variations in projected separation, as listed in WDS (from 10.0 to 11.1 arcsec
in 23 years), seem to argue against a physical association.
Also, when blinking images from different epochs, the candidate secondary does not appear to exhibit co-movement with the primary.

\item {\bf HIP 94235 = HD 178085}: (ASAS 191057$-$6016.0) 
New close visual binary ($\Delta H=3.8$ mag at 0.51 arcsec).
Proposed by \citet{2008hsf2.book..757T}
and \citet{2013ApJ...762...88M} as  a bona fide member  of AB Dor, while \citet{2014ApJ...783..121G}
on-line tool yields a membership probability of 33\%.  At  the $<$1.0
sigma level, its  kinematics are only consistent with AB Dor. Age indicators
are consistent with youth, reinforcing the  AB Dor  membership assignment.
\citet{2010A&A...520A..15M} report a rotation period P=2.24d, subsequently confirmed by \cite{2012AcA....62...67K}.
The most important periods derived from a Hipparcos photometric data analysis are P=4.28d and P=2.38d.
Considering the $v \sin i=$ 24 kms$^{-1}$ and the stellar radius R = 1.085~R$_{\odot}$,
we infer that the only consistent period is P=2.24d.

\item {\bf TYC 6893-1391-1 = CD -25 14224}: (ASAS 193919$-$2539.0; 1SWASP+J193918.66-253901.3) 
Early K-dwarf for which age indicators yield an age slightly older than the Pleiades.
Applying both Lomb-Scargle and Clean analyses to ASAS and SuperWASP data, we find a period P=7.49d, confirming the period determination by  \cite{2012AcA....62...67K}. 
The target's kinematics are only consistent with the Young, Nearby
population.

\item {\bf TYC 5736-0649-1 = BD -14 5534} (ASAS 194536$-$1427.9) 
This star has extremely high v$\sin{i}$ for a G6V star (206 km/s). The only period consistent with v$\sin{i}$ 
and the stellar radius is P=0.2516d.  \cite{2012AcA....62...67K} reports a double period P=0.5077d, too long for the measured v$\sin{i}$.
The star is found to be a relatively hard X-ray source (HR1 =0.61$\pm$0.25).
The target's  kinematics and age  are only consistent with  the Corona
Australis  association.  
The star was flagged in SACY as a possible SB,  having an RV rms of 3.5 km/s. However, the very large $v \sin i$ 
makes the RV error large.
H\&K flux is typically underestimated with such a large $v \sin i$.
In our FEROS spectrum, an extra absorption is seen in the core
of the Na D doublet. As it has a RV similar to that of the star, and is characterized by a single-component very narrow
line profile, we favour a circumstellar origin.

\item {\bf HD 189285 = BD -04 4987 = TYC 5155-1500-1} (ASAS 195924$-$0432.1)
 Proposed   as  a  90\%
probability member  of AB Dor  by \cite{2008hsf2.book..757T}.  At  the $<$2.0
sigma level,  its kinematics  match only AB  Dor.  The target's  age is
also consistent  with expectations for  the AB Dor group,  although it
does appear slightly older.  The analysis is therefore consistent with the
AB  Dor  membership  assignment.  It  should be  noted  that  \citet{2013ApJ...762...88M} and \citet{2014ApJ...783..121G}
Bayesian  analysis  tools, using our adopted parameters as inputs, predict  only a 6\% and $<1\%$, respectively,
probability  of AB  Dor membership,  at a  distance of  83.5  pc .  
Given the age of the star  resulting from age indicators, the AB
Dor  membership  assignment  is still possible.   
 \citet{2010A&A...520A..15M} report a rotation period P=4.85d.

\item {\bf HIP 98470 = HD 189245 = HR 7631} 
Fast rotating F-star, proposed by \citet{2012AJ....143....2N} as a  member of the AB Dor moving group.  
Its kinematics do match any of the considered groups at the $<2.0$ sigma level.
A low membership probability to AB Dor (3.9\%) is obtained using the 
\citet{2013ApJ...762...88M} Bayesian  analysis  tools.
The target's age from lithium, X-ray  and  Ca H\&K appears quite young, comparable to the loci
of AB Dor MG and Pleiades. However, considering these methods' limited age sensitivity for F-type stars, 
an age as old as 200-300 Myr is also possible.
\citet{1997A&A...323..429B} flagged the star as SB, but  without further details. 
In contrast, \citet{2009A&A...495..335L} found an RV dispersion of only 86 m/s, for 18 epochs over 259 days.
The other sparse RV measurements, on longer time baselines, are consistent with a constant RV
within errors. The CCF from our FEROS spectrum does not show signatures of additional components.
No period is found either in ASAS or in SuperWASP photometric data. 
A tentative period P=1.88d is inferred from the Hipparcos photometric data. 

\item {\bf TYC 5164-0567-1 = BD -03 4778}: (ASAS 200449$-$0239.2) 
Proposed by  \cite{2008hsf2.book..757T} as  a  100\%  probability  member  of AB  Dor.  
At the $<1.0$ sigma level, its kinematics match only AB Dor.
The ages derived from various methods are also broadly consistent with
the AB Dor group, but Li EW is larger than that of other members of similar color.  
\cite{2013ApJ...762...88M}  Bayesian  analysis  tools  
predict a 98.5\% probability of  AB Dor membership at a distance of
66.0 pc.   This predicted distance  for membership is  consistent with
the  photometric  distance  of  63.3  pc.  On the other hand, \citet{2014ApJ...783..121G} 
on-line tool yields a very low membership probability. 
\citet{2010A&A...520A..15M} report a rotation period P=4.68d, which was later confirmed by \cite{2012AcA....62...67K}.

\item {\bf HIP 99273 = HD 191089}  (ASAS 200905$-$2613.4; 1SWASP+J200905.21-261326.5)
Proposed by \cite{2006ApJ...644..525M}, \cite{2008hsf2.book..757T},
and \cite{2013ApJ...762...88M}  as  a  bona-fide member of $\beta$  Pic.
At  the $<1.0$ sigma level,
its  kinematics match  only $\beta$  Pic. 
The membership is also supported by \citet{2014ApJ...783..121G} on-line tool.
The age indicators are somewhat ambiguous, as expected for an F-type star.
The Lithium EW measured on our FEROS spectrum is compatible with a $\beta$ Pic age, but also
with significantly older ages; chromospheric and coronal emission are rather low.
No period is found in ASAS or SuperWASP data.
A tentative period P=0.488d is derived from the Hipparcos data analysis, but this period is absent in the other datasets.
We adopt $\beta$  Pic membership and age, but allow for a maximum age of 200 Myr.
\cite{2007ApJ...660.1556R} identify the star as SB, but  without further details. The literature RVs show
a modest scatter, which seems compatible with measurement errors, considering the early spectral type
and the rotational velocity of the star. 
The star hosts a debris disk, with $T_{dust}=95$~K, $R_{dust}=15$~AU and  
$M_{dust}=3.4 \times 10^{-2}~M_{\oplus}$	 
\citep{2007ApJ...660.1556R}. The disk was recently spatially resolved by \citet{2011MNRAS.410....2C}.

\item {\bf HD 199058}: (ASAS 205421+0902.4) 
New close visual binary ($\Delta H=2.5$ mag at 0.47 arcsec) identified by 
NaCo-LP observations. It also has a probable wide (245 arcsec projected separation) companion
{\bf TYC-1090-0543-1}, having very similar kinematic parameters
and indicators supporting a young age. 
If a physical association is real, the projected separation is about 16000~AU.
Both stars are included in the list of AB Dor members by \cite{2008hsf2.book..757T}.
At the $<$2.0 sigma level, the kinematics of HD 199058 match also Tuc/Hor, Carina, and $\epsilon$  Cha.
However, membership to Tuc/Hor or Carina is unlikely because of the star's northern declination.
The star's  sky position  is inconsistent with $\epsilon$  Cha
membership.
The age indicators of both components are consistent with AB Dor membership.
There is a discrepancy in the photometric distances of the two stars (55.0 pc for 
HD 199058, including binary correction, and 77.2 pc for  TYC-1090-0543-1).
It should  be noted  that \cite{2013ApJ...762...88M} Bayesian  analysis tools
predict only a  6\% probability of AB Dor membership at
a distance  of 55.0  pc.  This probability changes to $>$98\% at 74 pc, a distance nearly identical to
the photometric distance of TYC-1090-0543-1. 
On the other hand, \citet{2014ApJ...783..121G} on-line tool yields a very low membership probability.
We adopt the mean of the photometric distances of the two objects. 
Applying both Lomb-Scargle and Clean analyses in two seasons, we find an uncertain period P=3.47d for HD 199058.

\item {\bf TYC 5206-0915-1 = BD -07 5533} (ASAS 211834$-$0631.7) 
In ACVS, the star is listed with a period P=3.4447d and classified as a detached/semi-detached eclipsing binary. 
However, from our analysis of the photometric time-series we did not found convincing evidences
for eclipses using this proposed period or their harmonics. 
\cite{2012AcA....62...67K} reports a period P=2.37d, which our analysis retrieves in all the seasons.
From CCF in our FEROS spectrum, there are no indications of additional components. But the 
large RV scatter (about 40 km/s peak-to-valley), from our FEROS and CORALIE spectra and SACY, makes
the star a short period spectroscopic binary (SB1).
The fast rotational period and large activity levels may be due to youth or due to a tidally locked binary.
Lithium EW is consistent with a star of age intermediate between Pleiades and Hyades, 
but the possibility that this an RS CVn variable with detectable lithium can not be ruled out.
The  target has  very large uncertainties in UVW velocities because of a large spread in measured RVs.  
Until a systemic velocity  can be established and used to
calculate UVW, determining group membership is not possible.
Overall, age determination remains highly uncertain until the orbital period of the binary
can be established.

\item {\bf HIP 105384 = HD 203019}: (ASAS 212047$-$3846.9; 1SWASP+J212046.59-384653.2) 
The lack of lithium  in the spectrum of this star (from SACY) indicates an age older than
the Hyades. Coronal and chromospheric activity (the latter based on several independent sources)
instead suggest a younger age (about 250 Myr). 
RV monitoring indicates a single star 
 and the CCF does not show 
additional components. Therefore, we are not considering a tidally locked binary.
No rotation period is detected in our analysis, but the star exhibits non-periodic photometric variability.
The kinematics  do not  match  any   of  the  groups   and  associations
investigated at  the $<2.0$  sigma level.  The  kinematics are  also not
consistent with  the Nearby, Young population.  
An age of about 500 Myr might be roughly compatible with the various indicators, considering
the observed scatter at fixed age.

\item {\bf HIP 105612 = HD 202732}: (ASAS 212327$-$7529.6) 
Lithium and chromospheric activity indicate an age similar to the Hyades. X-ray emission suggests a slightly
younger age. We find an uncertain period P=12.9d in the complete series and in two seasons. 
The resulting gyrochronology age from this period would
be about 1100 Myr, older than the other indicators.
The kinematics  do   not  match  any   of  the  groups   and  associations
investigated at  the $<2.0$  sigma level and are  also not
consistent with  the Nearby, Young population.  The CCF does not show additional components
and RV suggests a single star.
Taking all these indications into account, we adopt an age of 700 Myr.

\item {\bf HIP 107684 = HD 207278}: (ASAS 214849$-$3929.1) 
New close visual binary ($\Delta H=2.8$ mag at 0.33 arcsec).
Proposed by \cite{2008hsf2.book..757T} as  a  100\%
probability member of the AB Dor moving group.
At  the  $<2.0$ sigma  level,  its kinematics  match  only  AB Dor.   Its
estimated age  is also  consistent with AB  Dor membership.   
\citet{2013ApJ...762...88M} Bayesian  analysis tools  predict a 92\%
probability  of AB  Dor membership  at a  distance of  81.0  pc.  This
predicted  distance for  membership  is consistent  with the  measured
distance of  90.2 pc. On the other hand, \citet{2014ApJ...783..121G} on-line tool yields a 
very low membership probability.
\citet{2010A&A...520A..15M,2011A&A...532A..10M} report a rotation period 
P=4.14d.

\item {\bf HIP 108422 = HD 208233}: (ASAS 215751$-$6812.8) 
Close visual binary discovered by \citet{2003A&A...404..157C}; the physical association
of the components was confirmed with our NaCo observations (Chauvin et al.~2013). 
The star was included as a member of the Tuc association by \citet{2004ARA&A..42..685Z}, but 
rejected by \citet{2008hsf2.book..757T}.
\citet{2013ApJ...762...88M} also propose this star as a bona fide member of Tuc/Hor
and the analysis based on \citet{2014ApJ...783..121G} confirms membership.
The star is characterized by a very fast rotation, identified both spectroscopically and
photometrically.
We find the period  P=0.4352d, that is also reported by 
\citet{2002MNRAS.331...45K} from Hipparcos data analysis and by \cite{2012AcA....62...67K} from ASAS data analysis. 
In our analysis, the kinematics match only Tuc/Hor but at the $>2.0$ sigma level. However, the kinematics
might somewhat altered by binarity.
The estimated age from lithium is consistent with the $\sim$30 Myr age of Tuc/Hor.
The star's isochrone age of 13$\pm$3 Myr is formally even younger than the nominal Tuc/Hor group age.
Taking all of these facts into consideration, we adopt a Tuc-Hor membership, but acknowledge the possibility 
of an even younger age as suggested by the CMD position. 

\item {\bf TYC 8004-0083-1 = CD-40 14901}: (ASAS 224633$-$3928.8; 1SWASP+J224633.48-392845.2) 
All the age indicators point to a rather young star (best estimate of 90 Myr).
However, the kinematics  do not match  any of the groups  and associations
investigated at  the $<2.0$  sigma level.  The  kinematics are  also not
consistent with  the Nearby, Young population.  
We find the period P=3.22d in all the ASAS 
and SuperWASP seasons.  This period is also reported by  \cite{2012AcA....62...67K}.

\item {\bf HIP\,114046 = HD\,217987 = GJ 887 } (ASAS 230552$-$3551.2) is the only M dwarf in our sample.
It was included because of the tentative lithium detection 
in the SACY database (EW=50 m\AA), which originally comes from \citet{1981MNRAS.194..829D}. However, 
our higher quality FEROS spectrum shows no lithium. 
The star is above the main sequence in the $M_{V}$ vs B-V color-mag diagram, which would imply a pre-MS
star.   But the 2MASS magnitudes
are most compatible with a main sequence star. \citet{2012A&A...539A..72D} estimated an isochrone 
age of 100 Myr - 10 Gyr from VJHK photometry.
The stellar radius, as derived by \citet{2009A&A...505..205D} using VLTI, is not anomalous in their
mass-radius and luminosity-radius relationships.
The star has rather low coronal and chromospheric activity.
No rotation period is found in either ASAS (sparse data)  or Hipparcos data.
Using the kinematic population diagnostic by \cite{2003A&A...410..527B}, we conclude a probable 
membership of the thick disk. 
We therefore adopt a very old age for the star, 8 Gyr.   
We adopt the stellar mass and metallicity from \citet{2013A&A...551A..36N}.

\item {\bf TYC 9338-2016-1= HD 220054}: (ASAS 232153$-$6942.2) 
All age indicators are consistent with a young age (10-50 Myr). 
We find the period P=1.875d in ASAS and SuperWASP data.
The same period was reported in ACVS, by \citet{2009OEJV...98....1B}, and by \cite{2012AcA....62...67K}.
RV measurements show some scatter (0.8 km/s rms in SACY); we consider it as a suspected SB.
The star's kinematics match  Tuc-Hor and TWA
at the $<2.0$ sigma level.  The star's estimated age is also quite young.  
\cite{2013ApJ...762...88M} Bayesian  analysis tools predict a  very low  probability of
being a true member of one of those groups at the photometric distance
of 99.6 pc.  The kinematics are consistent with the Nearby,
Young population.

\item {\bf TYC 9529-0340-1 = CD-86 147}: (ASAS 232749$-$8613.3) 
Proposed by  \citet{2008hsf2.book..757T} as  an  85\%  probability  member  of  Tuc/Hor.   
At the  $<1.0$  sigma  level,  its kinematics  match  only
Tuc/Hor.  The  ages derived from  various methods are  also consistent
with the Tuc/Hor association.  
\cite{2013ApJ...762...88M} Bayesian  analysis  tools predict  100\%
probability  of Tuc/Hor  membership at  a distance  of 64.0  pc.  This
predicted distance  for membership is consistent  with the photometric
distance of  68.8 pc.  \citet{2014ApJ...783..121G} on-line tool yields a 74\% membership
probability.   
\citet{2010A&A...520A..15M} report a rotation period P=2.31d, which they note 
to be inconsistent with v$\sin{i}=$ 73.9 kms$^{-1}$. A new analysis, also made 
with the Clean method, reveals a period P=0.7025d,  which confirms the rotation 
period P=0.6975d found by \cite{2012AcA....62...67K} and P=0.7024d found in ACVS. 

\item {\bf TYC 9339-2158-1 = CD-69\,2101}: (ASAS  233101$-$6905.2) 
The star  shares common proper motion, position and radial velocity with {\bf HIP\,116063= HD 221231}.
Parallax of the companion is adopted for the system. No rotation period is found
(likely due to the blending of the components). 
The age of the components from several diagnostics is estimated to be ~300 Myr.

\item {\bf TYC 6406-0180-1 = HD 221545}: 
Lithium EW and X-ray luminosity of this K0 star indicate an age of about 200 Myr. 
Kinematics derived assuming an MS status are consistent with a moderately young star.
Non-typical of a star of this type, a period P=36.271d is reported by ACVS, and confirmed by \cite{2012AcA....62...67K}, 
who reports P=36.34d. Our analysis of ASAS data reveals two major power peaks in the seasonal periodograms, 
one at P=36d and another at P=1.026d. The first is in agreement with the mentioned literature values, whereas 
the latter, together with a K-type MS stellar radius and the small $v \sin i$ value, would imply a very inclined 
(almost pole-on) star. In that latter case, we would not expect any light rotational modulation, whereas we find a 
light curve amplitude up to $\Delta$V = 0.07 mag. We therefore consider the period P=36d as 'confirmed'. 
Such a rotation period being consistent with the measured $v \sin i$=5.5 km s$^{-1}$ requires
a stellar radius R$>$3 R$_{\odot}$.  Large radii like this are found among subgiant K stars, but not in the current V class
assignation. The long rotation period is also unusual among active main sequence stars, but typical of evolved stars.
There is only one RV measurement in the literature. Therefore, limited inferences on binarity can be made.
We adopt the main sequence age, but allow for an age of 4 Gyr, which is obtained for an evolved star matching 
the stellar radius that is most consistent with the rotation period and $v \sin i$.

\item {\bf HIP 116910 = HD 222575}: (ASAS 234154$-$3558.7) 
Proposed   by \cite{2008hsf2.book..757T} and \cite{2013ApJ...762...88M} as  a high-probability member  of AB Dor.
\citet{2014ApJ...783..121G} on-line tool yields a membership probability of 36\%.
At  the $<1.0$ sigma level, the  kinematics match only AB Dor.   The ages derived from
various methods are slightly younger  than the proposed age of the AB Dor
group, but since the kinematics only  match this group, and the star is
surely  young,  the  analysis   reinforces  the  AB   Dor  membership
assignment. 
\citet{2011A&A...532A..10M} report a rotation period P=1.787d, which 
was subsequently confirmed by \cite{2012AcA....62...67K}.

\end{description}


\section{S index calibration for FEROS}
\label{a:sindex}

\begin{figure}[h]
\includegraphics[width=8cm]{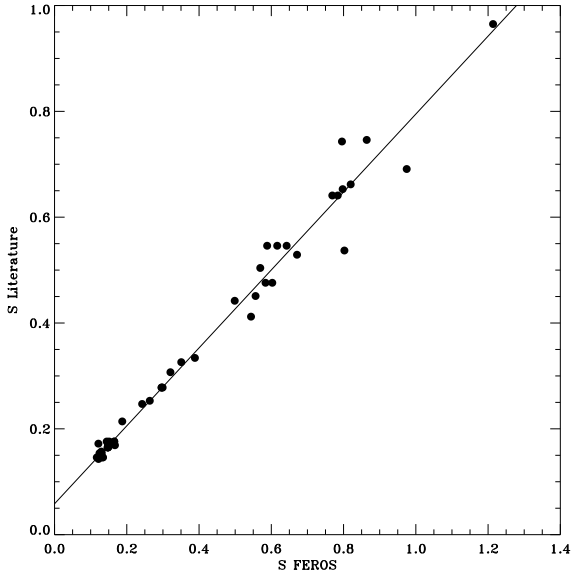}
\caption{Calibration of the instrumental S Index measured on FEROS spectra onto the standard Mt.~Wilson
scale}
\label{f:scalferos}
\end{figure}

\begin{figure}[h]
\includegraphics[width=8cm]{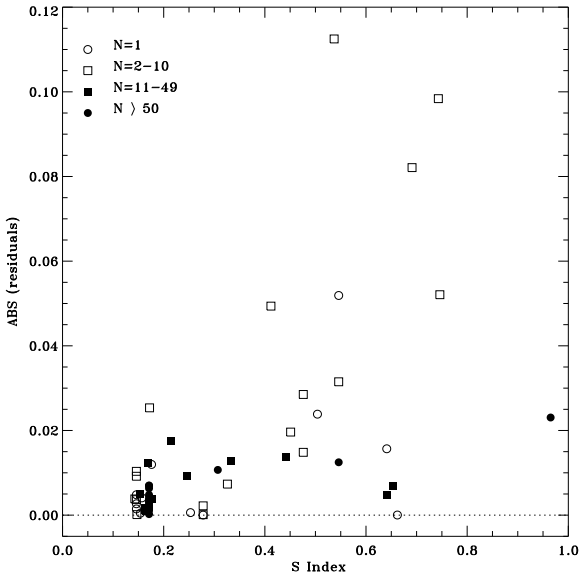}
\caption{Absolute value of the difference between calibrated FEROS S Index  and literature values as a function of 
literature S Index. Different symbols refer to stars with different number of measurements available in the
literature. The stars with the largest offsets all have small numbers of measurements in the literature, supporting
the intrinsic variability of chromospheric emission as the main source of the observed scatter in the calibration.}
\label{f:rescalferos}
\end{figure}

Instrumental S index was measured on FEROS spectra by integrating the
flux over rectangular bandpasses 1 \AA~ wide, as in \cite{2006A&A...454..553D}.
For calibration onto the standard Mount Wilson scale, we used 73 spectra
of 42 stars from the lists of  \cite{1995ApJ...438..269B} and \cite{2004ApJS..152..261W}. The relevant calibration stars were observed in the same 
nights  as were our program stars, and reduced in the same way. We determined that the inclusion of
additional calibration stars from other sources, which are typically characterized by
single-epoch measurements \citep[e.g.][]{1996AJ....111..439H}, increases the scatter of the calibration.
A linear fit between FEROS instrumental S index and the literature calibrated S index was found
to be fully adequate (Fig.~\ref{f:scalferos}). The rms of residuals from the calibration is about 5\%.  
To better understand the measurement errors, we plotted in Fig.~\ref{f:rescalferos} the absolute value
of the difference between our calibrated S index and the literature ones. 
The stars with larger differences are those that are more active and have a small
number of measurements in the calibration papers. This indicates that the intrinsic variability of chromospheric activity has  a significant impact on  our calibration.
For the low-activity, low-variability star $\tau$ Ceti, the observed rms of the S index for 15 spectra
is 0.0034  (2\%).
It should be further  noted that for stars with $v \sin i > 40$ km/s, the measured S index is underestimated,
as part of the H and K emission lies outside the defined integration windows \citep{2006A&A...454..553D}.
The calibrated S index was transformed into the chromospheric flux $\log R_{HK}$ following the prescriptions
by \citet{1984ApJ...279..763N}.

\section{Details on rotation analysis}
\label{a:rot}

We present here some additional details on the result of our search for rotation periods.
For the targets with newly 
determined rotation periods, or confirmed literature values, we list in Table D.1 the following: 
(1) the time interval spanned by each segment; (2) 
the number of V magnitude measurements; (3) the brightest V magnitude; (4) the light curve amplitude, which is measured 
as the amplitude of the sinusoid fit to the phased light curve; (5) the photometric accuracy; (6) the ratio of the average 
residuals from the fit on the light curve amplitude; (7) the rotation period; (8) the rotation uncertainty; (9) the normalized power 
of the periodogram peak, corresponding to the rotation period, and the power level corresponding to a 1\% false alarm 
probability; and (10) the sequential number of the timeseries segment (the number 1 indicates the complete un-segmented 
timeseries).

\begin{figure}[h]
\includegraphics[width=8.5cm]{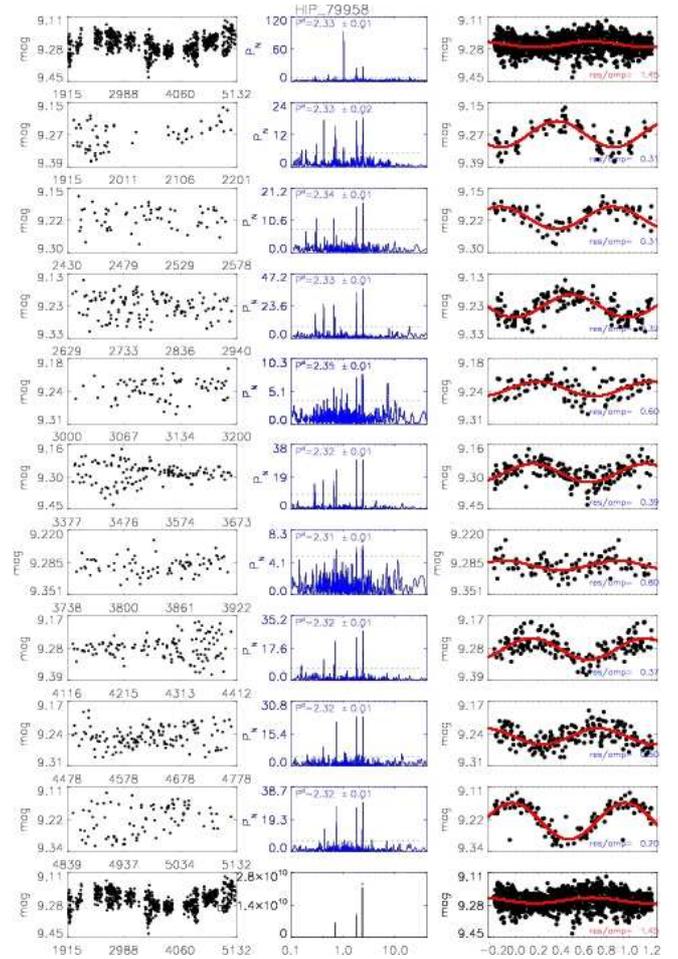}
\caption{Example of photometric analysis for the target HIP 79958.}
\label{f:rot_hip79958}
\end{figure}

Figure \ref{f:rot_hip79958} shows, as example, the results of our period search for HIP\,79958. 
In the left column we plot the V magnitude time series (complete in the top left panel, and segmented in the 
other lower panels). In the middle column, we plot the Lomb-Scargle periodograms with indication of the power 
level corresponding to the 1\% FAP, and the peak corresponding to the rotation period. In the right column, 
we plot the phased light curve with the rotation period. The initial epoch is the same for all the 
light curves, whereas the rotation period is the one found in the corresponding season. 
In the seasons where no period is detected, we used the rotation period found from the whole time series analysis. 
In the bottom panels we have multiplied all the periodograms and extracted the most significant periodicities. This represents 
 a different approach to reveal peaks that are low power, but persistent in each season. 
For all the targets that we newly determined, or confirmed, the rotation period, 
similar plots are available as online material.

\onecolumn
\include{tabella_total}

\end{document}